\documentclass[conference]{IEEEtran}
\usepackage{url}
\usepackage[cmex10]{amsmath}
\usepackage{amsthm}

\newtheorem{theorem}{Theorem}

\usepackage{changepage}
\usepackage[noend]{algorithmic}
\usepackage{algorithm2e}
\usepackage{multirow}
\usepackage{xcolor}
\usepackage{color}
\usepackage{times}
\usepackage{mathptmx}
\usepackage{mathtools}
\usepackage{mathrsfs}
\usepackage{enumerate}
\usepackage{amssymb}
\usepackage{subfigure, float}

\DeclareMathAlphabet{\mathcal}{OMS}{cmsy}{m}{n}

\newcommand{\nop}[1]{}

\newcommand{\todo}[1]{\textcolor{red}{[\sc #1]}}
\newcommand{\updates}[1]{\textcolor{black}{#1}}

\newcommand{\argmax}{\arg\!\max}

\newtheorem{lemma}{Lemma}
\newtheorem{corollary}{Corollary}

\usepackage{cite}

\ifCLASSINFOpdf
\else
\fi
\begin{document}
%
\title{Tracking Top-K Influential Vertices in Dynamic Networks}

\author{\IEEEauthorblockN{Yu Yang$^{1}$, Zhefeng Wang$^{2}$, Tianyuan Jin$^{2}$, Jian Pei$^{1}$ and Enhong Chen$^{2}$}
\IEEEauthorblockA{
$^{1}$Simon Fraser University, Burnaby, Canada\\
$^{2}$University of Science and Technology of China, Hefei, China\\
yya119@sfu.ca, \{zhefwang,jty123\}@mail.ustc.edu.cn, jpei@cs.sfu.ca, cheneh@ustc.edu.cn}
}

\maketitle

\begin{abstract}
Influence propagation in networks has enjoyed fruitful applications and has been extensively studied in literature. However, only very limited preliminary studies tackled the challenges in handling highly dynamic changes in real networks. In this paper, we tackle the problem of tracking top-$k$ influential vertices in dynamic networks, where the dynamic changes are modeled as a stream of edge weight updates. Under the popularly adopted linear threshold (LT) model and the independent cascade (IC) model, we address two essential versions of the problem: tracking the top-$k$ influential individuals and finding the best $k$-seed set to maximize the influence spread (Influence Maximization). We adopt the polling-based method and maintain a sample of random RR sets so that we can approximate the influence of vertices with provable quality guarantees. It is known that updating RR sets over dynamic changes of a network can be easily done by a reservoir sampling method, so the key challenge is to efficiently decide how many RR sets are needed to achieve good quality guarantees. We use two simple signals, which both can be accessed in $O(1)$ time, to decide a proper number of RR sets. We prove the effectiveness of our methods. For both tasks the error incurred in our method is only a multiplicative factor to the ground truth. For influence maximization, we also propose an efficient query algorithm for finding the $k$ seeds, which is one order of magnitude faster than the state-of-the-art query algorithm in practice. In addition to the thorough theoretical results, our experimental results on large real networks clearly demonstrate the effectiveness and efficiency of our algorithms.
\end{abstract}


%
\IEEEpeerreviewmaketitle

\section{Introduction}\label{sec:intro}

Influence propagation in large networks, such as social networks, has enjoyed fruitful applications.  There are two typical kinds of application scenarios. The first type is the well known influence maximization (IM) problem~\cite{kempe2003maximizing}, which finds a set of $k$ seed vertices that can collaboratively maximize the influence spread over the whole network. For example, in viral marketing, to promote a product in a social network, a company can use IM to use a small group of influential users to spread the influence in the social network.  The second type of application scenarios is to track the top-$k$ most influential nodes in a large network.  For example, consider cold-start recommendation in a social network, where we want to recommend to a new comer some existing users in a social network.  A new user may want to subscribe to the posts by some users in order to obtain hot posts (posts that are widely spread in the social network) at the earliest time. A strategy is to recommend to such a new comer some influential users in the current network. IM cannot find those influential users because IM assumes that all seed users have to be synchronized to spread the same content, while in reality online influential individuals often produce and spread their own contents in an asynchronized manner. The influential users needed in this scenario are those who have high individual influence.

More often than not, a network is highly dynamic.  For example, in a social network, each vertex is often a user and an edge captures the interaction from a user to another.  User interactions evolve continuously over time.  In an active social network, such as Twitter, Facebook, LinkedIn, Tencent WeChat, and Sina Weibo, the evolving dynamics, such as rich user interactions over time, indeed produce the most significant value. It is critical to capture the most influential users in an online manner.  To address the needs, we have to tackle two challenges at the same time, influence computation and dynamics in networks.

Although some aspects of influence propagation in large networks have been extensively investigated, such as IM on static networks~\cite{kempe2003maximizing, chen2013information, du2013scalable, borgs2014maximizing, tang2015influence,  chen2010scalableLT, goyal2011simpath, lucier2015influence, rossi2015spread}, there are only very limited studies on influence computation in dynamic networks~\cite{chen2015influential,aggarwal2012influential}, most being heuristic.  To the best of our knowledge,  Ohsaka~\textit{et~al.}~\cite{ohsaka2016dynamic} and we~\cite{yang2017tracking} are the first to address influence computation in dynamic networks with provable quality guarantees. However, in~\cite{ohsaka2016dynamic}, how to decide the maintained RR sets are enough to ensure good qualities of seed sets extracted is based on a flawed conclusion of an early version of~\cite{borgs2014maximizing}, which now is corrected in the latest version of~\cite{borgs2014maximizing}. Thus, the method in~\cite{ohsaka2016dynamic} is also heuristic.

We~\cite{yang2017tracking} gave solutions to tracking influential individuals with influence greater than a threshold $T$ and tracking top-$k$ influential individuals. Our solutions ensure that, with high probability, the recall is 100\% and the smallest influence of the returned set of individuals does not deviate from the true threshold (or the influence of the true $k$-th most influential individual) by an absolute value $\epsilon n$. A major disadvantage of~\cite{yang2017tracking} is that we need to carefully set the parameters to obtain meaningful results. If the threshold $T$ is too high, ~\cite{yang2017tracking} may return nothing. If the parameter $\epsilon$, which controls the tolerable absolute error in the top-$k$ task, is not set properly, $\epsilon n$ may be even greater than the influence of the $k$-th most influential individual, and~\cite{yang2017tracking} may return all vertices, which may be meaningless. Thus, to run algorithms in~\cite{yang2017tracking} properly, one needs some prior knowledge about the data, which is often not easy to get.

To tackle the obstinate fundamental challenges in influence computation in dynamic networks, in this study, we systematically tackle the two essential tasks of tracking top-$k$ influential vertices: (1) tracking the top-$k$ most influential vertices; and (2) supporting efficient influence maximization queries. For both tasks, our goal is to control the error incurred in our algorithms a multiplicative factor to the ground truth, so that even without prior knowledge about the data, we can still obtain meaningful results by setting a relative error threshold.

Similar to~\cite{ohsaka2016dynamic} and~\cite{yang2017tracking}, in this paper, we also maintain a sample of RR sets to approximate influence spread of vertices. Both~\cite{ohsaka2016dynamic} and~\cite{yang2017tracking} point out that updating existing RR sets against an update of the network can be easily done by a reservoir sampling method, and after updating, a critical step is to decide if the current sample size (number of RR sets) is proper to achieve good quality guarantees. Due to the high-speed updates in real networks, such decision has to be made efficiently. Meanwhile the decision should help us maintain as few RR sets as possible.

In this paper, we make several technical contributions. Our foremost contributions are at the theoretical side. We adopt two simple signals, which both can be accessed in $O(1)$ time, to decide a proper sample size for the two tasks. For the influential individual tracking task, our sample size is very close to the minimum sample size for estimating the $k$ largest individual influence with a small relative error. For influence maximization, our sample size is not big when the seed set size $k$ of IM queries is small but becomes too large to use when $k$ is large. Thus, we also give a practical solution that normally is more efficient than the method in~\cite{ohsaka2016dynamic}, and also effective in practice. In addition to the thorough theoretical results, our experimental results on real networks clearly demonstrate the effectiveness and efficiency of our algorithms. Our methods are more efficient than the state-of-the-art methods as the baselines.  The largest network used in the experiments has $41.7$ million vertices, almost $1.5$ billion edges and 0.3 billion edge updates.

The rest of the paper is organized as follows. We review the related work in Section~\ref{sec:related}. In Section~\ref{sec:pre}, we recall the Linear Threshold model and the Independent Cascade model, review the polling-based method for computing influence spread on dynamic networks, and introduce two major tools from probability theory that we will use in this paper. In Section~\ref{sec:individual}, we tackle the problem of tracking top-$k$ influential individuals with limited relative error rates. In Section~\ref{sec:im}, we analyze how to maintain a number of RR sets to support influence maximization queries with good quality guarantees. We also devise an efficient implementation of the greedy algorithm for IM queries. We report the experimental results in Section~\ref{sec:exp} and conclude the paper in Section~\ref{sec:con}.

\nop{For the interest of space, we skip all mathematical proofs, which can be found in the full version of this paper~\footnote{\url{http://www.sfu.ca/~yya119/papers/Tracking_Top-K.pdf}}.} 

\section{Related Work}\label{sec:related}

In this section, we briefly review the state of the art about influence computation on both static and dynamic networks.

\subsection{Influence Computation in Static Networks}
\nop{
Domingos~\textit{et~al.}~\cite{domingos2001mining} proposed to take advantage of peer influence between users in social networks for marketing. Kempe~\textit{et~al.}~\cite{kempe2003maximizing} formulated the problem using two discrete influence models, namely the Independent Cascade model and the Linear Threshold model. Since then, influence computation, especially influence maximization, has drawn much attention from both academia and industry~\cite{chen2013information, du2013scalable, borgs2014maximizing, tang2015influence, chen2010scalableLT, lucier2015influence, rossi2015spread}.
}

One major difficulty in influence computation is that computing influence spread is \#P-hard under both the LT and IC models~\cite{chen2010scalable,chen2010scalableLT}.
Besides some heuristic methods for estimating influence spread~\cite{goyal2011simpath, chen2010scalableLT, cohen2014sketch}, recently, a polling-based method~\cite{borgs2014maximizing, tang2015influence, tang2014influence} was proposed for influence maximization (IM) under general triggering models. The key idea is to use some ``Reverse Reachable'' (RR) sets~\cite{tang2015influence, tang2014influence} to approximate the true influence spread of vertices. The error of approximation can be bounded with a high probability if the number of RR sets is large enough. Nguyen~\textit{et~al.}~\cite{nguyen2016stop} exploits the Stopping Rule~\cite{dagum2000optimal} for sampling to further improve the efficiency of the polling based IM algorithms.

\subsection{Heuristics for Tracking Influential Vertices}

For evolving networks~\cite{leskovec2005graphs}, rather than re-computing from scratch, incremental algorithms are more desirable in analytic tasks. Some studies propose heuristic methods for influence computation on dynamic networks. Aggarwal~\textit{et~al.}~\cite{aggarwal2012influential} explored how to find a set of vertices that has the highest influence within a time window $[t_0,t_0+h]$ by modeling influence propagation as a non-linear system. Chen~\textit{et~al.}~\cite{chen2015influential} investigated incrementally updating a seed set for influence maximization under the Independent Cascade model. Both~\cite{aggarwal2012influential,chen2015influential} cannot deal with real-time updates of dynamic networks, since in~\cite{aggarwal2012influential} the snapshot of the network in the time window $[t_0,t_0+h]$ needs to be extracted before the mining process, and~\cite{chen2015influential} has the time complexity $O(m)$ to deal with one update of the network, where $m$ is the number of edges.

Recently, Ohsaka~\textit{et~al.}~\cite{ohsaka2016dynamic} studied maintaining a collection of RR sets $\mathcal{R}$ over a stream of network updates under the IC model such that $(1-1/e-\epsilon)$-approximation IM queries can be achieved with probability at least $1-\frac{1}{n}$. Specifically, ~\cite{ohsaka2016dynamic} maintains the invariant $C(\mathcal{R})=\Theta(\frac{(m+n)\log{n}}{\epsilon^3})$, where $C(\mathcal{R})$ is the number of edges traveled for generating all RR sets in $\mathcal{R}$. Unfortunately, setting $C(\mathcal{R})=\Theta(\frac{(m+n)\log{n}}{\epsilon^3})$ is based on a flawed conclusion in an early version of~\cite{borgs2014maximizing} and the correct one is to set $C(\mathcal{R})=\Theta(k\frac{(m+n)\log{n}}{\epsilon^2})$. Thus, ~\cite{ohsaka2016dynamic} does not have provable quality guarantees. Moreover, even $\Theta(\frac{(m+n)\log{n}}{\epsilon^3})$ is a huge number so in the experiments of~\cite{ohsaka2016dynamic}, $C(\mathcal{R})$ is empirically set to $32(m+n)\log{n}$, which is far less than $\Theta(\frac{(m+n)\log{n}}{\epsilon^3})$ but still a very large number.

\subsection{Tracking Influential Vertices with Quality Guarantees}
Our previous work~\cite{yang2017tracking} is the first to track influential vertices in dynamic networks with quality guarantees. We~\cite{yang2017tracking} tackled two versions of tracking influential vertices, namely tracking vertices with influence greater than a threshold $T$ and tracking the top-$k$ influential vertices. With high probability, ~\cite{yang2017tracking} achieves 100\% recall and a bounded error of all false positive vertices, where the error of a false positive vertex $u$ is the threshold (the $k$-th largest influence spread in the top-$k$ task) minus the influence of $u$. The major disadvantage of~\cite{yang2017tracking} is that the parameters have to be carefully set, otherwise the returned result may be overwhelming and undesirable. More specifically, for the threshold-based task, if the threshold $T$ is set to much larger than the greatest individual influence, the algorithm may return nothing. For the top-$k$ task, since the error in~\cite{yang2017tracking} is an absolute error $\epsilon n$, and we do not know the value of $I^k$ beforehand, where $I^k$ is the $k$-th greatest individual influence, one may set $\epsilon n$ to even greater than $I^k$ and then~\cite{yang2017tracking} may return all vertices.

Recently, Wang~\textit{et~al.}~\cite{wang2017real} proposed to maintain an $\frac{\epsilon(1-\beta)}{2}$-approximation solution of influence maximization dynamically. However, rather than a propagation model like the IC or the LT model, ~\cite{wang2017real} employs the simple reachability in a social action graph as influence. Moreover, the input of~\cite{wang2017real} is a stream of social actions which is different from edge propagation probability updates in~\cite{ohsaka2016dynamic, yang2017tracking} and this paper. Thus, ~\cite{wang2017real} cannot be used to track influential vertices under propagation models like the IC or LT model.

\section{Preliminaries}\label{sec:pre}
In this section, we review the two most widely used influence models and the polling method for estimating influence spread.  We also briefly illustrate how to update RR sets in the polling method over a stream of edge weight updates. The two major probabilistic methods used to derive sample size in this paper are introduced at the end of this section. For readers' convenience, Table~\ref{tab:notation} lists some frequently used notations.

\begin{table}[t]
\centering
\begin{tabular}{|c|p{55mm}|}
\hline
Notation & Description \\ \hline
$I_u$ & The influence spread of vertex $u$ \\ \hline
$\mathcal{R}$ & A set of $M$ random RR sets \\ \hline
$\mathcal{D}(u)$ & The degree of $u \in V$ in $\mathcal{R}$, or equivalently the number of RR sets containing $u$ \\ \hline
$\mathcal{D}^*$ & $\mathcal{D}^* = \max_{u \in V}{\mathcal{D}(u)}$ \\ \hline
$\mathcal{D}(S)$ & The degree of $S$ in $\mathcal{R}$, the number of RR sets containing at least one vertex from the set $S$ \\ \hline
$I^k$ & Influence spread of the $k-$th most influential individual vertex \\ \hline
$S_k$ & The $k$-seed set produced by running the greedy algorithm (Algorithm~\ref{alg:greedy}) on RR sets \\ \hline
$S^*_k$ & The optimal $k$-seed set, $S^*_k=\argmax_{S \subseteq V, |S|=k}{I(S)}$ \\ \hline
$\Upsilon(\epsilon, \delta)$ & $\Upsilon(\epsilon, \delta)=\frac{4(e-2)\ln{\frac{2}{\delta}}}{\epsilon^2}$\\ \hline
$\Upsilon_1(\epsilon, \delta)$ & $\Upsilon_1(\epsilon, \delta)=1+(1+\epsilon)\Upsilon(\epsilon,\delta)$ \\ \hline
\end{tabular}
\caption{Frequently used notations.}
\label{tab:notation}
\end{table}

\subsection{The LT and IC Models}

Consider a directed social network $G = \langle V,E,w \rangle$, where $V$ is a set of vertices, $E \subseteq V \times V$ is a set of edges, and each edge $(u,v) \in E$ is associated with an influence weight $w_{uv} \in [0,+\infty)$.

\subsubsection{The LT Model}
In the Linear Threshold (LT) model~\cite{kempe2003maximizing}, each vertex $v \in V$ also carries a weight $w_v$, which is called the \emph{self-weight} of $v$. Denote by $W_v=w_v+\sum_{u \in N^{in}(v)}{w_{uv}}$ the total weight of $v$, where $N^{in}(v)$ is the set of $v$'s in-neighbors. The \emph{influence probability} of an edge $(u,v)$ is $p_{uv}=\frac{w_{uv}}{W_v}$. Clearly, for $v \in V$, $\sum_{u \in N^{in}(v)}{p_{uv}} \leq 1$.

In the LT model, given a seed set $S \subseteq V$, the influence is propagated in $G$ as follows. First, every vertex $u$ randomly selects a threshold $\lambda_u \in [0,1]$, which reflects our lack of knowledge about users' true thresholds. Then, influence propagates iteratively. Denote by $S_i$ the set of vertices that are active in step $i$ $(i=0, 1, \ldots)$. Set $S_0=S$. In each step $i \geq 1$, an inactive vertex $v$ becomes active if
$
    \sum_{u \in N^{in}(v) \cap S_{i-1}}{p_{uv}} \geq \lambda_v
$.
The propagation process terminates in step $t$ if $S_t=S_{t-1}$.

Let $I(S)$ be the expected number of vertices that are finally active when the seed set is $S$. We call $I(S)$ the \emph{influence spread} of $S$. Let $I_u$ be the influence spread of a single vertex $u$. Kempe~\textit{et~al.}~\cite{kempe2003maximizing} proved that the LT model is equivalent to a ``live-edge'' process where each vertex $v$ picks at most one incoming edge $(u,v)$ with probability $p_{uv}$.  Consequently, $v$ does not pick any incoming edges with probability $1-\sum_{u \in N^{in}(v)}{p_{uv}}=\frac{w_v}{W_v}$. All edges picked are ``live'' and the others are ``dead''. Then $I(S)$ is the expected number of vertices reachable from $S \subseteq V$ through live edges.

\subsubsection{The IC Model}

In the Independent Cascade (IC) model~\cite{kempe2003maximizing}, for an edge $(u,v)$, $w_{u,v}$ is the \emph{influence probability} ($0 \leq w_{uv} \leq 1$). Given a seed set $S$, the influence propagates in $G$ in a way different from the LT model. Denote by $S_i$ the set of vertices that are active in step $i$ $(i=0, 1, \ldots)$. Set $S_0=S$. In each step $i \geq 1$, each vertex $u$ that is newly activated in step $i-1$ has a single chance to influence its inactive neighbor $v$ with an independent probability $w_{uv}$. The  propagation process terminates in step $t$ when $S_t=S_{t-1}$.

Similar to the LT model, the influence spread $I(S)$ is the expected number of vertices that are finally active when the seed set is $S$. The equivalent ``live-edge'' process~\cite{kempe2003maximizing} of the IC model is to keep each edge $(u,v)$ with a probability $w_{uv}$ independently. All kept edges are ``live'' and the others are ``dead''. Then $I(S)$ is the expected number of vertices reachable from $S$ via live edges.

\subsection{Polling Estimation of Influence Spread}\label{sec:poll}
Computing influence spread is \#P-hard under both the LT model and the IC model~\cite{chen2010scalableLT, chen2010scalable}. Recently, a polling-based method~\cite{borgs2014maximizing, tang2014influence, tang2015influence} was proposed for approximating influence spread of triggering models~\cite{kempe2003maximizing} like the LT model and the IC model. Here we briefly review the polling method for computing influence spread.

Given a social network $G=\langle V,E,w \rangle$, a \emph{poll} is conducted as follows: we pick a vertex $v \in V$ in random and then try to find out which vertices are likely to influence $v$. We run a Monte Carlo simulation of the equivalent ``live-edge'' process.  The vertices that can reach $v$ via live edges are considered as the potential influencers of $v$. The set of influencers found by each poll is called a \emph{random RR} (for \emph{Reverse Reachable}) \emph{set}.

Let $\mathcal{R}=\{R_1, R_2, ..., R_M\}$ be a set of random RR sets generated by $M$ polls. For a set of vertices $S$, denote by $\mathcal{D}(S)$ the degree of $S$ in $\mathcal{R}$, which is the number of RR sets that contain at least one vertex in $S$. By the linearity of expectation, $\frac{n\mathcal{D}(S)}{M}$ is an unbiased estimator of $I(S)$~\cite{borgs2014maximizing}. Thus, in the polling method, $\frac{n\mathcal{D}(S)}{M}$ is used to approximate $I(S)$. How to decide the sample size $M$ often is the key point for efficient computation in tasks related to influence.

\subsection{Updating RR Sets on Dynamic Networks}\label{sec:update}
We~\cite{yang2017tracking} modeled the updates of an influence network as a stream of edge weight updates, where an edge weight update is depicted as a 5-tuple $(u,v,+/-,\Delta,t)$. $(u,v)$ denotes the edge to be updated, $+$ and $-$, respectively, indicate whether $w_{uv}$ is increased or decreased, $\Delta>0$ is the amount of change and $t$ is the time stamp. ~\cite{yang2017tracking} pointed out that a RR set under the LT model is a random path, and both~\cite{ohsaka2016dynamic} and~\cite{yang2017tracking} illustrated that a RR set in the IC model is a random connected component. When an edge weight update $(u,v,+/-,\Delta,t)$ comes, Algorithm~\ref{alg:framework} gives a framework that updates RR sets correspondingly so that $\frac{n\mathcal{D}(S)}{M}$ remains an unbiased estimator of $I(S)$ for any $S \subseteq V$.

\begin{algorithm}[t]\small
\caption{Framework of Updating RR Sets}
\label{alg:framework}
\begin{algorithmic}[1]
\STATE retrieve RR Sets affected by the updates of the graph \label{line:retrieve}
\STATE update retrieved RR sets \label{line:update}
\IF {the current RR sets are insufficient} \label{line:size1}
	\STATE add new RR sets
\ELSE
	\IF {the current RR sets are redundant} \label{line:size2}
		\STATE delete the redundant RR sets
	\ENDIF
\ENDIF
\end{algorithmic}
\end{algorithm}

Line~\ref{line:retrieve} can be easily achieved by maintaining an inverted index on all RR sets so that we can access all RR sets passing a specific vertex. Line~\ref{line:update} can also be easily tackled by a method similar to Reservoir Sampling~\cite{vitter1985random}, no matter the influence model is LT or IC. The key point of Algorithm~\ref{alg:framework} is Lines~\ref{line:size1} and~\ref{line:size2}, that is, how to decide the proper current sample size $M$. $M$ varies in different tasks of influence computation. Also, ~\cite{yang2017tracking} analyzed that the cost of dealing with an edge weight update is proportional to the sample size $M$ for both the LT and IC model. Thus, we remark that \textbf{efficiently deciding a small and proper sample size $M$} is the core of influence computation tasks on dynamic networks.

\subsection{Martingale Inequalities and Stopping Rule Theorem}\label{sec:inequality}
In this section, we introduce two major tools from probability theory that help us derive the proper sample size $M$.

Let $R_1$, $R_2$, ..., $R_M$ be a sequence of random RR sets generated by $M$ polls, where $M$ can also be a random variable. Tang~\textit{et~al.}~\cite{tang2015influence} proved that the corresponding sequence $Z_1=\sum_{i=1}^{1}{(x_i-\frac{I(S)}{n})}$, $Z_2=\sum_{i=1}^{2}{(x_i-\frac{I(S)}{n})}$, ..., $Z_M=\sum_{i=1}^{M}{(x_i-\frac{I(S)}{n})}$ is a martingale~\cite{chung2006concentration}, where $x_i=1$ if $S \cap RR_i \neq \emptyset$ and $x_i=0$ otherwise. We have $E[\sum_{i=1}^M{x_i}]=E[\mathcal{D}(S)]=\frac{MI(S)}{n}$. The following results~\cite{tang2015influence} show how $E[\sum_{i=1}^M{x_i}]$ is concentrated around $\frac{M\cdot I(S)}{n}$, when \updates{variables $Z_1,Z_2,...,Z_M$} may be weakly dependent due to the stopping condition on $M$. 

\begin{corollary}[\cite{tang2015influence}]\label{cor:c1}
For any $\xi>0$,
\small{$$
\begin{aligned}
    \textup{Pr}\Big[\sum_{i=1}^M{x_i}-Mp \geq \xi Mp \Big] & \leq & \textup{exp}\Big ( -\frac{\xi^2}{2+\frac{2}{3} \xi} Mp \Big)\\
    \textup{Pr}\Big[\sum_{i=1}^M{x_i}-Mp \leq -\xi Mp \Big] & \leq & \textup{exp}\Big ( -\frac{\xi^2}{2} Mp \Big)
\end{aligned}
$$}
where $p=\frac{I(S)}{n}$.
\end{corollary}

Besides the above martingale inequalities, we also introduce a Stop-Rule Sampling method for obtaining an $(\epsilon,\delta)$ estimation\footnote{$\hat{I}(S)$ is an $(\epsilon,\delta)$-estimation of $I(S)$ if $\textup{Pr}\{|\hat{I}(S)-I(S)| \leq \epsilon I(S)\} \geq 1-\delta$.} of the expectation of a Bernoulli random variable.

Define $\Upsilon(\epsilon,\delta)=\frac{4(e-2)\ln{\frac{2}{\delta}}}{\epsilon^2}$ and $\Upsilon_1(\epsilon,\delta)=1+(1+\epsilon)\Upsilon(\epsilon,\delta)$. Dagum~\textit{et al.}~\cite{dagum2000optimal} proposed a Stopping Rule algorithm (Algorithm~\ref{alg:sra}) to obtain an $(\epsilon,\delta)$ estimation of the mean of a Bernoulli random variable.

\begin{algorithm}[t]\small
\caption{Stopping Rule Algorithm}
\label{alg:sra}
\KwIn{$(\epsilon,\delta)$}
\KwOut{$\hat{\mu}_Z$}
\begin{algorithmic}[1]
\STATE $\Upsilon_1(\epsilon,\delta)=1+(1+\epsilon)\Upsilon(\epsilon,\delta)$
\STATE $M \leftarrow 0$, $A \leftarrow 0$
\WHILE {$A < \Upsilon_1$}
	\STATE $M \leftarrow M+1$; $A \leftarrow A+Z_M$
\ENDWHILE
\RETURN $\hat{\mu}_Z \leftarrow \Upsilon_1/M$
\end{algorithmic}
\end{algorithm}

\begin{corollary}[Stopping Rule Theorem~\cite{dagum2000optimal}]\label{cor:sr}
	Let $Z$ be a Bernoulli random variable with $\mu_Z=E[Z] > 0$. Let $\hat{\mu}_Z$ be the estimate produced and let $M$ be the number of experiments that the Stopping Rule algorithm runs with respect to $Z$ on the input $\epsilon$ and $\delta$. Then, (1) $\textup{Pr}\{\hat{\mu}_Z \geq (1-\epsilon)\mu_Z\} \geq 1-\frac{\delta}{2}$ and $\textup{Pr}\{\hat{\mu}_Z \leq (1+\epsilon)\mu_Z\} \geq 1-\frac{\delta}{2}$, (2) $E[M] \leq \Upsilon_1(\epsilon,\delta)/\mu_Z$.
\end{corollary}

Corollary~\ref{cor:sr} implies that with probability at least $1-\delta$, $\frac{\Upsilon_1(\epsilon,\delta)}{(1+\epsilon)\mu_Z} \leq M \leq \frac{\Upsilon_1}{(1-\epsilon)\mu_Z}$. It is obvious that $\Upsilon(\epsilon,\delta)$ is monotonically decreasing with respect to $\epsilon$, thus $\lceil \Upsilon_1 \rceil/M$ is a slightly better estimate of $\mu_Z$ because $\lceil \Upsilon_1(\epsilon,\delta) \rceil=1+(1+\bar{\epsilon})\Upsilon(\bar{\epsilon},\delta)$, where $\bar{\epsilon} \leq \epsilon$.

Note that in the Algorithm~\ref{alg:sra}, in the last experiment the random variable $Z_M$ must be 1. Thus, the stopping time of Algorithm~\ref{alg:sra} is the first time when the number of positive samples equals $\lceil \Upsilon_1(\epsilon,\delta) \rceil$. We prove that by relaxing this condition a little bit, that is, we only need a sufficient number of positive samples but the last sample does not have to be 1, we still can bound the error in the estimation tightly.
\begin{theorem}\label{th:sample}
	Let $Z$ be a Bernoulli random variable with $\mu_Z=E[Z] > 0$. Let $Z_1,Z_2,...$ be independently and identically distributed according to $Z$. Let $M$ be a stopping time ($M$ is a random variable) for this sequence. Given $(\epsilon,\delta)$, if $\sum_{i=1}^M{Z_i}=\lceil \Upsilon_1(\epsilon,\delta) \rceil$, then $\textup{Pr}\{\frac{\sum_{i=1}^M{Z_i}}{M} \leq (1+\epsilon)\mu_Z\} \geq 1-\frac{\delta}{2}$ and $\textup{Pr}\{\frac{\sum_{i=1}^M{Z_i}}{M} \geq\frac{\lceil \Upsilon_1(\epsilon,\delta) \rceil}{\lceil \Upsilon_1(\epsilon,\delta) \rceil+1}(1-\epsilon)\mu_Z\} \geq 1-\frac{\delta}{2}$.
\end{theorem}

As illustrated, the random variable $x_i$ ($x_i=1$  if $S \cap R_i \neq \emptyset$ and $x_i=0$ otherwise) is clearly a Bernoulli random variable with mean $\frac{I(S)}{n}$. Thus, we can use Theorem~\ref{th:sample} to analyze the quality of estimations of influence spreads based on the maintained RR sets.

\section{Tracking Top-$k$ Individuals}\label{sec:individual}

\subsection{Algorithm}\label{ssec:alg_ind}
First of all, tracking influential individuals is not an instance of the Heavy Hitters problem~\cite{cormode2008finding}. Detailed reasons are illustrated in~\cite{yang2017tracking}.

Due to the \#P-hardness of computing influence spread~\cite{chen2010scalable, chen2010scalableLT}, it is unlikely that we can find in polynomial time the exact set of top-$k$ influential individual vertices. Thus, we turn to algorithms that allow controllable small errors. Specifically, we ensure that the recall of the set of vertices found by our algorithm is 100\% and we tolerate some false positive vertices. Moreover, the influence spreads of those false positive vertices should take a high probability to have a lower bound that is not much smaller than $I^k$, the influence spread of the $k$-th most influential vertex. We call $\max_{u \in S}{(I^k-I_u)}$ the error of $S$, where $S$ is returned by a top-$k$ influential vertices finding algorithm.

We give a simple solution to track top-$k$ individuals by a collection of RR sets $\mathcal{R}$. We split $\mathcal{R}$ into two disjoint parts $\mathcal{R}_1$ and $\mathcal{R}_2$.
$\mathcal{R}_1$ is for deriving an upper bound and a lower bound of $I^k$, and a proper sample size of $\mathcal{R}_2$. We use $\mathcal{R}_2$ and the lower bound of $I^k$ to return a set of influential individuals. Denote by $\mathcal{D}_1(u)$ and $\mathcal{D}_2(u)$ the degrees of $u$ in $\mathcal{R}_1$ and $\mathcal{R}_2$, respectively. Let $M_1=|\mathcal{R}_1|$ and $M_2=|\mathcal{R}_2|$. Our algorithm works as follows,
\begin{enumerate}
	\item Update the RR sets in $\mathcal{R}=\mathcal{R}_1 \cup \mathcal{R}_2$ when the network structure changes using the method in~\cite{yang2017tracking}.
	\item Maintain the invariant $\mathcal{D}_1^k=\lceil \Upsilon_1(\epsilon,\frac{\delta}{n}) \rceil$, where $\mathcal{D}_1^k$ is the $k$-th largest degree of the vertices in $\mathcal{R}_1$.
	\item Also maintain the invariant $M_2=M_1$.
	\item \updates{When a top-$k$ individuals query is issued}, return all vertices $u$ such that $\mathcal{D}_2(u) \geq T=\frac{1-\epsilon}{1+\epsilon}\mathcal{D}_1^k$.
\end{enumerate}

In our method, $\mathcal{D}_1^k$ is the signal to decide if the current sample size is appropriate. When $\mathcal{D}_1^k$ is not $\lceil \Upsilon_1(\epsilon,\frac{\delta}{n}) \rceil$, we adjust the sizes of $\mathcal{R}_1$ and $\mathcal{R}_2$ by adding or deleting some RR sets. In the rest of this section, we show that our algorithm can achieve the goals of 100\% recall and an $O(\epsilon)$ relative error with high probability. We also propose an efficient way to retrieve $\mathcal{D}_1^k$ in $O(1)$ time such that whether the current sample size is proper can be efficiently decided. 

\subsection{Analysis of Sample Size}\label{ssec:size_ind}

The key point is to show that $M_2$, the sample size of $\mathcal{R}_2$, is enough to achieve our goals.

Our analysis consists of two major steps. The first step is to bound $I^k$ using $\mathcal{D}_1^k$. Once we can bound $I^k$ within a small range, we can set a safe threshold $T$ on $\mathcal{D}_2(v)$, the degree of $v$ in $\mathcal{R}_2$ (or equivalently the number of RR sets containing $v$ in $\mathcal{R}_2$), to find the top-$k$ influential individuals and filter out the vertices that very likely are not ranked within top-$k$. As introduced above, we set the threshold $T=\frac{(1-\epsilon)}{1+\epsilon} \mathcal{D}_1^k$. The second step is to use the upper bound and the lower bound of $I^k$ to derive the error rate of the false positive vertices, that is, how much the influence of a false positive vertex may be smaller than $I^k$. 

First, we prove that for every $v \in V$, $\frac{n\mathcal{D}_1(v)}{M_1}$ can be used to bound its influence $I_v$ with high probability. For each vertex $v$, we can calculate $\epsilon_v$ such that $\mathcal{D}_1(v)=\Upsilon_1(\epsilon_v,\frac{\delta}{n})$. We split the vertices in $V$ into two parts, $V_1$ and $V_2$, where $V_1=\{v \mid \epsilon_v \leq 0.6\}$ and $V_2=\{v \mid \epsilon_v > 0.6\}$. We prove that with high probability for every $v \in V_1$, its influence $I_v$ has an upper bound and a lower bound, and for every $v \in V_2$, $I_v$ has an upper bound. \nop{The reason we do not care the lower bound of $I_v$ when $v \in V_2$ is that \updates{if $v \in V_2$, the lower bound of $I_v$ is not utilized in our analysis.}}

\begin{lemma}\label{lemma:rnd}
	Suppose in $\mathcal{R}_1$, $\mathcal{D}_1^k=\lceil \Upsilon_1(\epsilon,\frac{\delta}{n}) \rceil$. For any vertex $v \in V$, we calculate a corresponding $\epsilon_v$ such that $\mathcal{D}_1(v)=\Upsilon_1(\epsilon_v,\frac{\delta}{n})$. When $\epsilon \leq \frac{1}{3}$ and $\delta \leq \frac{1}{4}$, (1) if $\epsilon_v \leq 0.6$, $\textup{Pr}\{I_v \geq \frac{n\mathcal{D}_1(v)}{(1+\epsilon_v)M_1}\} \geq 1-\frac{\delta}{2n}$ and $\textup{Pr}\{I_v \leq \frac{n(\mathcal{D}_1(v)+1)}{(1-\epsilon_v)M_1}\} \geq 1-\frac{\delta}{2n}$, (2) if $\epsilon_v > 0.6$, $\textup{Pr}\{ I_v < \frac{n\mathcal{D}_1^k}{M_1} \} \geq 1-\frac{\delta}{2n}$.
\end{lemma}

We now derive a lower bound of $I^k$ when $\mathcal{D}_1^k=\lceil \Upsilon_1(\epsilon,\frac{\delta}{n}) \rceil$ in our maintained $\mathcal{R}_1$. The intuition is if we can find at least $k$ vertices whose influence spreads probably are no smaller than $T$, then $T$ is a lower bound of $I^k$ with high probability.
\begin{lemma}\label{lemma:lower}
    If $\epsilon \leq \frac{1}{3}$, $\delta \leq \frac{1}{4}$, and $\mathcal{D}_1^k=\lceil \Upsilon_1(\epsilon,\frac{\delta}{n}) \rceil$, then $\textup{Pr}\{I^k \geq \frac{n\mathcal{D}_1^k}{(1+\epsilon)M_1}\} \geq 1-\frac{\delta}{2}$.
\end{lemma}

Now we show that with high probability, $I^k \leq \frac{n(\mathcal{D}_1^k+1)}{(1-\epsilon)M_1}$. Similar to Lemma~\ref{lemma:lower}, the intuition is that with high probability, there are at least $n-k+1$ vertices whose influence spreads are no greater than $\frac{n(\mathcal{D}_1^k+1)}{(1-\epsilon)M_1}$.

\begin{lemma}\label{lemma:upper}
    Suppose $\epsilon \leq \frac{1}{3}$ and $\delta \leq \frac{1}{4}$. If $\mathcal{D}_1^k=\lceil \Upsilon_1(\epsilon,\frac{\delta}{n}) \rceil$, we have $\textup{Pr}\{I^k \leq \frac{n(\mathcal{D}_1^k+1)}{(1-\epsilon)M_1}\} \geq 1-\frac{\delta}{2}$.
\end{lemma}

Putting Lemmas~\ref{lemma:lower} and~\ref{lemma:upper} together, we have the following.

\begin{corollary}\label{cor:bound_Ik}
When $\epsilon \leq \frac{1}{3}$ and $\delta \leq \frac{1}{4}$, if $\mathcal{D}_1^k=\lceil \Upsilon_1(\epsilon,\frac{\delta}{n}) \rceil$, with probability at least $1-\delta$, $\frac{n\mathcal{D}_1^k}{(1+\epsilon)M_1} \leq I^k \leq \frac{n(\mathcal{D}_1^k+1)}{(1-\epsilon)M_1}$.
\end{corollary}

Note that $\textup{Pr}\{I^k \geq \frac{n\mathcal{D}_1^k}{(1+\epsilon)M_1}\} \geq 1-\frac{\delta}{2}$ implies $\textup{Pr}\{M_1 \geq \frac{n\lceil \Upsilon_1(\epsilon,\frac{\delta}{n}) \rceil}{(1+\epsilon)I^k}\} \geq 1-\frac{\delta}{2}$. Recall that in our algorithm, we make $|\mathcal{R}_2|=M_2=M_1$. Thus, $\textup{Pr}\{M_2 \geq \frac{n\lceil \Upsilon_1(\epsilon,\frac{\delta}{n}) \rceil}{(1+\epsilon)I^k}\} \geq 1-\frac{\delta}{2}$. We show that when $M_2 \geq \frac{n\lceil \Upsilon_1(\epsilon,\frac{\delta}{n}) \rceil}{(1+\epsilon)I^k}$, for every true top-$k$ vertex $u$, it is very likely that $\frac{n\mathcal{D}_2(u)}{M_2}$ is not much smaller than $I^k$. We also show that, for every vertex $u$ such that $I_u$ is sufficiently smaller than $I^k$, its estimation $\frac{n\mathcal{D}_2(u)}{M_2}$ does not deviate from the expectation $I_u$ by a value decided by $I^k$.

\begin{lemma}\label{lemma:deviate}
    For $M_2 \geq \frac{n\lceil \Upsilon_1(\epsilon,\frac{\delta}{n}) \rceil}{(1+\epsilon)I^k}$ RR sets in $\mathcal{R}_2$, with probability at least $1-\frac{\delta}{2}$, we have (1) for every $v$ such that $I_v \geq I^k$, $\frac{n\mathcal{D}_2(v)}{M_2} \geq (1-\epsilon)I^k$; and (2) for every $v$ such that $I_v \leq (1-2\epsilon) I^k$, $\frac{n\mathcal{D}_2(v)}{M_2} \leq I_v+\epsilon I^k$.
\end{lemma}

Based on Corollary~\ref{cor:bound_Ik} and Lemma~\ref{lemma:deviate}, we have the following theorem that shows the effectiveness of our algorithm.
\begin{theorem}\label{th:individual}
When $\epsilon \leq \frac{1}{3}$ and $\delta \leq \frac{1}{4}$, our algorithm described in Section~\ref{ssec:alg_ind} returns a set of vertices $S$ such that all real top-$k$ vertices are included, and $\min_{u \in S}{I_u} \geq [(1-\frac{\epsilon^2}{B})\frac{(1-\epsilon)^2}{1+\epsilon}-\epsilon]I^k \geq (1-\frac{61}{15}\epsilon)I^k$, where $B=4(e-2)\ln{\frac{2n}{\delta}}$.
\end{theorem}

Theorem~\ref{th:individual} shows the effectiveness of our method. Note that we prove $\min_{u \in S}{I_u} \geq (1-\frac{61}{15}\epsilon)I^k$ is just for showing there is a relative error bound. In practice, we use $[(1-\frac{\epsilon^2}{B})\frac{(1-\epsilon)^2}{1+\epsilon}-\epsilon]$ to calculate the upper bound of relative error, since it is tighter than $1-\frac{61}{15}\epsilon$.

\smallskip
\noindent \textbf{Remark.} One may ask why we do not apply Lemma~\ref{lemma:deviate} on $\mathcal{R}_1$ and return all vertices $u$ such that $\mathcal{D}_1(u) \geq T$. In such a case, we do not need $\mathcal{R}_2$. However, we cannot do this because probabilistic support (implication) is not transitive~\cite{shogenji2003condition}. Even when we know $M_1 \geq \frac{n\lceil \Upsilon_1(\epsilon,\frac{\delta}{n}) \rceil}{(1+\epsilon)I^k}$ with high probability, we cannot establish the conditions (1) and (2) in Lemma~\ref{lemma:deviate} for $\mathcal{R}_1$. Lemma~\ref{lemma:deviate} holds for $\frac{n\lceil \Upsilon_1(\epsilon,\frac{\delta}{n}) \rceil}{(1+\epsilon)I^k}$ randomly generated RR sets without any prior knowledge, while we do have some prior knowledge about $\mathcal{R}_1$ that $\mathcal{D}_1^k=\lceil \Upsilon_1(\epsilon,\frac{\delta}{n}) \rceil$. To fix this issue, we generate $\mathcal{R}_2$ that contains another $M_2=M_1=|\mathcal{R}_1|$ independent RR sets to find influential vertices. Then Lemma~\ref{lemma:deviate} holds for $\mathcal{R}_2$. 

To analyze the efficiency, we investigate the number of RR sets needed, since Section~\ref{sec:update} already indicates that the maintenance cost for computing influence dynamically is proportional to the sample size. Corollary~\ref{cor:bound_Ik} also implies that, when $\epsilon \leq \frac{1}{3}$, $\delta \leq \frac{1}{4}$ and $\mathcal{D}_1^k=\lceil \Upsilon_1(\epsilon,\frac{\delta}{n}) \rceil$, with probability at least $1-\frac{(n+k+2)\delta}{2n}$, $\frac{n\mathcal{D}_1^k}{(1+\epsilon)I^k} \leq M_1 \leq \frac{n(\mathcal{D}_1^k+1)}{(1-\epsilon)I^k}$. It is easy to verify that both sides of this inequality are $\Theta(\frac{n\ln{\frac{n}{\delta}}}{\epsilon^2I^k})$. Thus, with high probability, $M=|\mathcal{R}|=M_1+M_2=\Theta(\frac{n\ln{\frac{n}{\delta}}}{\epsilon^2I^k})$. It is worth noting that, according to Dagum~\textit{et~al.}~\cite{dagum2000optimal}, even when we know which vertex is the $k$-th most influential individual, to obtain an $(\epsilon,\delta)$-estimation of $I^k$, at least $\Theta(\frac{\ln{\frac{1}{\delta}}}{\epsilon^2I^k}\max\{n-I^k,\epsilon I^k\})$ RR sets are needed. Considering normally $I^k \ll n$, the minimum number of RR sets to achieve an $(\epsilon,\delta)$-estimation of $I^k$ is $\Theta(\frac{n\ln{\frac{1}{\delta}}}{\epsilon^2I^k})$, \updates{which is only smaller than our sample size $\Theta(\frac{n\ln{\frac{n}{\delta}}}{\epsilon^2I^k})$ by at most a factor of $\ln{n}$.}

\smallskip
\noindent \textbf{Comparison to~\cite{yang2017tracking}} For the top-$k$ tracking algorithm in~\cite{yang2017tracking}, we set the absolute error $\epsilon_1 n$ to the same value of the error in our method, which is less than $4\epsilon I^k$ according to Theorem~\ref{th:individual} (suppose magically we know the value of $I^k$ beforehand). Then the number of RR sets is $O(\frac{I^*}{I^k}\frac{n\ln{\frac{n}{\delta}}}{\epsilon^2I^k})$, where $I^*$ is the maximum individual influence. It is obvious that $I^* \geq I^k$ and in real social networks, the gap between $I^*$ and $I^k$ can be large. Thus, our algorithm is normally more efficient than the top-$k$ tracking algorithm in~\cite{yang2017tracking}, when the errors controlled by the two methods are the same in absolute value.  

\subsection{Maintaining Ranks of Vertices and $\mathcal{D}_1^k$ Dynamically}\label{sec:ds}

\begin{figure}
    \centering
    \includegraphics[width=.32\textwidth]{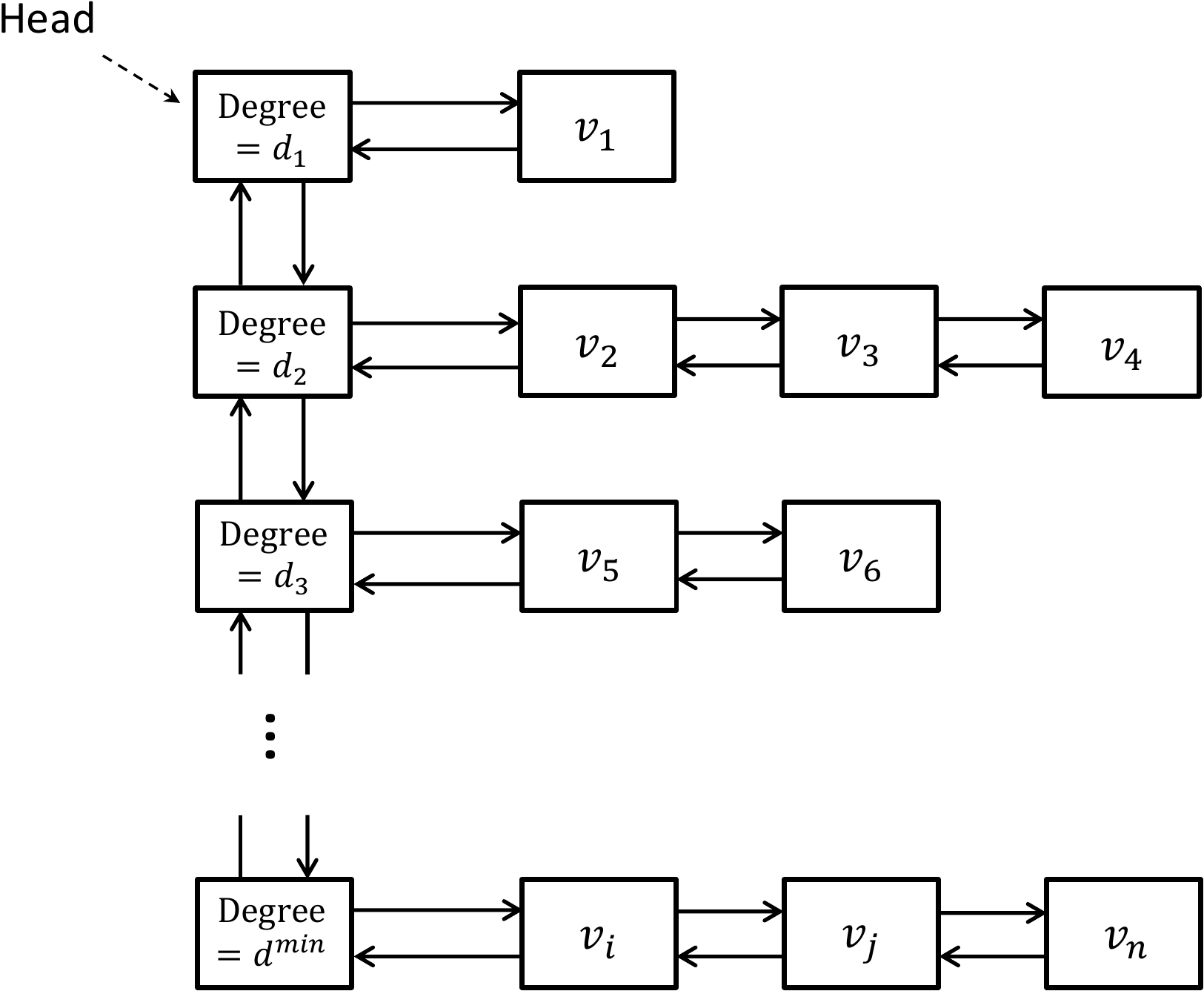}
    \caption{Linked List Structure, where $d_1 > d_2 > d_3 >...>d^{min}$}
    \label{fig:linkedlist}
\end{figure}

\updates{By applying the Linked List structure~\cite{yang2017tracking} on $\mathcal{R}_2$, we dynamically maintain all vertices sorted by their degrees in $\mathcal{R}_2$. As demonstrated in Fig.~\ref{fig:linkedlist}, the vertices with the same degree in $\mathcal{R}_2$ are grouped together in a doubly linked list whose first node is a special node called a head node. All head nodes are sorted in a doubly linked list (the vertical one in Fig.~\ref{fig:linkedlist}). A nice property of the Linked List structure is that maintaining it does not increase the complexity of updating RR sets. For the interest of space, we skip the details of how to maintain the Linked List, which can be found in our previous work~\cite{yang2017tracking}. When a top-$k$ individual query is issued, by the Linked List structure on $\mathcal{R}_2$, we only need $O(k')$ time to return all vertices $u$ such that $\mathcal{D}_2(u) \geq \frac{1-\epsilon}{1+\epsilon}\mathcal{D}_1^k$, if there are $k'$ such vertices.}

\updates{
Besides dynamically maintaining ranks of vertices by their degrees in $\mathcal{R}_2$, we still need to maintain $\mathcal{D}_1^k$ efficiently and dynamically. This is because $\mathcal{D}_1^k$ is the signal used in deciding sample size. Moreover, we need the value of $\mathcal{D}_1^k$ in order to return all vertices $u$ such that $\mathcal{D}_2(u) \geq \frac{1-\epsilon}{1+\epsilon}\mathcal{D}_1^k$ when a top-$k$ individual query is issued. Unfortunately, ~\cite{yang2017tracking} did not discuss how to maintain $\mathcal{D}_1^k$.
}

To maintain the value of $\mathcal{D}_1^k$ against updates, \updates{we also apply the Linked List structure on $\mathcal{R}_1$}. We record $H_k$, the head node of the doubly linked list containing vertices whose degrees in $\mathcal{R}_1$ are all $\mathcal{D}_1^k$. We also need a bias $b$ which indicates there are $k-b$ vertices $u$ such that $\mathcal{D}_1(u) > \mathcal{D}_1^k$. Suppose due to an update, $\mathcal{D}_1(u)$ increases by 1. Let $\mathcal{D}_1(u)^{old}$ be the value before the update and $\mathcal{D}_1(u)^{new}$ the value after the update. Denote by $H_k.degree$ the degree of vertices that $H_k$ is their head node and $H_k.num$ the number of such vertices. Let $H_k.up$ be the head node above it and $H_k.down$ the head node below it. To update $H_k$ and $b$, we have two cases depending on $\mathcal{D}_1(u)^{old}$.
\begin{enumerate}
    \item If $\mathcal{D}_1(u)^{old} \neq H_k.degree$, we do not need to update $H_k$ or $b$; and
    \item If $\mathcal{D}_1(u)^{old}=H_k.degree$, there are two subcases according to $b$. If $b=1$ before the update, we set $H_k$ as $H_k.up$, and set $b$ to the number of vertices in the linked list whose head is the updated $H_k$. If $b>1$ before the update, we do not update $H_k$ but we decrease $b$ by 1.
\end{enumerate}

Similarly, if $\mathcal{D}_1(u)$ of a vertex $u$ decreases by 1, to update $H_k$ and $b$, we have three cases depending on $\mathcal{D}_1(u)^{old}$.
\begin{enumerate}
    \item If $\mathcal{D}_1(u)^{old} < H_k.degree$ or $\mathcal{D}_1(u)^{old} > H_k.degree+1$, we do not need to update $H_k$ and $b$;
    \item If $\mathcal{D}_1(u)^{old}=H_k.degree+1$, we do not update $H_k$ but we increase $b$ by 1; and
    \item If $\mathcal{D}_1(u)=H_k.degree$, there are two subcases. If $b=H_k.num$ before the update, we set $H_k$ as $H_k.down$ and set $b=1$. If $b<H_k.num$ before the update, we do not update $H_k$ or $b$.
\end{enumerate}

Clearly, the updates on $H_k$ and $b$ only take $O(1)$ time when $\mathcal{D}_1(u)$ changes by 1 for a vertex $u$. After the maintenance, $\mathcal{D}_1^k=H_k.degree$.

When a RR set is updated/inserted/deleted, $\mathcal{D}_1(u)$ changes at most by 1 for each $u$. We need $\Omega(1)$ time to update its inverted index and only $O(1)$ time to update the linked list data structure, $H_k$ and $b$. Thus, the linked list and our maintenance of $\mathcal{D}_1^k$ do not increase the complexity of updating RR sets. Also, $\mathcal{D}_1^k$ can be retrieved in $O(1)$ time.

Moreover, if we set $k=1$, by employing the method described above, we can maintain $\mathcal{D}_1^*=\max_{u \in V}{\mathcal{D}_1(u)}=\mathcal{D}_1^1$, which is used in deciding a proper sample size for maintaining RR sets against network updates for IM queries in Section~\ref{ssec:size_im}.

\section{Efficient Influence Maximization Queries}\label{sec:im}
In this section, we tackle a different version of the top-$k$ influential vertex tracking task, that is, influence maximization (IM). Similar to Section~\ref{sec:individual}, we first develop an efficient method to decide if we have a proper amount of RR sets. Then we propose efficient implementations of the greedy algorithm for IM queries on the maintained RR sets, which work well in practice, especially when the number of RR sets is small.

\subsection{Algorithm and Sample Size}\label{ssec:size_im}
Like IM in static networks~\cite{borgs2014maximizing, tang2014influence, tang2015influence, nguyen2016stop}, our goal is to achieve $1-\frac{1}{e}-\epsilon$ optimal IM queries on dynamic networks with high probability.

\begin{algorithm}[t]\small
\caption{Greedy Algorithm}
\label{alg:greedy}
\KwIn{seed set size $k$ and $\mathcal{R}$ (a collection of RR sets)}
\KwOut{$S_k$}
\begin{algorithmic}[1]
\STATE $S_k \leftarrow \emptyset$
\FOR {$i \leftarrow 1$ to $k$}
	\STATE $S_k \leftarrow \argmax_{v \in V \setminus S_k}{\mathcal{D}(S_k \cup \{v\})}$
\ENDFOR
\RETURN $S_k$
\end{algorithmic}
\end{algorithm}

Unlike tracking top-$k$ influential individuals, which is more in the flavor of a ranking problem, IM is a combinatorial optimization problem. Besides sampling a number of RR sets to approximate influence spreads of vertices or vertex sets, an IM query also needs to execute a greedy algorithm, such as Algorithm~\ref{alg:greedy}, on the sampled RR sets to find the best $k$-seed set. In Algorithm~\ref{alg:greedy}, $\mathcal{D}(S_k \cup \{v\})$ is the degree of $S_k \cup \{v\}$ in $\mathcal{R}$, which equals the number of RR sets in $\mathcal{R}$ that contains at least one vertex from $S_k \cup \{v\}$. In the state-of-the-art static network IM algorithms~\cite{tang2015influence,nguyen2016stop}, the value of $\mathcal{D}(S_k)$, where $S_k$ is returned by Algorithm~\ref{alg:greedy}, is used as a signal to decide if the current sample size of $\mathcal{R}$ is enough to achieve $1-\frac{1}{e}-\epsilon$ optimal IM queries with high probability. Unfortunately, for IM on dynamic networks, we cannot use $\mathcal{D}(S_k)$ as the signal because running Algorithm~\ref{alg:greedy} on $\mathcal{R}$ may take time linear to the size of $\mathcal{R}$ (the sum of numbers of vertices in all RR sets in $\mathcal{R}$), which is unaffordable for real-time updates.

Dinh~\textit{et~al.}~\cite{dinh2015social} proposed to use $\mathcal{D}^*=\max_{u \in V}{\mathcal{D}(u)}$ as the signal to decide if we have enough RR sets for IM on static networks. \updates{Note that by using our Linked List structure, $\mathcal{D}^*$ can be accessed in $O(1)$ time. The algorithm in~\cite{dinh2015social} keeps sampling RR sets until $\mathcal{D}^*= \lceil \Upsilon_1(\frac{\epsilon}{2(1-1/e)-\epsilon},\delta) \rceil$}. The intuition of~\cite{dinh2015social} and some other polling based algorithms~\cite{tang2015influence,nguyen2016stop} is to make sure $\textup{Pr}\{\frac{n\mathcal{D}(S_k)}{M} \leq (1+\epsilon_1)I(S_k) \} \geq 1-\delta_1$ and $\textup{Pr}\{\frac{n\mathcal{D}(S^*_k)}{M} \geq (1-\epsilon_2)I(S^*_k)\} \geq 1-\delta_2$, where $S_k$ is the seed set extracted by running the greedy algorithm on $\mathcal{R}$, and $S_k^*$ is the optimal $k$-seed set. In such a case, due to the submodularity of $\mathcal{D}(S)$ with respect to $S$, we have $\textup{Pr}\{ I(S_k) \geq \frac{1-\epsilon_2}{1+\epsilon_1}(1-\frac{1}{e})I(S_k^*) \} \geq 1-\delta_1-\delta_2$. Unfortunately the proof of  $\textup{Pr}\{\frac{n\mathcal{D}(S_k)}{M} \leq (1+\epsilon_1)I(S_k) \} \geq 1-\delta_1$ in~\cite{dinh2015social} is incorrect because a union bound is missed. Detailed discussion of the mistakes made by~\cite{dinh2015social} can be found Appendix.

\nop{Since using the data structure in Section~\ref{sec:ds} $\mathcal{D}^*$ can be retrieved in $O(1)$ time, we also use $\mathcal{D}^*$ as the signal to decide a proper sample size, and we give the correct stopping condition.}

We give a theoretically sound method to maintain RR sets, where the signal used for deciding sample size can be accessed in $O(1)$ time. First, for the maintained collection of RR sets $\mathcal{R}$, we also split it into two disjoint parts $\mathcal{R}_1$ and $\mathcal{R}_2$. Denote by $\mathcal{D}_1(S)$ and $\mathcal{D}_2(S)$, respectively, the degrees of $S$ in $\mathcal{R}_1$ and $\mathcal{R}_2$. Let $M_1=|\mathcal{R}_1|$ and $M_2=|\mathcal{R}_2|$. Let $N=\binom{n}{k_{max}}$, where $k_{max}$ is the maximum $k$ of an IM query. Our algorithm works as follows,
\begin{enumerate}
	\item Update the RR sets in $\mathcal{R}=\mathcal{R}_1 \cup \mathcal{R}_2$ when the network structure changes using the method in~\cite{yang2017tracking}.
	\item Maintain the invariant that $\mathcal{D}_1^*$, the greatest degree in $\mathcal{R}_1$, always equals $\lceil \Upsilon_1(\epsilon,\frac{2\delta}{3n}) \rceil$.
	\item Maintain the invariant $M_2=\frac{M_1\ln{\frac{3N}{2\delta}}}{\ln{\frac{2n}{\delta}}}$, where $N=\binom{n}{k_{max}}$.
	\item \updates{When an IM query with parameter $k \leq k_{max}$ is issued, run Algorithm~\ref{alg:greedy} on $\mathcal{R}_2$ to return $S_k$.}
\end{enumerate}
When $\mathcal{D}_1^* \neq \lceil \Upsilon_1(\epsilon,\frac{\delta}{2n}) \rceil$, we adjust the sizes of $\mathcal{R}_1$ and $\mathcal{R}_2$ by adding or deleting some RR sets. As illustrated in Section~\ref{sec:ds}, by applying the Linked List structure on $\mathcal{R}_1$, $\mathcal{D}_1^*$ can be efficiently maintained against network updates and it can be accessed in $O(1)$ time when needed. 
 
\begin{theorem}\label{th:im}
	Our algorithm returns a seed set $S_k$ for an IM query with $k \leq k_{max} \leq \frac{n}{2}$, such that $\textup{Pr}\{I(S_k) \geq [1-\frac{1}{e}-(2-\frac{1}{e})\epsilon]I(S_k^*)\} \geq 1-\delta$, where $S^*_k$ is the optimal $k$-seed set.
\end{theorem}

Based on Theorem~\ref{th:im}, to achieve $(1-1/e-\epsilon)$ optimal IM queries with probability at least $1-\frac{1}{n}$ (similar to the quality guarantees in the state-of-the-art IM algorithms~\cite{tang2015influence,nguyen2016stop} on static networks), under any seed set size $k \leq k_{max}$, we maintain the invariants $\mathcal{D}_1^*=\lceil \Upsilon_1(\frac{\epsilon}{2-1/e}, \frac{2}{3n^2}) \rceil$ and $M_2=\frac{M_1\ln{\frac{3Nn}{2}}}{\ln{\frac{3n^2}{2}}}$. Similar to tracking influential individuals, with probability at least $1-\frac{n}{3}$, we have $M_1=O(\frac{n\ln{n}}{\epsilon^2 I^*})$ and $M_2=O(\frac{n\ln{Nn}}{\epsilon^2 I^*})=O(\frac{k_{max}n\ln{n}}{\epsilon^2 I^*})$. Thus, with high probability, in total $M=|\mathcal{R}|=M_1+M_2=O(\frac{k_{max}n\ln{n}}{\epsilon^2 I^*})$. Since the average number of edges and vertices traversed when generating an RR set is no more than $\frac{(m+n)I^*}{n}$, we have the cost $C(\mathcal{R})=O(\frac{k_{max}(m+n)\ln{n}}{\epsilon^2})$, where $C(\mathcal{R})$ is the total number of edges and vertices traversed when generating $\mathcal{R}$,
which is in the same order of the correct bound of $C(\mathcal{R})$ in the method in~\cite{ohsaka2016dynamic} and also the bound in~\cite{tang2015influence}.
 
In practice, $O(\frac{k_{max}n\ln{n}}{\epsilon^2 I^*})$ is too large because it is linear to $k_{max}$ and $k_{max}$ may easily be in the order of hundreds. \updates{Thus, in our experiments, in stead of strictly implementing our algorithm, we sample fewer RR sets by ignoring the factor $k_{max}$. This is similar to~\cite{ohsaka2016dynamic} where $C(\mathcal{R})$ is set to $O(\frac{(m+n)\ln{n}}{\epsilon^2})$ by ignoring $k_{max}$ in the correct bound $O(\frac{k_{max}(m+n)\ln{n}}{\epsilon^2})$. Specifically, our practical solution is to just maintain the invariant $\mathcal{D}^*=\frac{1}{2} \lceil \Upsilon_1(\frac{\epsilon}{2-1/e}, \frac{2}{3n^2}) \rceil$ (we do not split $\mathcal{R}$ into $\mathcal{R}_1$ and $\mathcal{R}_2$ in the practical solution). It is not difficult to derive that in this way the sample size $M$ is roughly $\frac{1}{k_{max}}$ times the theoretically sound sample size $O(\frac{k_{max}n\ln{n}}{\epsilon^2 I^*})$. We demonstrate in experiments that our practical solution normally leads to much fewer RR sets than maintaining $C(\mathcal{R})=32(m+n)\log{n}$ in~\cite{ohsaka2016dynamic}, and the quality of seed set mined is not compromised.}

\nop{
Although the method in~\cite{dinh2015social} does not have its claimed quality guarantee, it works well in practice as demonstrated by experiments~\cite{dinh2015social}. Thus, in our experiments, in stead of strictly implementing our algorithm, we adopt the method in~\cite{dinh2015social} where we do not split the maintained collection of RR sets $\mathcal{R}$ and we maintain the invariant $\mathcal{D}^*=\lceil \Upsilon_1(\epsilon_1,\frac{1}{n}) \rceil$, where $\frac{1-\epsilon_1}{1+\epsilon}(1-1/e)=1-1/e-\epsilon$. We will demonstrate in experiments that maintaining $\mathcal{D}^*=\lceil \Upsilon_1(\epsilon_1,\frac{1}{n}) \rceil$ normally leads to much fewer RR sets than maintaining $C(\mathcal{R})=32(m+n)\log{n}$ in~\cite{ohsaka2016dynamic}, and the quality of seed set mined is not compromised.   
}

\subsection{Speeding Up the Greedy Algorithm for IM Queries}\label{im:query}

In addition to fast maintenance of the RR sets, we also need to run the greedy algorithm to find seed vertices. Efficient implementation of the greedy algorithm becomes critical. In the following discussion, we assume that the greedy algorithm is conducted on a collection of RR sets $\mathcal{R}$. Note that if one strictly implements our method as described in Section~\ref{ssec:size_im}, which has theoretical guarantees, then $\mathcal{R}$ should be $\mathcal{R}_2$.

\updates{Our Linked List data structure in Fig.~\ref{fig:linkedlist} can be used to implement the greedy algorithm (Algorithm~\ref{alg:greedy}). Algorithm~\ref{alg:llgreedy} shows the implementation. In line~\ref{line:dif1}, we take $O(n)$ time to copy the Linked List $\mathcal{D}$ to $\mathcal{D}'$. This step is essential because $\mathcal{D}$ cannot be modified during an IM query. If we modify $\mathcal{D}$ when executing an IM query, when the query ends $\mathcal{D}(u)$ may not equal the number of RR sets containing $u$ such that $\mathcal{D}$ cannot be used in the greedy algorithm (Algorithm~\ref{alg:greedy}) for a new IM query. It is easy to see that the cost of the rest part (after line~\ref{line:dif1}) of Algorithm~\ref{alg:llgreedy} is $O(\sum_{i=1}^M{|R_i|})$, where $|R_i|$ is the number of vertices in the $i$-th RR set in $\mathcal{R}$. Thus, the computational cost of Algorithm~\ref{alg:llgreedy} is $O(n+\sum_{i=1}^M{|R_i|})$.}

\nop{
But Algorithm~\ref{alg:greedy} cannot be directly used to answer influence maximization queries on dynamic networks, because on dynamic networks, \todo{The following two sentences are unclear.  My understanding is that the greedy algorithm locks the Linked List and the maintained RR sets so that the dynamic updates cannot be allowed.  Is that right?  If so, these two sentences have to be rewritten.} a key point is that during the execution of the greedy algorithm, the Linked List should be locked and read only. Also the maintained RR sets cannot be modified during the execution of an IM query. Our implementation is shown in Algorithm~\ref{alg:llgreedy} and is similar to the one described in~\cite{borgs2014maximizing}. The differences are in Lines~\ref{line:dif1} and~\ref{line:dif}, where we copy the Linked List first. We do not delete the RR sets containing the selected seeds.  Instead, we mark them as ``covered by selected seeds so far''. It is easy to see that our implementation has the complexity $O(n+\sum_{i=1}^M{|R_i|})$, where $O(n)$ is due to copying the Linked List data structure.
}
\begin{algorithm}[t]\small
\caption{Greedy Algorithm using Linked List}
\label{alg:llgreedy}
\KwIn{$k$, $\mathcal{R}$ (a collection of RR sets) and a Linked List $\mathcal{D}$}
\KwOut{$S_k$}
\begin{algorithmic}[1]
\STATE Copy $\mathcal{D}$ to $\mathcal{D}'$ \label{line:dif1}
\STATE $S_k \leftarrow \emptyset$
\STATE Mark all RR sets as uncovered
\FOR {$i \leftarrow 1$ to $k$}
	\STATE $u \leftarrow \argmax_{v \in V}{\mathcal{D}'(u)}$
	\STATE $S_k \leftarrow S_k \cup \{u\}$
	\FOR {each $R_j$ containing $u$}
		\IF {$R_j$ is not covered yet} \label{line:dif}
			\STATE Mark $R_j$ as covered
			\FOR {$v \in R_j$}
				\STATE DecreaseDegree($\mathcal{D}'(v)$)
			\ENDFOR
		\ENDIF
	\ENDFOR
	\STATE Remove $\mathcal{D}'(u)$ from $\mathcal{D}'$
\ENDFOR
\RETURN $S_k$
\end{algorithmic}
\end{algorithm}

Ohsaka~\textit{et~al.}~\cite{ohsaka2016dynamic} reported that employing the Lazy Evaluation techniques~\cite{leskovec2007cost} can achieve better performance in practice. However, the method in~\cite{ohsaka2016dynamic} still needs to take $O(n)$ time to copy $\mathcal{D}$ to $\mathcal{D}'$. In large networks, if the maintained RR sets are not many, it is possible that $\sum_{i=1}^M{|R_i|}$ is even smaller than $n$, the number of vertices. In such a case, copying the whole $\mathcal{D}$ may be a waste of time. 

Intuitively, in the greedy algorithm, when $k$ is small, most vertices are useless because their degrees in $\mathcal{R}$ are too small and probably they are never be picked as a seed. Exploiting this intuition, we design more efficient algorithms by only copying part of $\mathcal{D}$ to $\mathcal{D}'$. Specifically, we use a threshold $T_{\mathcal{D}}$. When iterating $\mathcal{D}$ from the top head node, $\mathcal{D}(v)$ is copied to $\mathcal{D}'$ only if $\mathcal{D}(v) > T_{\mathcal{D}}$. We show that this filtering strategy can achieve good performance with provable guarantees.

Suppose Algorithm~\ref{alg:llgreedy} returns a seed set $S_k$, and $S'_k$ is the seed set returned by the greedy algorithm where only vertices $v$ such that $\mathcal{D}(v) > T_{\mathcal{D}}$ are copied to $\mathcal{D}'$. Let $U=\{v \mid \mathcal{D}(v) > T_{\mathcal{D}}\}$ and $L=\{v \mid \mathcal{D}(v) \leq T_{\mathcal{D}}\}$. Suppose $S'_k=\{u_1,...,u_k\}$, where $u_i$ is the $i$-th seed added to $S_k$. Define $\mathcal{M}(u_i;S'_k)=\mathcal{D}(\{u_1,...,u_{i-1}\} \cup \{u_i\})-\mathcal{D}(\{u_1,...,u_{i-1}\})$ the marginal gain of $u_i$. Suppose $u_q$ is the first seed added to $S'_k$ such that $\mathcal{M}(u_q;S'_k) \leq T_{\mathcal{D}}$. If such $q$ does not exist, we set $q=k+1$.

\begin{theorem}\label{th:filter_greedy}
    $\mathcal{D}(S'_k) \geq \mathcal{D}(S_k)-(k-q+1)T_{\mathcal{D}}$.
\end{theorem}

Apparently, when $q=k+1$, we have $\mathcal{D}(S'_k)=\mathcal{D}(S_k)$. Also, by setting $T_\mathcal{D}=\min\{\frac{\mathcal{D}^*}{k-1},\mathcal{D}^*-1\}$, we have $\mathcal{D}(S'_k) \geq \frac{1}{2}\mathcal{D}(S_k)$.

\begin{corollary}\label{cor:filter}
	If $\mathcal{D}^* \geq 2$ and $T_\mathcal{D}=\min\{\frac{\mathcal{D}^*}{k-1},\mathcal{D}^*-1\}$, then $\mathcal{D}(S'_k) \geq \frac{1}{2}\mathcal{D}(S_k)$.
\end{corollary}

Based on our analysis, we propose Algorithm~\ref{alg:newgreedy}, the New Greedy Algorithm, which also exploits the lazy evaluation method for maximizing monotone and submodular functions. The returned $t$ by Algorithm~\ref{alg:newgreedy} can tell us if the seed set found has the same quality as the seed set returned by Algorithm~\ref{alg:llgreedy}. In our implementation, we set $T_\mathcal{D}=\min\{\frac{\mathcal{D}^*}{k-1},\mathcal{D}^*-1\}$ and run the New Greedy algorithm. If the returned $q$ is not $k+1$, we reset $T_\mathcal{D}$ to $-1$ and re-run the New Greedy algorithm. Setting $T_{\mathcal{D}}=-1$ is equivalent to the query algorithm in~\cite{ohsaka2016dynamic}. By doing so we can guarantee that the returned seed set $S_k$ has $\mathcal{D}(S_k)$ at least $(1-1/e)\mathcal{D}(S^{\circ}_k)$, where $\mathcal{D}(S^{\circ}_k)=\max_{|S|=k}{\mathcal{D}(S)}$. In our experiments we find that $T_\mathcal{D}=\frac{\mathcal{D}^*}{k-1}$ is an effective and efficient threshold, because we never need to reset $T_\mathcal{D}$ to re-run the New Greedy algorithm.

\begin{algorithm}[t]\small
\caption{New Greedy}
\label{alg:newgreedy}
\KwIn{$k$, $\mathcal{R}$ which is a set of random RR sets and a Linked List $\mathcal{D}$, and a threshold $T_{\mathcal{D}}$}
\KwOut{$S_k$ and $q$}
\begin{algorithmic}[1]
\STATE Copy all $\mathcal{D}(u)$ in $\mathcal{D}$ such that $\mathcal{D}(u) > T_{\mathcal{D}}$ to $\mathcal{D}'$, and set $update(u) \leftarrow 1$    \label{line:copy}
\STATE $S_k \leftarrow \emptyset$
\STATE Mark all RR sets as uncovered
\STATE $q \leftarrow k+1$
\FOR {$i \leftarrow 1$ to $k$}
	\WHILE {true}
		\STATE $u \leftarrow \argmax_{v \in V}{\mathcal{D}'(u)}$ \label{line:improve}
		\IF {$update(u) = i$}
		    \STATE $MarginCover \leftarrow 0$
			\FOR {each $R_j$ containing $u$}
			    \IF {$R_j$ is not covered}
				    \STATE Mark $R_j$ as covered
				    \STATE $MarginCover \leftarrow MarginCover+1$
				\ENDIF
			\ENDFOR
			\STATE $S_k \leftarrow S_k \cup \{u\}$
			\STATE Remove $\mathcal{D}'(u)$ from $\mathcal{D}'$
			\IF {$MarginCover < T_{\mathcal{D}} \wedge q > k$} \label{line:stime1}
			    \STATE $q \leftarrow i$ \label{line:stime2}
			\ENDIF
			\STATE \textbf{break while loop}
		\ELSE
			\STATE $MarginInf \leftarrow 0$
			\FOR {each $R_j$ containing $u$}
				\IF {$R_j$ is not covered yet}
					\STATE $MarginInf \leftarrow MarginInf + 1$
				\ENDIF
			\ENDFOR
			\STATE $Dif \leftarrow \mathcal{D}'(u)-MarginInf$
			\FOR {$r \leftarrow 1$ to $Dif$}
				\STATE DecreaseDegree($\mathcal{D}'(u)$)
			\ENDFOR
			\STATE $update(u) \leftarrow i$
		\ENDIF
	\ENDWHILE
\ENDFOR
\RETURN $S_k$ and $q$
\end{algorithmic}
\end{algorithm}

\section{Experiments}\label{sec:exp}
In this section, we report a series of experiments on five real networks to verify our algorithms and our theoretical analysis. The experimental results demonstrate that our algorithms are both effective and efficient.

\subsection{Experimental Settings}\label{sec:exp_setting}
We used five real network data sets that are publicly available online (\url{http://snap.stanford.edu}, \url{http://www.cs.ubc.ca/~welu/} and \url{https://an.kaist.ac.kr/traces/WWW2010.html}). Table~\ref{tab:Datasets} shows the basic statistics of the five data sets.

\begin{table}
\centering
\begin{tabular}{|c|c|c|c|c|c|c|}
\hline
Network & Vertices & Edges & $\tau$ & IM Queries & $k_{max}$ \\ \hline
wiki-Vote & 7K & 104K & $10^2$ & 212 & 50 \\ \hline
Flixster & 99K & 978K & $10^3$ & 223 & 100 \\ \hline
soc-Pokec & 1.6M & 31M & $10^4$ & 636 & 200\\ \hline
flickr-growth & 2.3M & 33M & $10^4$ & 779 & 200 \\ \hline
Twitter & 41.6M & 1.5G & $10^5$ & 3011 & 500\\ \hline
\end{tabular}
\caption{The statistics of the data sets.}
\label{tab:Datasets}
\end{table}

\begin{table*}
\centering
\caption{Recall and Maximum Error Rate. Theoretical values hold with high probability.} \label{tab:recall_error}
\footnotesize
\begin{tabular}{|c|c|c|c|c|c|c|}
	\hline
		\multirow{2}{*}{} & \multicolumn{3}{c|}{wiki-Vote} & \multicolumn{3}{c|}{Flixster} \\ \cline{2-7}
			& Theoretical Value & Ave.$\pm$SD (LT) & Ave.$\pm$SD (IC) & Theoretical Value & Ave.$\pm$SD (LT) & Ave.$\pm$SD (IC) \\ \hline
		Recall & 100\%   &  100\%   &  100\%   &  100\%  &  100\%  &  100\% \\ \hline
		Max Error Rate & $36.5\%$ & 18.4\%$\pm$0.97\% & 18.5\%$\pm$1.17\% & $36.5\%$  & 20.2\%$\pm$0.90\% & 19.9\%$\pm$0.51\% \\ \hline
\end{tabular}
\end{table*}

\begin{table}
\centering \footnotesize
\caption{Running time (s) on static networks.}
\label{tab:time_static}
\begin{tabular}{*{6}{|c}|}
    \hline
    	\multirow{2}{*}{Dataset} & \multirow{2}{*}{\#Updates} & \multicolumn{2}{c|}{LT} & \multicolumn{2}{c|}{IC} \\ \cline{3-6}
			& &	Total	&	ST	&	Total	&	ST	\\ \hline
	wiki-Vote	& $2.1 \times 10^4$ &  11.3 & 2.3 & 29.3 & 8.2 \\ \hline
	Flixster	& $2.2 \times 10^5$ & 266 & 28 & 522 & 85 \\ \hline
	soc-Pokec   & $6.4 \times 10^6$ & 3165 & 311 & 4461 & 735 \\ \hline
	flickr-growth & $7.8 \times 10^6$ & 1908 & 201 & 3223 & 935 \\ \hline
	Twitter & $3.0 \times 10^8$ & 15369 & 375 & 19803 & 4770 \\ \hline
\end{tabular}
\end{table}

To simulate dynamic networks, for each data set, we randomly partitioned all edges exclusively into 3 groups: $E_1$ (85\% of the edges), $E_2$ (5\% of the edges) and $E_3$ (10\% of the edges). We used $B=\langle V,E_1 \cup E_2\rangle$ as the base network. $E_2$ and $E_3$ were used to simulate a stream of updates.

For the LT model, for each edge $(u,v)$ in the base network, we set the weight to 1. For each edge $(u,v) \in E_3$, we generated a weight increase update $(u,v,+,1, \_)$ (timestamps ignored at this time). For each edge $(u, v) \in E_2$, we generated one weight decrease update $(u,v,-,\Delta, \_)$ and one weight increase update $(u,v,+,\Delta, \_)$, where $\Delta$ was picked uniformly at random in $[0,1]$. We randomly shuffled those updates to form an update stream by adding random time stamps. For each data set, we generated 10 different instances of the base network and the update stream, and thus ran the experiments 10 times. Note that for the 10 instances, although the base networks and the update streams are different, the final snapshots of them are identical to the data set itself.

For the IC model, we first assigned propagation probabilities of edges in the final snapshot, that is, the whole graph. We set $w_{uv}=\frac{1}{\text{in-degree}(v)}$, where $\text{in-degree}(v)$ is the number of in-neighbors of $v$ in the whole graph. Then, for each edge $(u,v)$ in the base network, we set $w_{uv}$ to $\frac{1}{\text{in-degree}(v)}$. For each edge $(u,v) \in E_3$, we generated a weight increase update $(u,v,+,\frac{1}{\text{in-degree}(v)}, \_)$ (again, timestamps ignored at this time). For each edge $(u, v) \in E_2$, we generated one weight decrease update $(u,v,-,\Delta\frac{1}{\text{in-degree}(v)}, \_)$ and one weight increase update $(u,v,+,\Delta\frac{1}{\text{in-degree}(v)}, \_)$, where $\Delta$ was picked uniformly at random in $[0,1]$. We randomly shuffled those updates to form an update stream by adding random time stamps. For each dataset we also generated 10 instances.

For the parameters of tracking top-$k$ influential individuals, that is, the parameters in Theorem~\ref{th:individual}, we set $k=50$, $\delta=0.001$ and $\epsilon=0.1$, which means that the relative error rate is roughly bounded by 36.5\% according to the proof of Theorem~\ref{th:individual}. For influence maximization (IM), \updates{as illustrated at the end of Section~\ref{ssec:size_im}, we implement a practical solution that maintains $\mathcal{D}^*=\frac{1}{2} \lceil \Upsilon_1(\frac{\epsilon}{2-1/e}, \frac{2}{3n^2}) \rceil$. We set $\epsilon=\frac{\frac{1}{2}-\frac{1}{e}}{2-\frac{1}{e}}$ such that $1-\frac{1}{e}-(2-\frac{1}{e})\epsilon=0.5$. Remember that $1-\frac{1}{e}-(2-\frac{1}{e})\epsilon$ is the approximation ratio of the theoretical sound algorithm in Section~\ref{ssec:size_im} which roughly maintains $k_{max}$ times the number of RR sets as that of the practical solution.}

\nop{
since our theoretically correct algorithm, which maintains $O(\frac{k_{max}n\ln{n}}{\epsilon^2 I^*})$ RR sets, is costly in practice, we employ the strategy that maintain the invariant $\mathcal{D}^*=\lceil \Upsilon_1(\epsilon_1,\delta) \rceil$, which tends to generate a small number of RR sets but works well in practice as illustrated at the end of Section~\ref{ssec:size_im}. We set $\epsilon_1=0.12$ and $\delta=\frac{1}{n}$.
}

\updates{To mimic the real application environment, we inserted an IM query in the update stream every $\tau$ edge weight updates. The reason we did not insert top-$k$ individual queries is because outputting top individual vertices is very efficient (thanks to our Linked List data structure that maintains the ranking of vertices). To better simulate the queries of users, the seed set size constraint $k$ is randomly drawn from $[1,k_{max}]$. The values of $\tau$, $k_{max}$ and the number of inserted IM queries for each network are also shown in Table~\ref{tab:Datasets}.}

\nop{
Since the greedy algorithm for IM queries is costly, we also test our implementation of the greedy algorithm. We inserted an IM query in the update stream every $\tau$ edge weight updates. To better simulate the queries of users, the seed set size constraint $k$ is randomly drawn from $[1,k_{max}]$. The values of $\tau$, $k_{max}$ and the number of inserted IM queries for each network are also shown in Table~\ref{tab:Datasets}.
}

We also compare our algorithms with baselines. For tracking influential individuals, we compare our algorithm with the top-$k$ influential individual tracking algorithm in~\cite{yang2017tracking}, which controls an absolute error $\epsilon'$ with probability at least $\delta$. We set $\epsilon'=0.0005$ for all datasets except for Twitter. For Twitter, we set $\epsilon'=0.0025$. We set $\delta=0.001$ for all datasets. We compare the running time and demonstrate the limitation of absolute error in tracking influential individuals. For influence maximization, we compare with the algorithm in~\cite{ohsaka2016dynamic} that maintains $C(\mathcal{R}) \geq 32(m+n)\log{n}$. We report the ratio of $C(\mathcal{R})$ of this baseline to the $C(\mathcal{R})$ in our algorithm. Note that this ratio is approximately the ratio of the number of RR sets of the baseline to the number of RR sets maintained by our algorithm. We also compare the query algorithm of~\cite{ohsaka2016dynamic} (Lazy Evaluation) to our New Greedy algorithm.

All algorithms were implemented in Java and run on a Linux machine of an Intel Xeon 2.00 GHz CPU and 1 TB main memory.

\subsection{Tracking Influential Individuals}

\subsubsection{Verifying Provable Quality Guarantees}
A challenge in evaluating the effectiveness of our algorithms is that the ground truth is hard to obtain. The existing literature of influence maximization~\cite{kempe2003maximizing, chen2010scalable, goyal2011simpath, tang2014influence, tang2015influence, cohen2014sketch} always uses the influence spread estimated by 20,000 times Monte Carlo (MC) simulations as the ground truth. However, such a method is not suitable for our tasks, because the ranking of vertices really matters here. Even 20,000 times MC simulations may not be able to distinguish vertices with close influence spread. As a result, the ranking of vertices may differ much from the true ranking. Moreover, the effectiveness of our algorithms has theoretical guarantees while 20,000 times MC simulations is essentially a heuristic. It is not reasonable to verify an algorithm with a theoretical guarantee using results obtained by a heuristic method without any quality guarantees.

In our experiments, we only used wiki-Vote and Flixster to run MC simulations and compare the results to those produced by our algorithms. We used 2,000,000 times MC simulations as the (pseudo) ground truth in the hope we can get more accurate results. According to our experiments, even so many MC simulations may generate slightly different rankings of vertices in two different runs but the difference is acceptably small. We only compare results on the identical final snapshot shared by all instances because running MC simulations on multiple snapshots is unaffordable (e.g., 10 days on the final snapshots of Flixster).

Tables~\ref{tab:recall_error} reports on the wiki-Vote and Flixster data sets the recall of the sets of influential vertices returned by our algorithms and the maximum error rates of the false positive vertices in absolute influence value. The results are obtained by taking the average of the results on the 10 runs on 10 instances. Our methods achieved 100\% recall every time as guaranteed theoretically.  Moreover, the true error rates were substantially smaller than the maximum error rate provided by our theoretical analysis.

For the other data sets, we did not run 2,000,000 times MC simulations to obtain the pseudo ground truth since the MC simulations are too costly. Instead, we compare the similarity between the results generated by different instances. Recall that the final snapshots of the 10 instances are the same. If the sets of influential vertices at the final snapshots of the 10 instances are similar, at least our algorithms are stable, that is, insensitive to the order of updates. The similarity between two sets of influential vertices is measure by the Jaccard similarity.

\begin{figure*}
  \centering
    \subfigure[\small{soc-Pokec (LT)}]{
      \includegraphics[width=.14\textwidth]{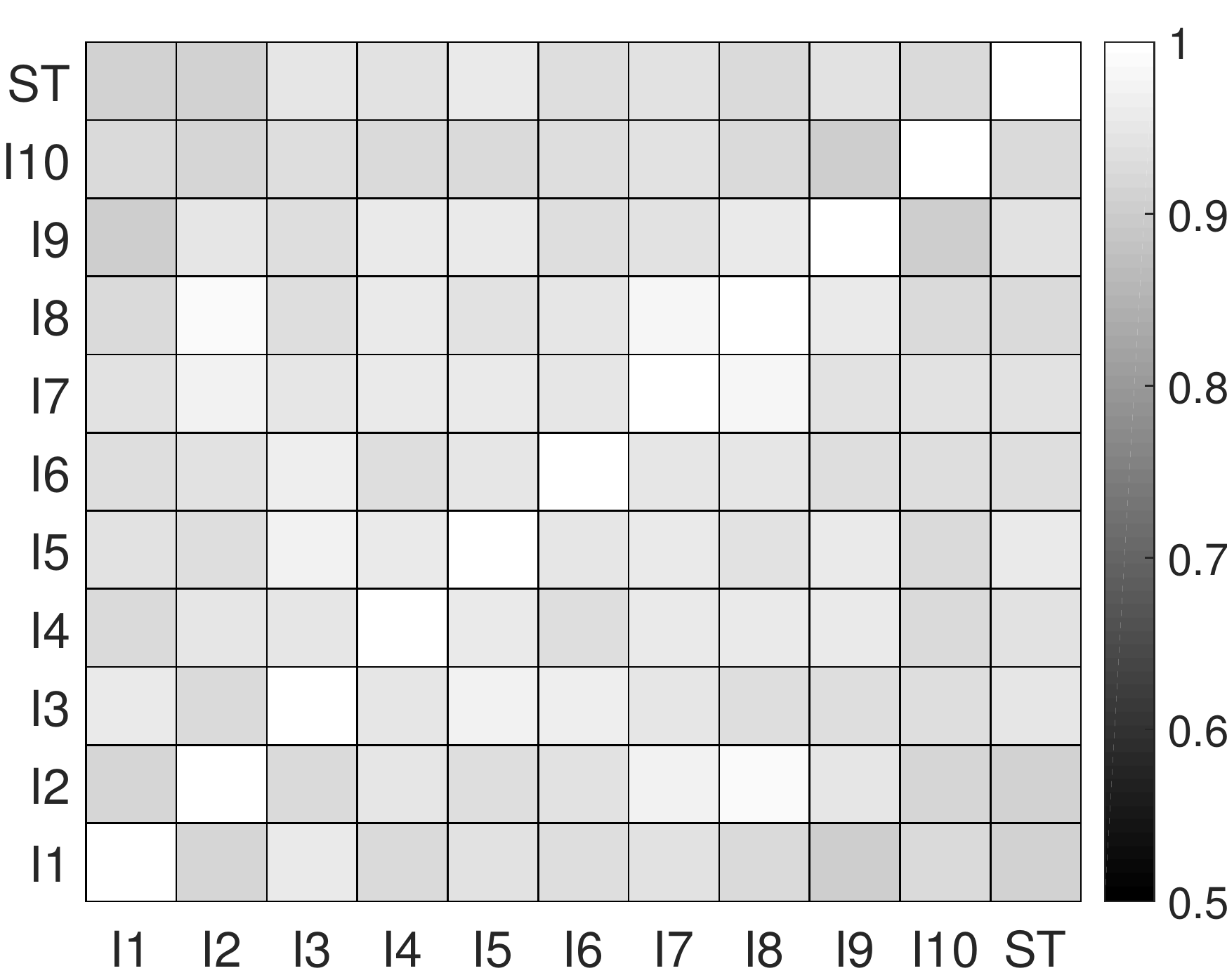}
    }
    \subfigure[\small{soc-Pokec (IC)}]{
      \includegraphics[width=.14\textwidth]{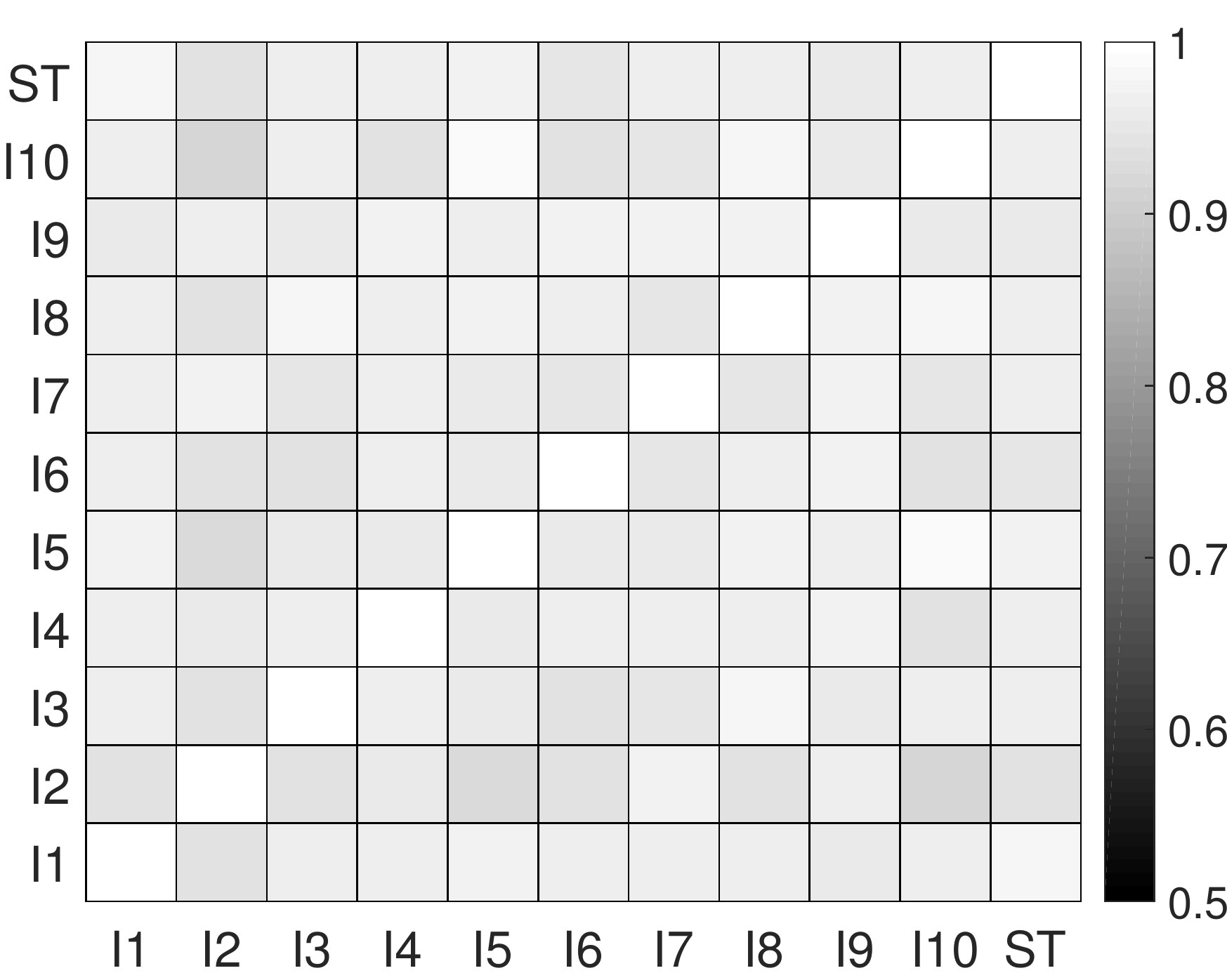}
    }
    \subfigure[\small{flickr-growth (LT)}]{
      \includegraphics[width=.14\textwidth]{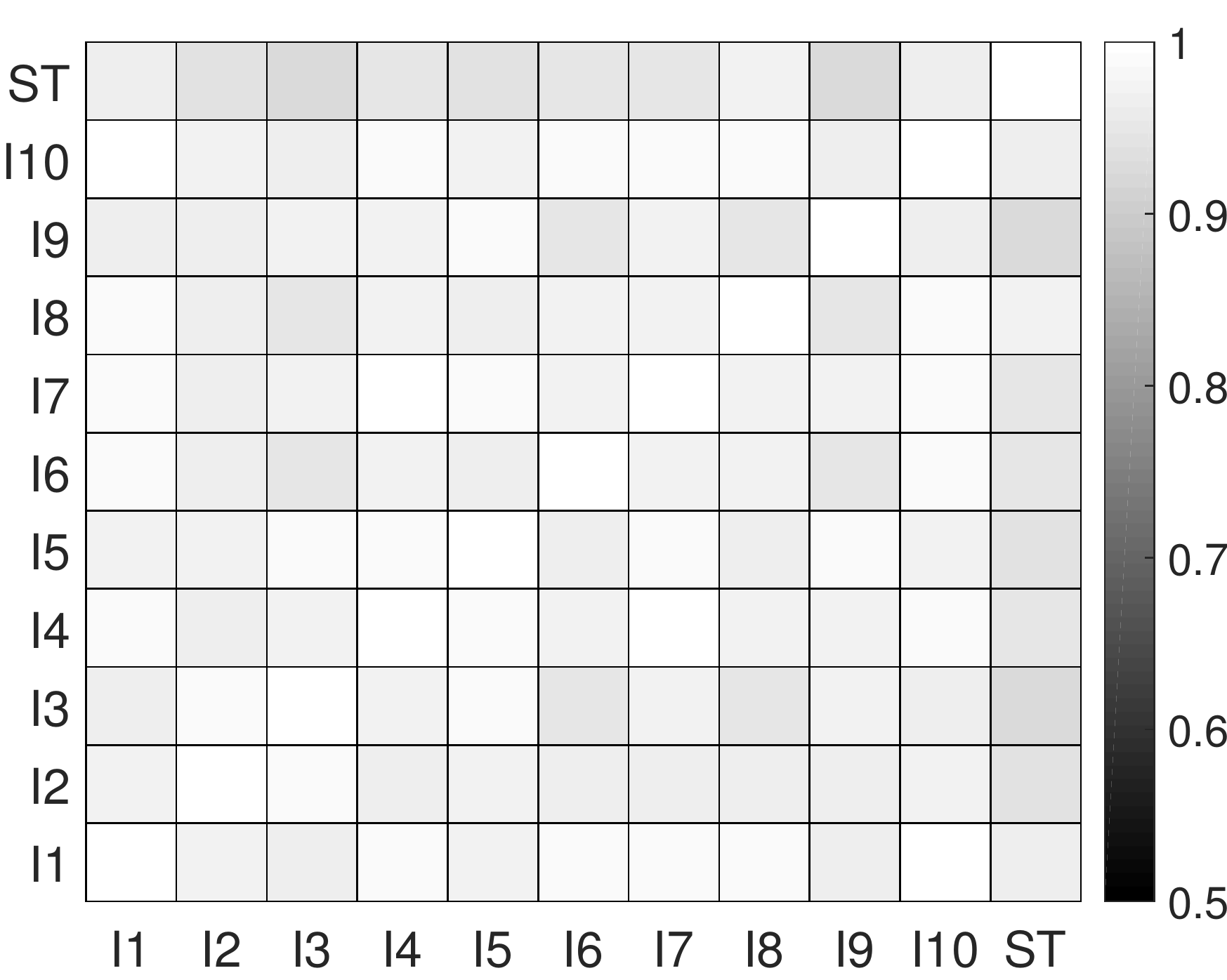}
    }
    \subfigure[\small{flickr-growth (IC)}]{
      \includegraphics[width=.14\textwidth]{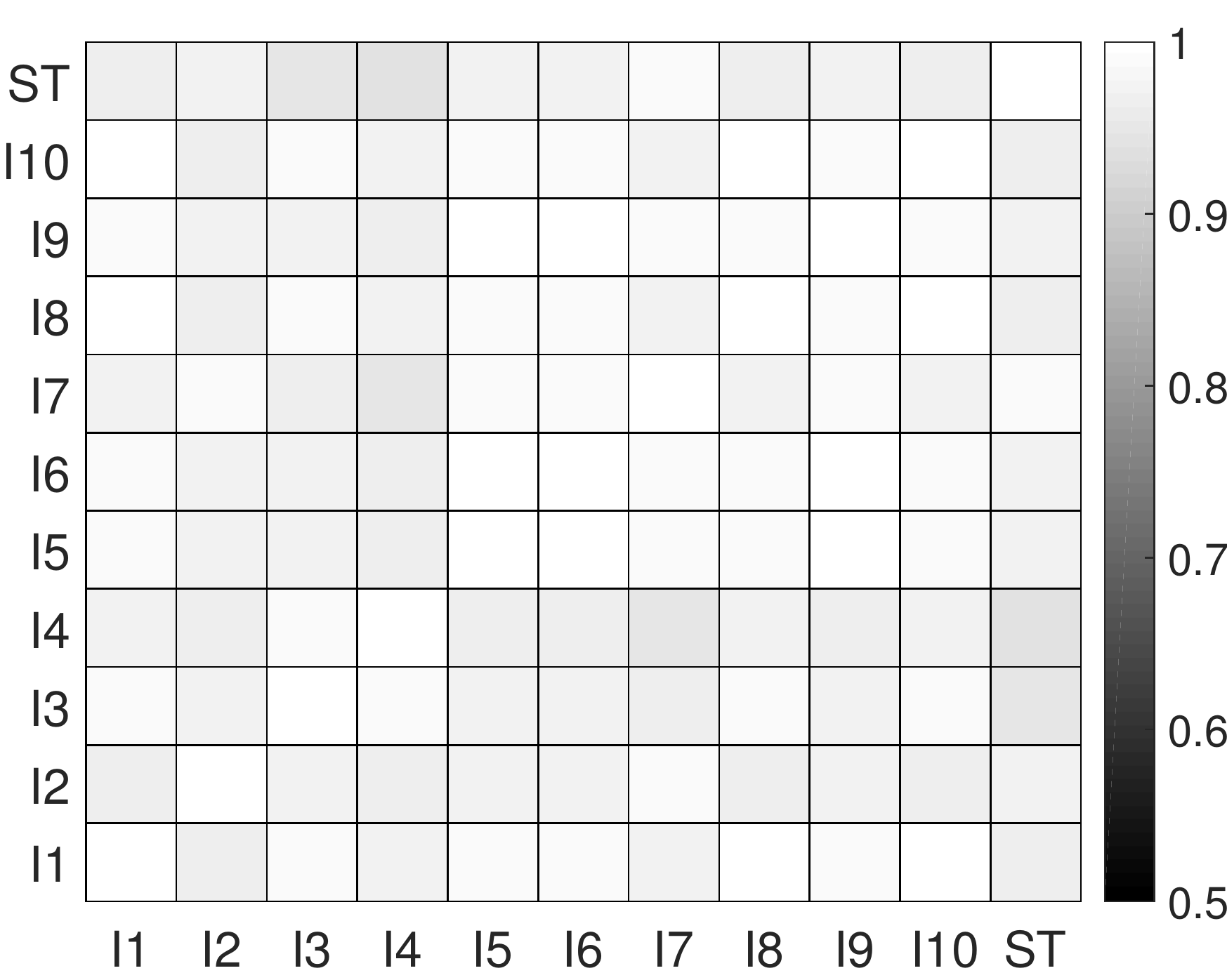}
    }
    \subfigure[\small{Twitter (LT)}]{
      \includegraphics[width=.14\textwidth]{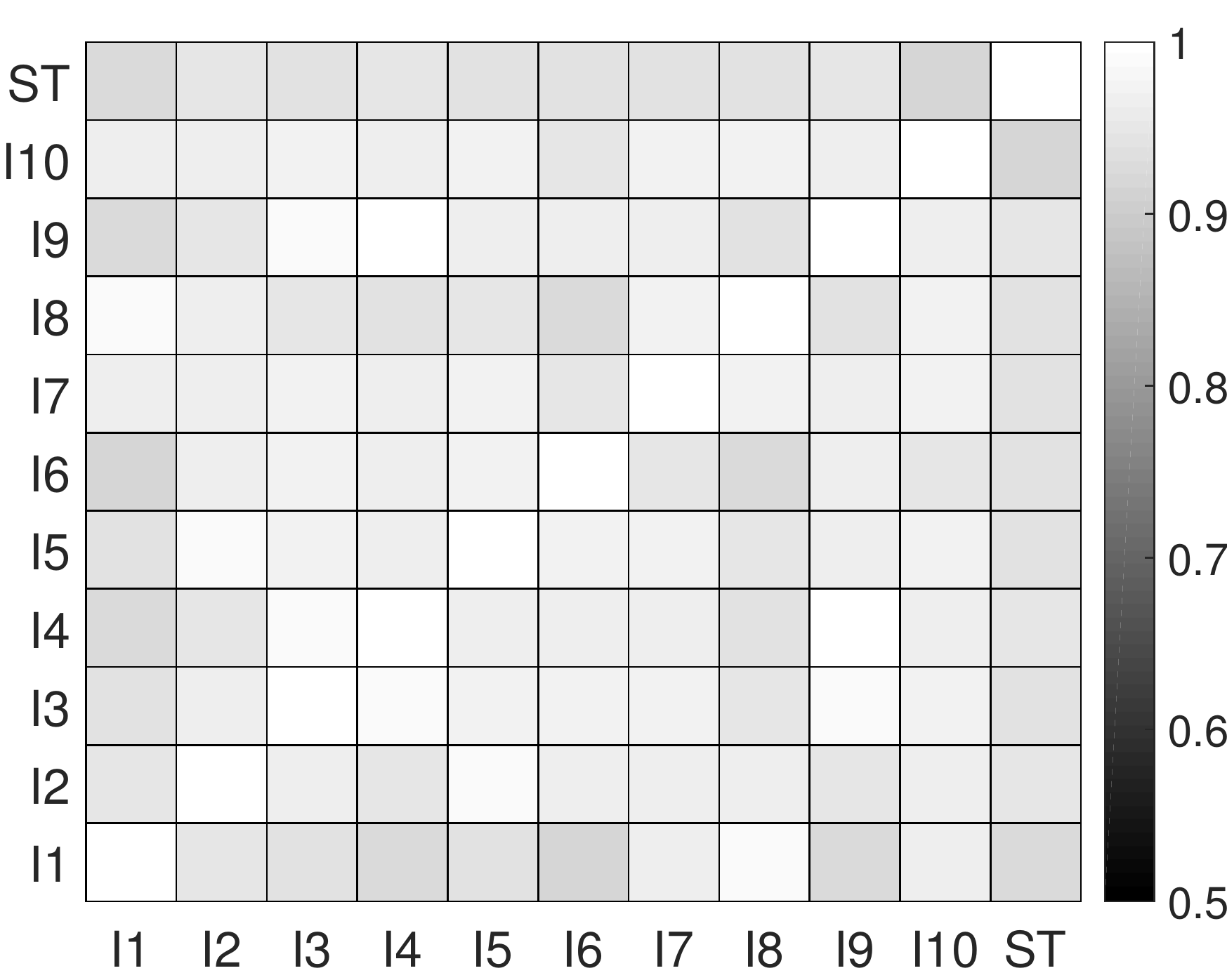}
    }
    \subfigure[\small{Twitter (IC)}]{
      \includegraphics[width=.14\textwidth]{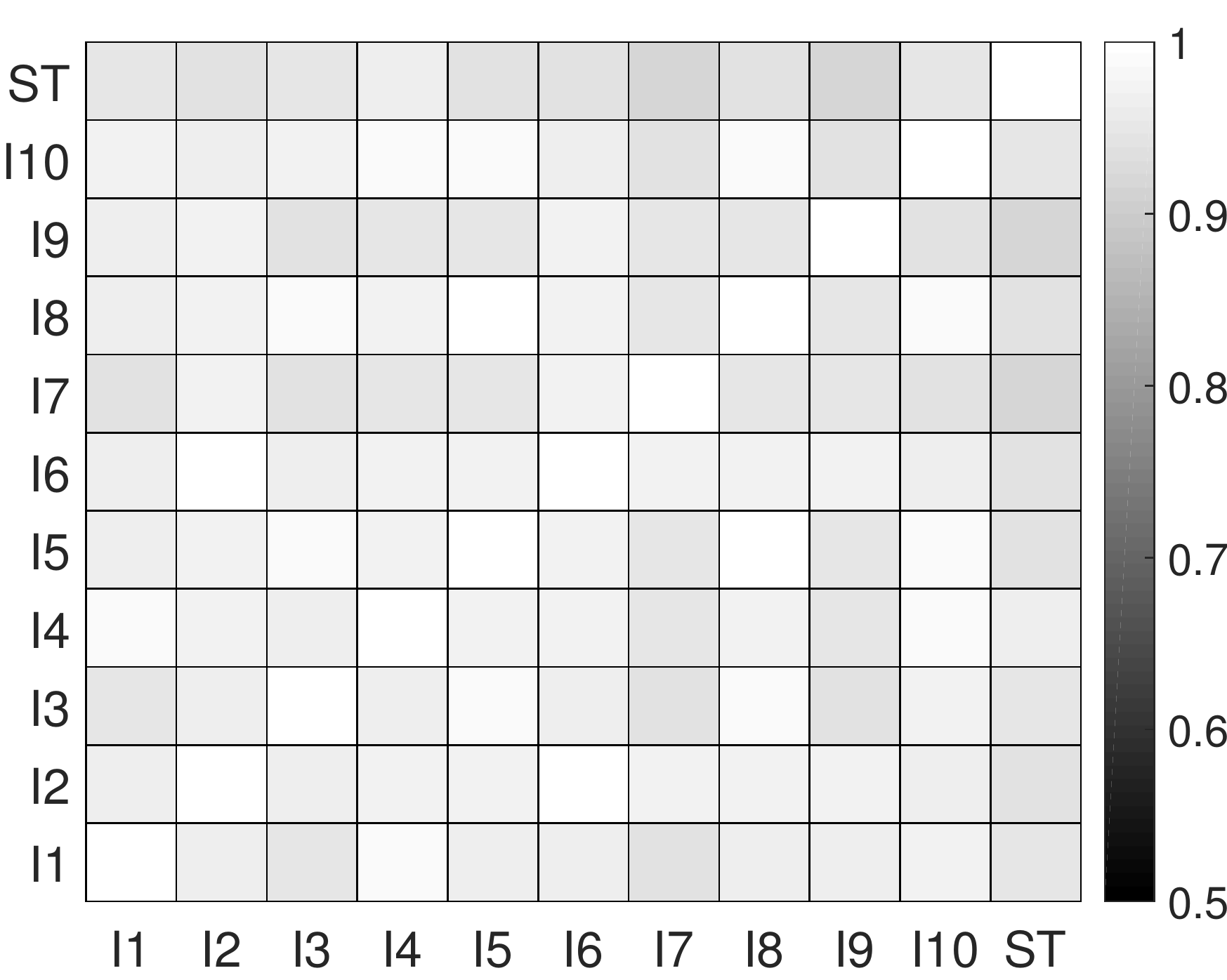}
    }
    \caption{Similarity among results in different instances.}
    \label{fig:effectiveness}
\end{figure*}

Fig.~\ref{fig:effectiveness} shows the results where I1, \ldots, I10 represent the results of the first, \ldots, tenth instances, respectively. ST denotes the result obtained by computing the influential vertices directly from the final snapshot using our sampling methods without any updates. Fig.~\ref{fig:effectiveness} shows that the outcomes from different instances are very similar, and they are similar to the outcome from ST, too. The minimum similarity is over 90\%.

\subsubsection{Scalability \& Comparison with~\cite{yang2017tracking}}\label{subsec:individual_sca}
\begin{figure*}[t]
  \centering
    \subfigure[wiki-Vote]{
      \includegraphics[width=.17\textwidth]{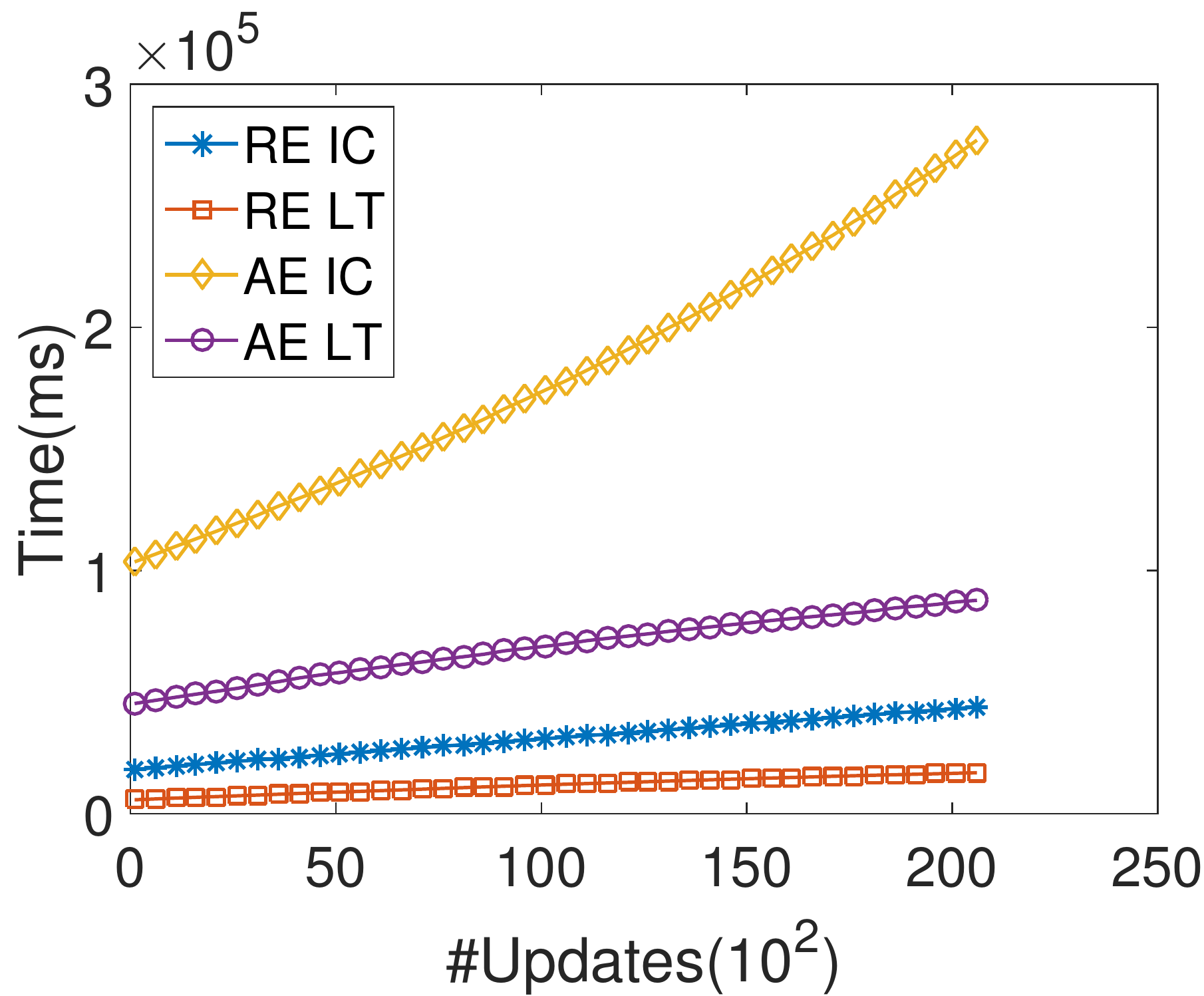}
    }
    \subfigure[Flixster]{
      \includegraphics[width=.17\textwidth]{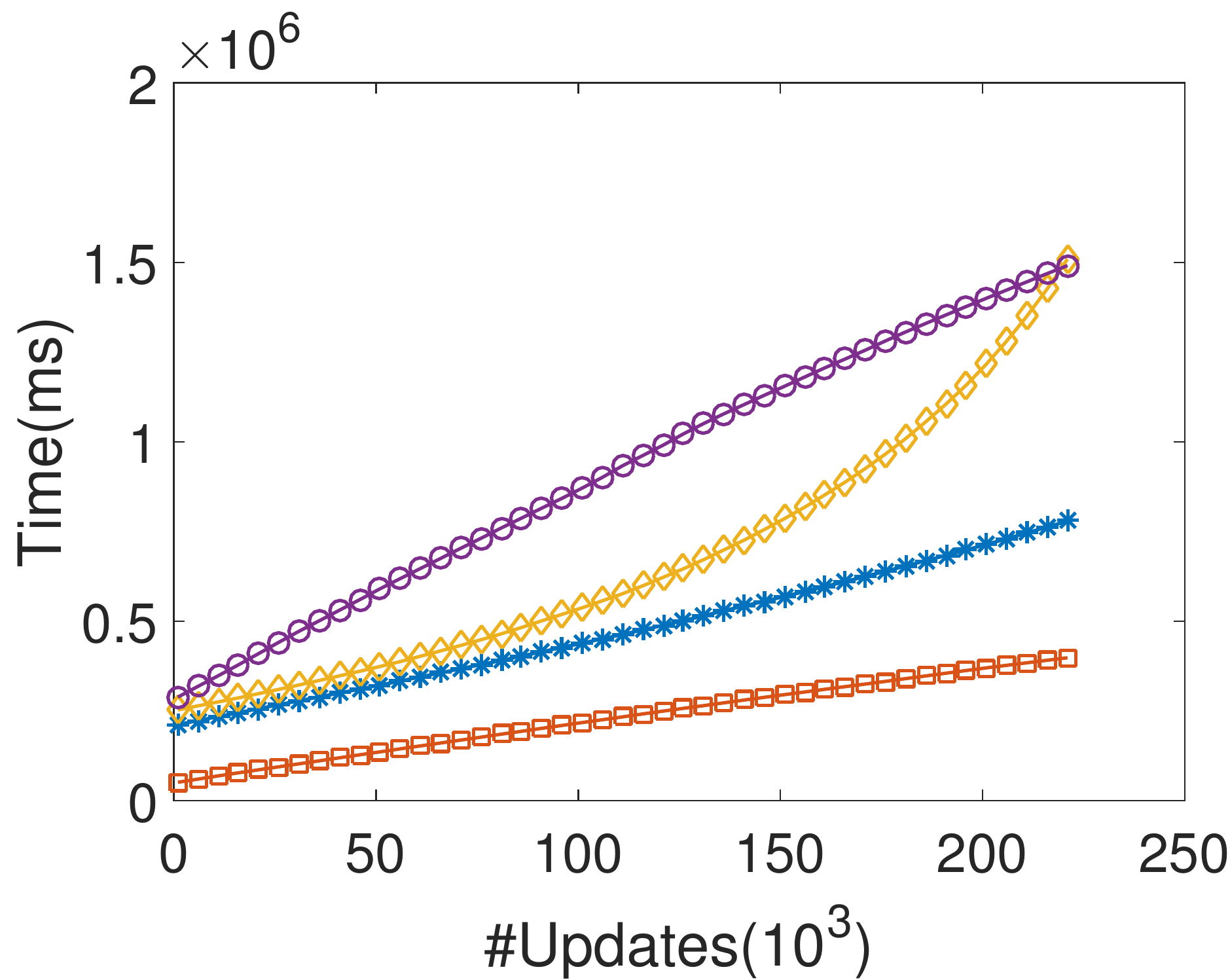}
    }
    \subfigure[soc-Pokec]{
      \includegraphics[width=.17\textwidth]{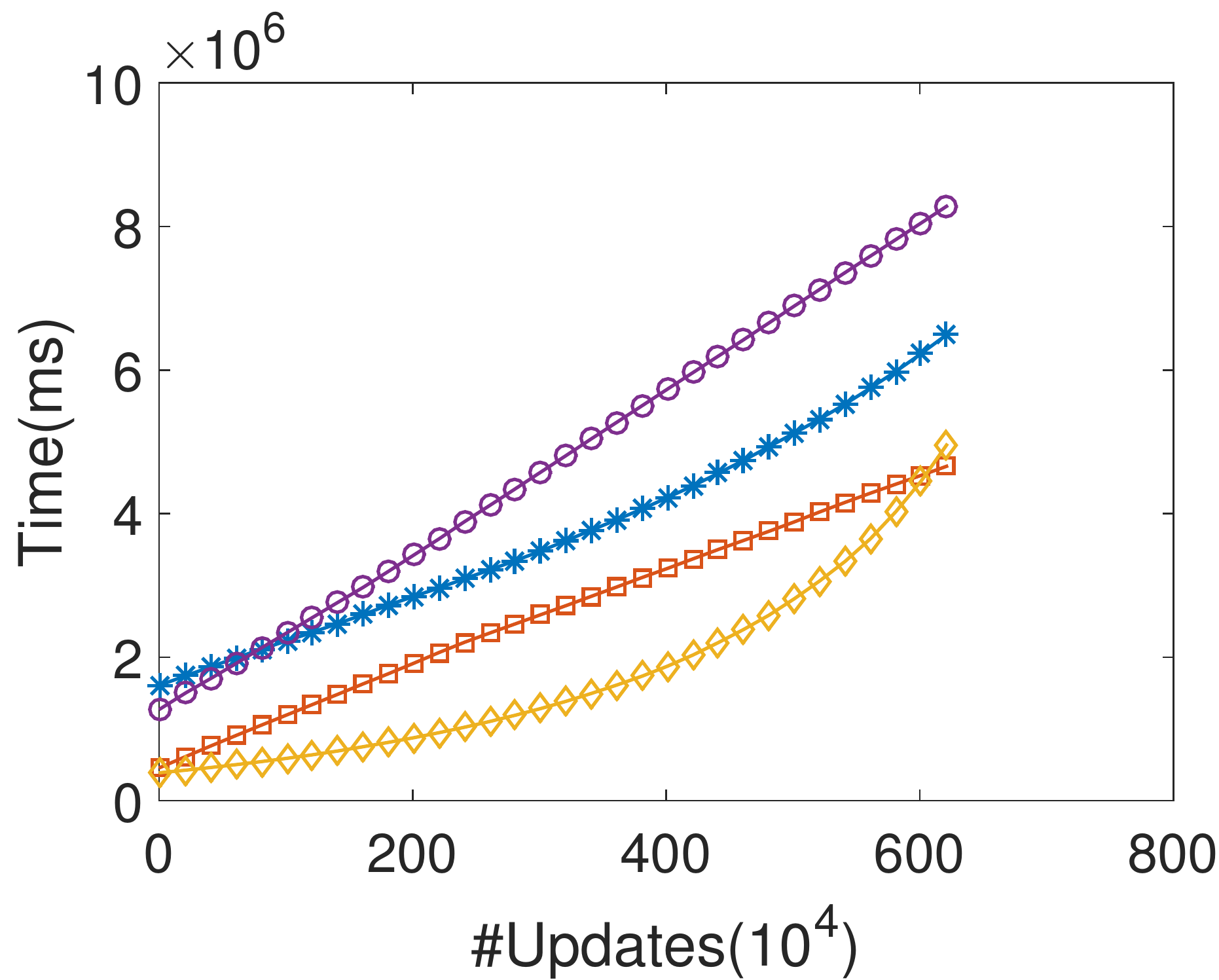}
    }
    \subfigure[flickr-growth]{
      \includegraphics[width=.17\textwidth]{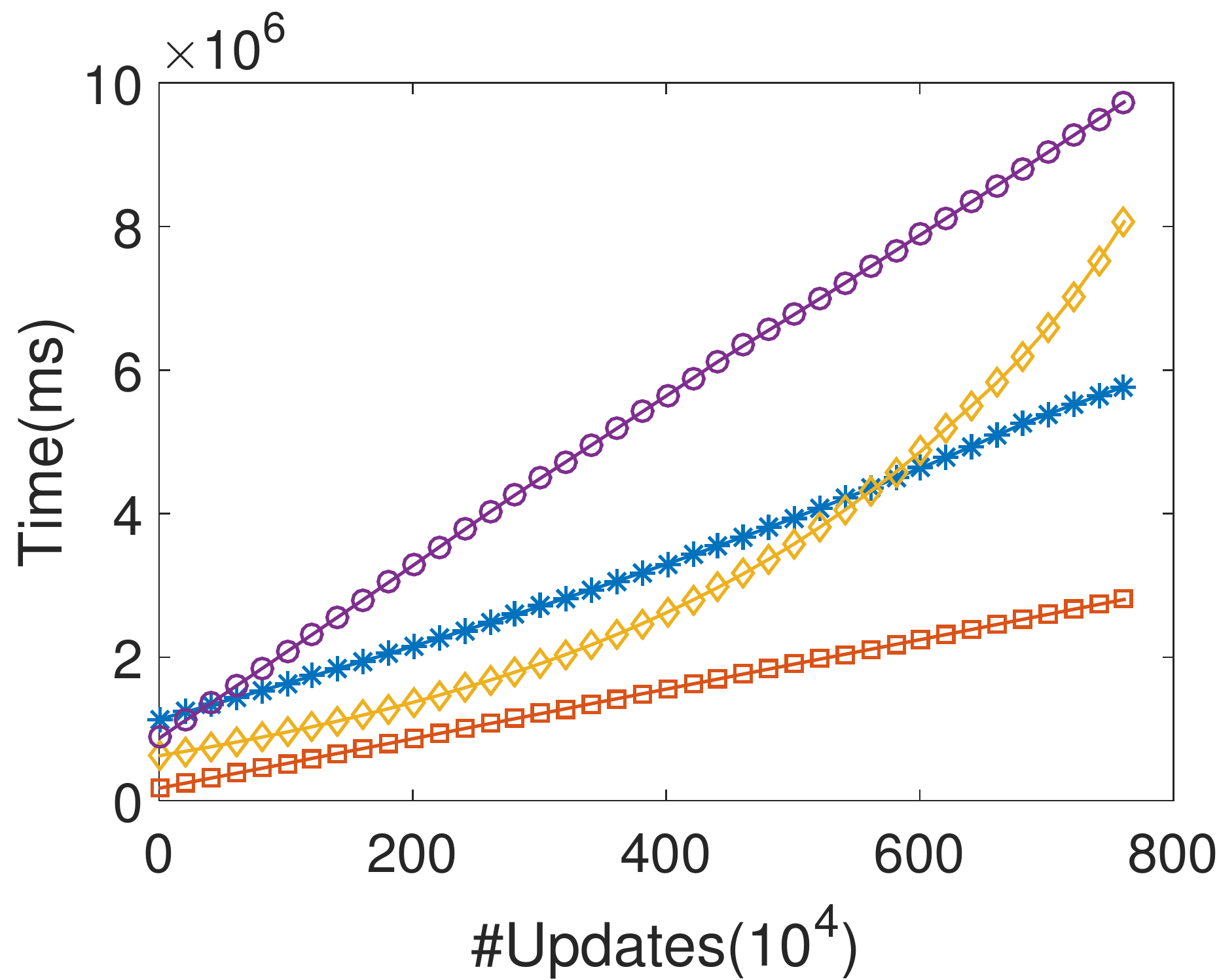}
    }
    \subfigure[Twitter]{
      \includegraphics[width=.17\textwidth]{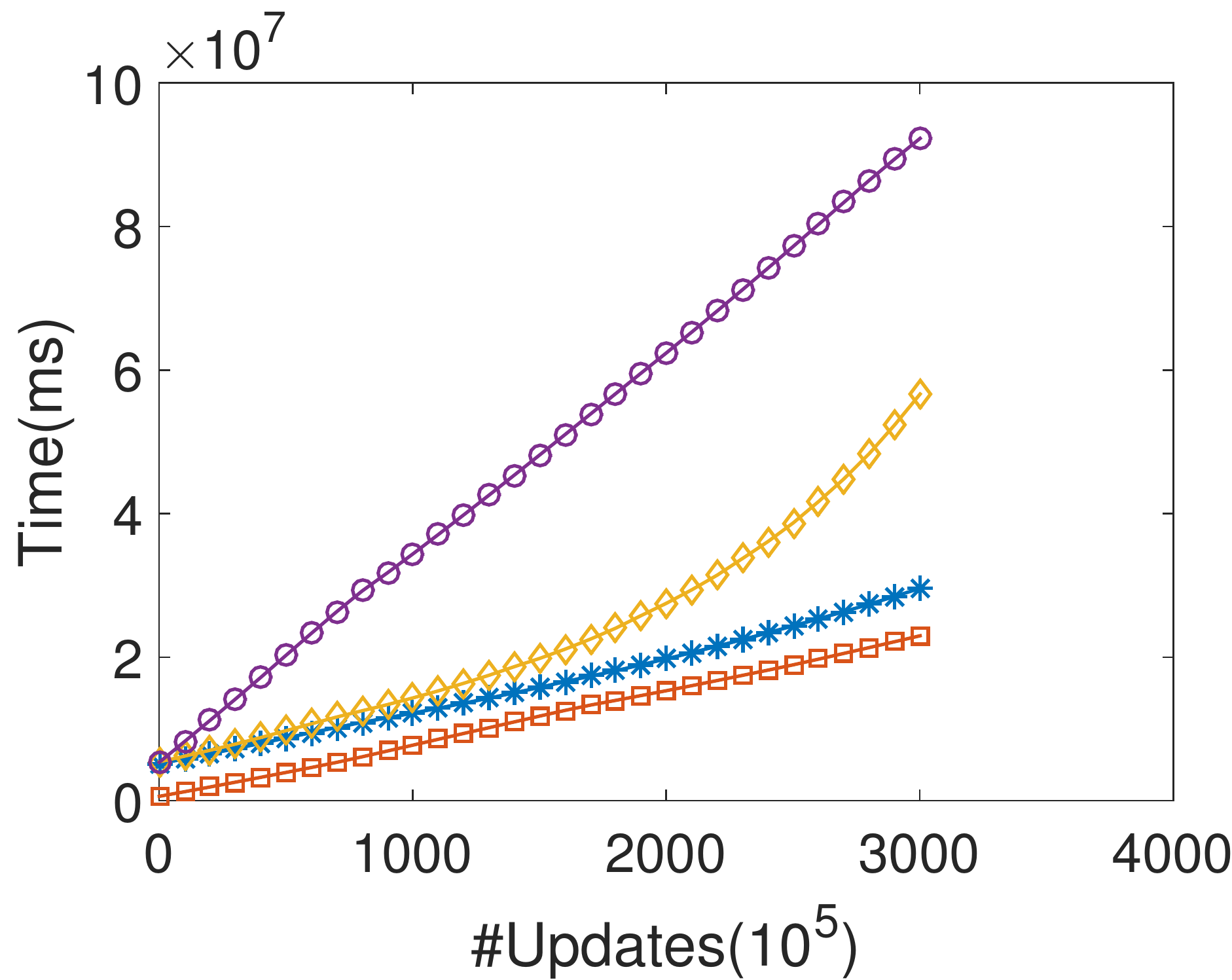}
    }
    \caption{Scalability (Tracking Top-$k$ Influential Individuals).}
    \label{fig:efficiency}
\end{figure*}

We also tested the scalability of our algorithm and the top-$k$ influential vertice tracking algorithm in~\cite{yang2017tracking}. Fig.~\ref{fig:efficiency} shows the average running time with respect to the number of updates processed, where ``RE IC'' (RE is short for relative error) is our algorithm under the IC model, while ``AE IC'' (AE is short for absolute error) stands for the algorithm in~\cite{yang2017tracking} under the IC model. The average is taken on the running times of the 10 instances. \updates{The point at \#Updates=0 of each curve in Fig.~\ref{fig:efficiency} represents the time spent by sampling enough RR sets on the base network.} In all cases, our algorithm handles the whole update stream in substantially shorter time than the baseline in~\cite{yang2017tracking}. Our algorithm under the LT model scales up roughly linearly. Under the IC model the running time sometimes increases more than linear. This is probably due to our experimental settings. For the LT model, the sum of propagation probabilities from all in-neighbors of a node is always 1, while in the IC model, at the beginning this value is roughly 0.9 but becomes 1 finally. So some updates of the IC model may lead to big change of the maximum influence or the average influence, and consequently the running time increases more than linearly.

\begin{figure*}[t]
  \centering
    \subfigure[wiki-Vote]{
      \includegraphics[width=.17\textwidth]{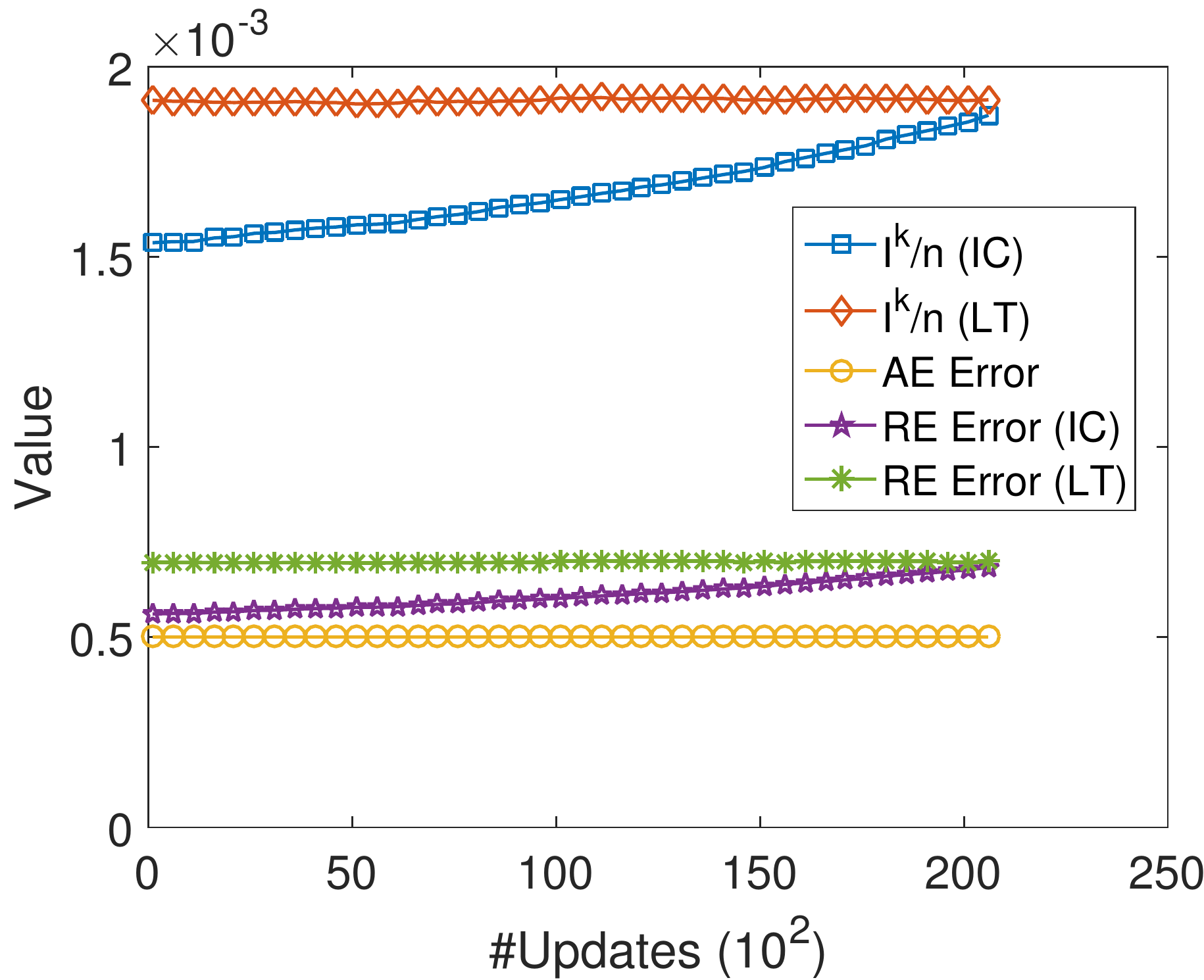}
    }
    \subfigure[Flixster]{
      \includegraphics[width=.17\textwidth]{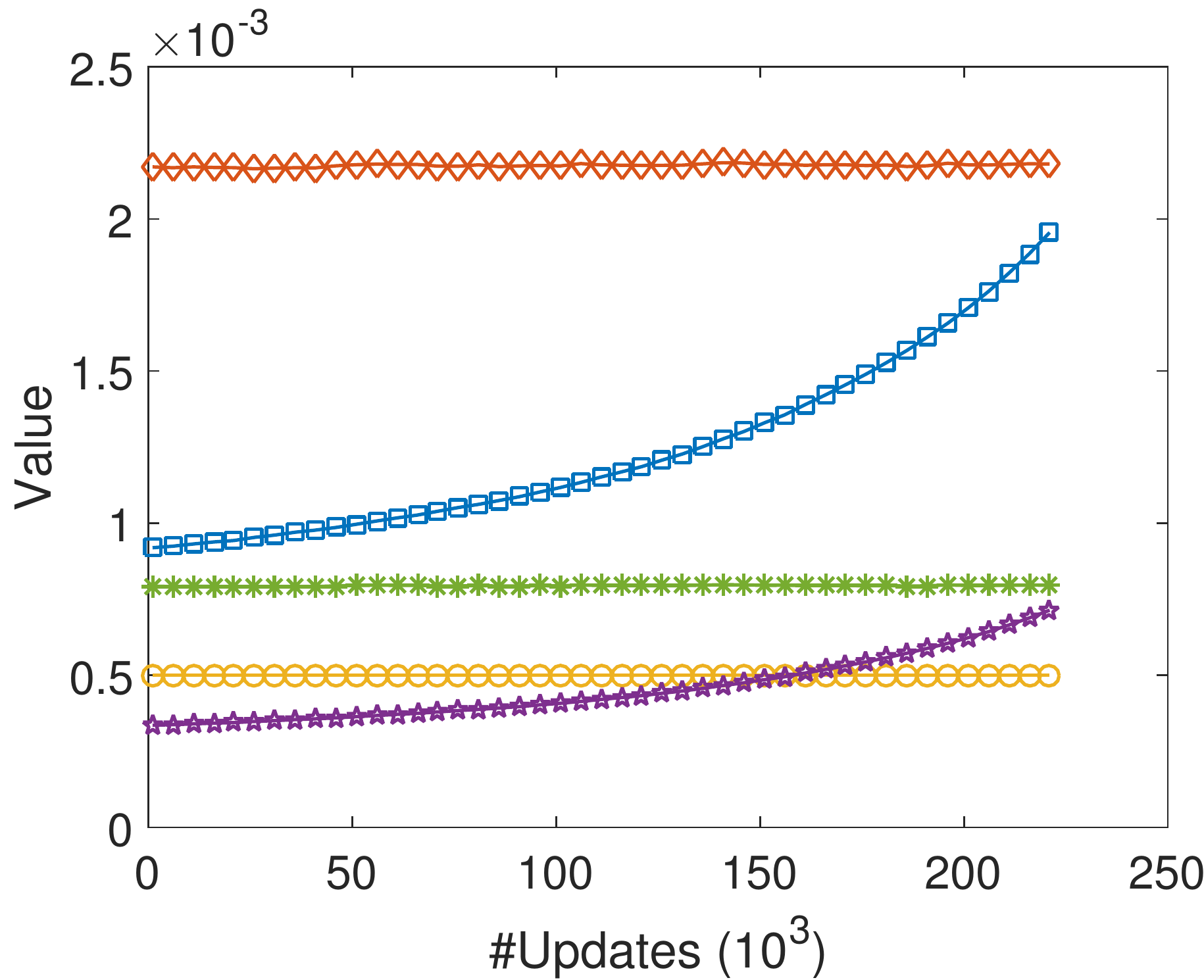}
    }
    \subfigure[soc-Pokec]{
      \includegraphics[width=.17\textwidth]{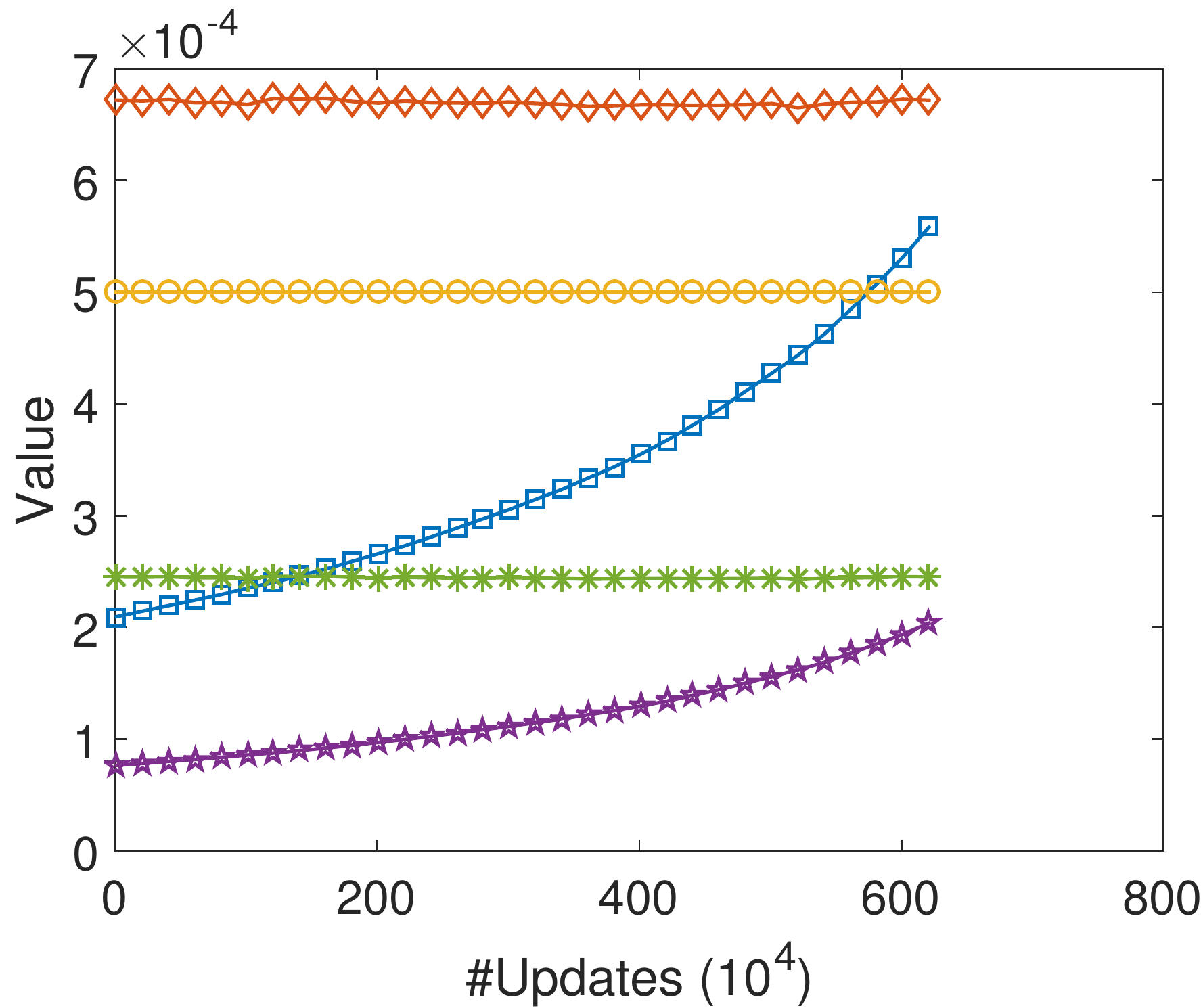}
    }
    \subfigure[flickr-growth]{
      \includegraphics[width=.17\textwidth]{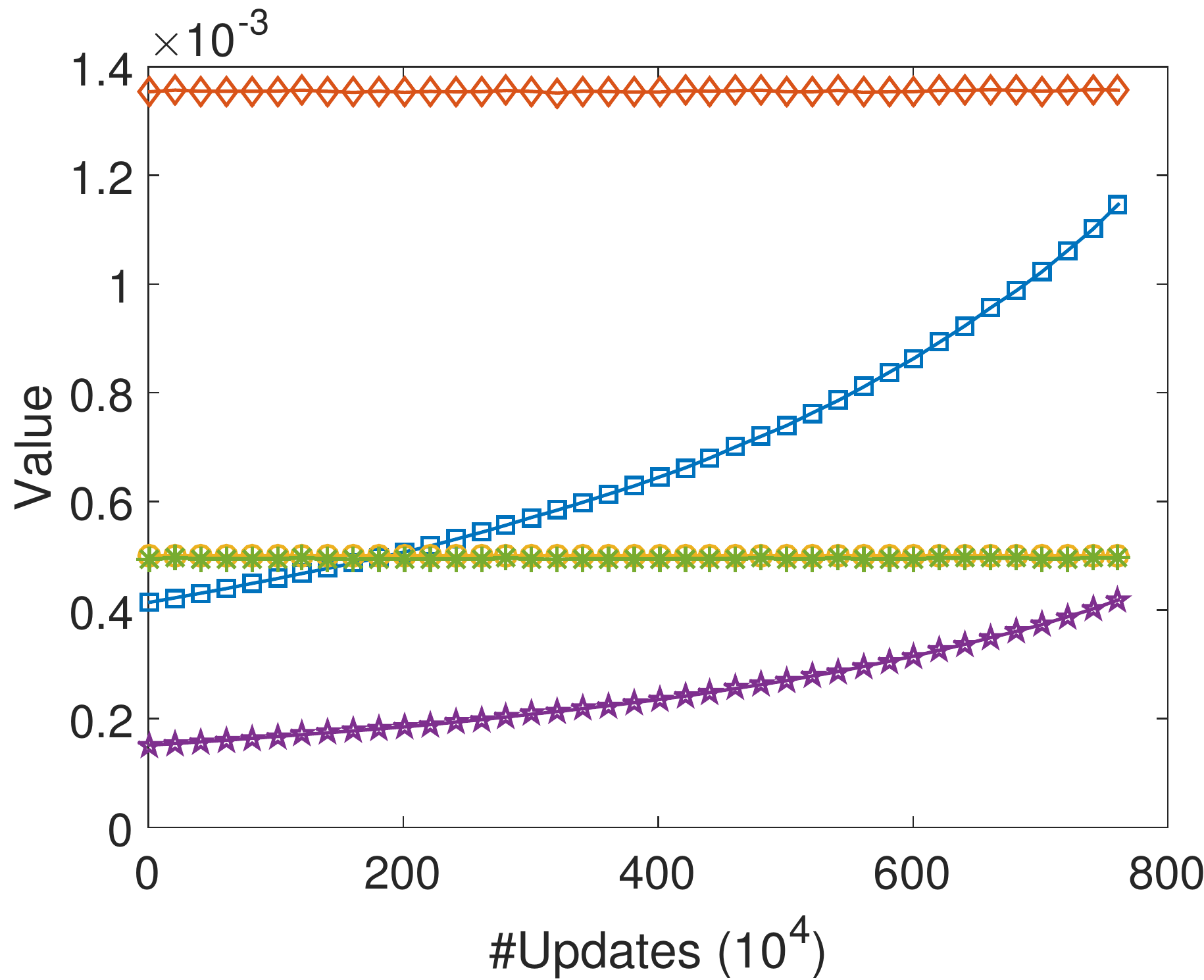}
    }
    \subfigure[Twitter]{
      \includegraphics[width=.17\textwidth]{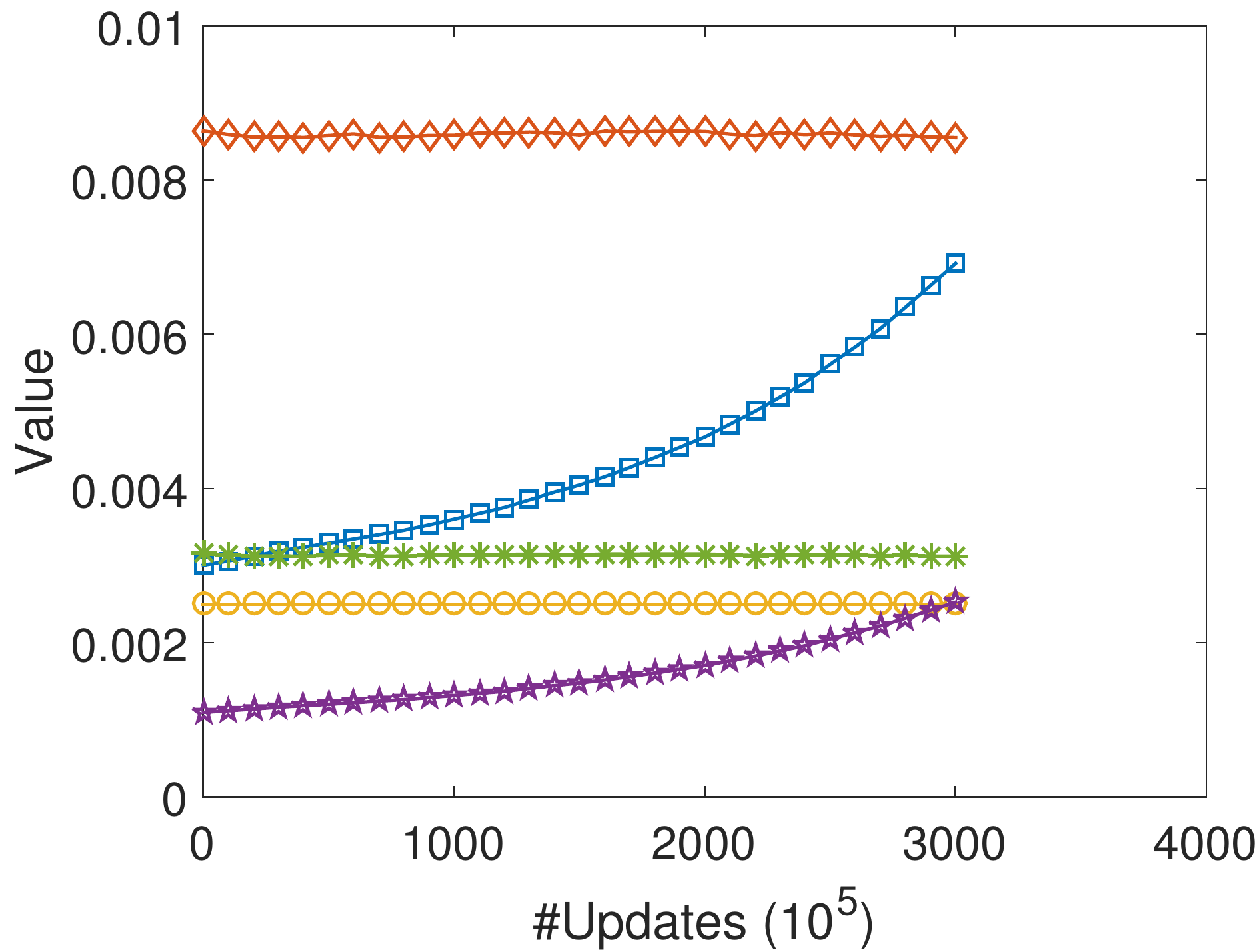}
    }
    \caption{Average $\frac{I^k}{n}$ over time.}
    \label{fig:topk_inf}
\end{figure*}

We also demonstrate the limitations of controlling $\max_{u \in S}{I^k-I_u}$ as an absolute error, where $S$ is the set of vertices mined. Fig.~\ref{fig:topk_inf} shows how the value $I^k$ varies over time, and so does the theoretically maximum error $\max_{u \in S}{\frac{I^k-I_u}{n}}$ over time. The value of $\frac{I^k}{n}$ is estimated by $\frac{\mathcal{D}_1^k}{M_1}$. Note that for the algorithm in~\cite{yang2017tracking}, we set $\epsilon_1$ to the same for both the IC and the LT models, where the parameter $\epsilon_1$ is exactly the maximum absolute error $\max_{u \in S}{\frac{I^k-I_u}{n}}$ we want to control. The maximum error of our algorithm (RE Error) is either only a little bigger or smaller than the maximum error of the baseline (AE Error). Moreover, we find that the value of $\frac{I^k}{n}$ varies over time, especially under the IC model. In Fig.~\ref{fig:topk_inf} (c), (d) and (e), sometimes the error of~\cite{yang2017tracking} (AE Error) is even greater than $\frac{I^k}{n}$, which makes the result meaningless because~\cite{yang2017tracking} will return all vertices as influential vertices. This demonstrates the limitation of~\cite{yang2017tracking}, which controls an absolute error.

We also report the running time of ST, which is directly applying our sampling methods without any updates on the final snapshot of each dataset to extract influential individuals. Table~\ref{tab:time_static} compares the running time of ST and the total time (denoted by ``Total'') that our algorithm samples RR sets on the base network and deals with the whole update stream. \updates{In Table~\ref{tab:time_static}, the running time of Total is at most $40$ times longer than that of ST. Thus, if we re-sample RR sets from scratch every time when the network updates, we probably can only deal with tens of updates within the same time as Total spends on all updates. However, the number of total updates is huge, tens of thousands or even hundreds of millions. This indicates that the non-incremental algorithm (re-sampling RR sets from scratch when the network updates) is not competitive at all.}

\begin{figure*}[t]
  \centering
    \subfigure[wiki-Vote (LT)]{
      \includegraphics[width=.17\textwidth]{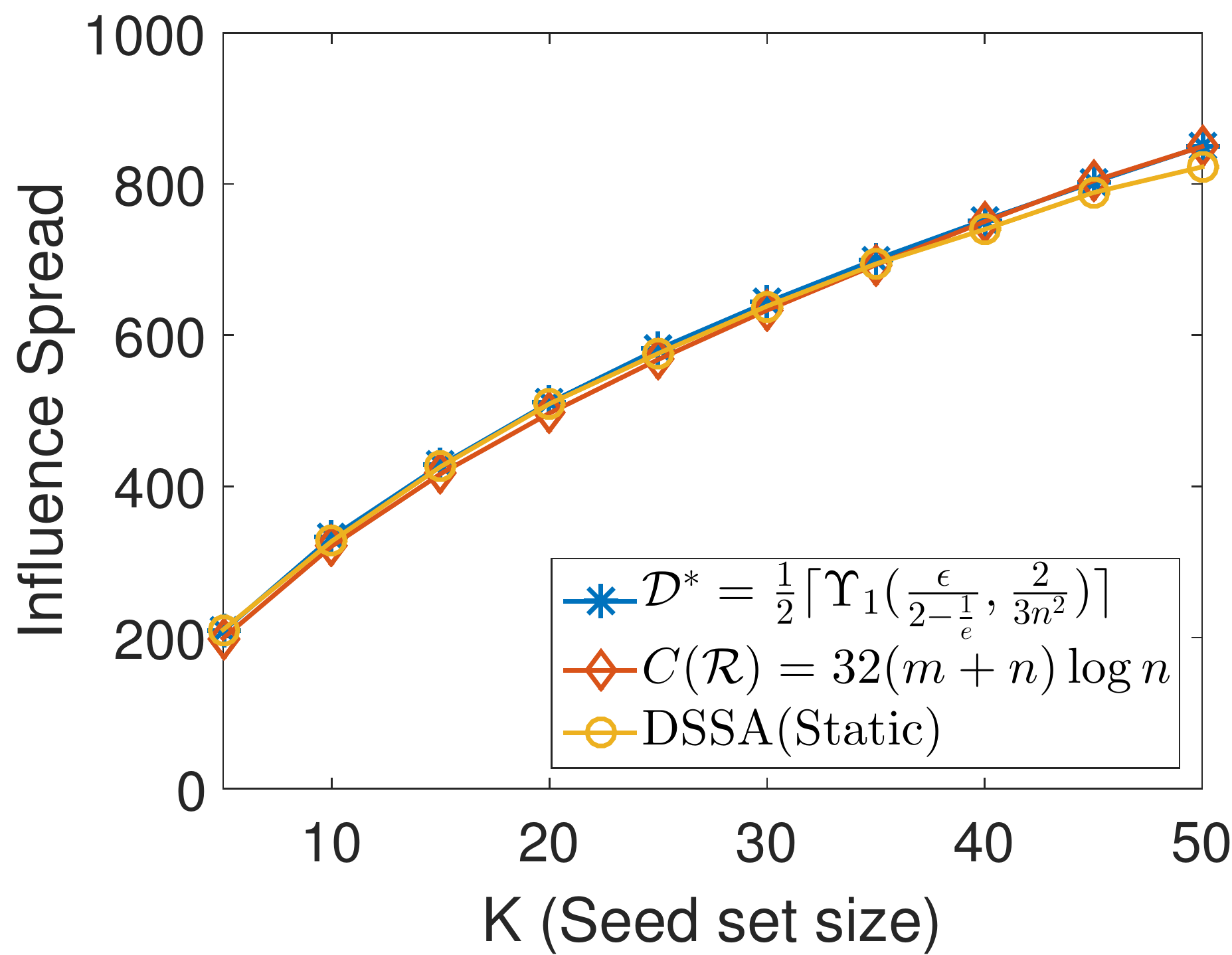}
    }
    \subfigure[Flixster (LT)]{
      \includegraphics[width=.17\textwidth]{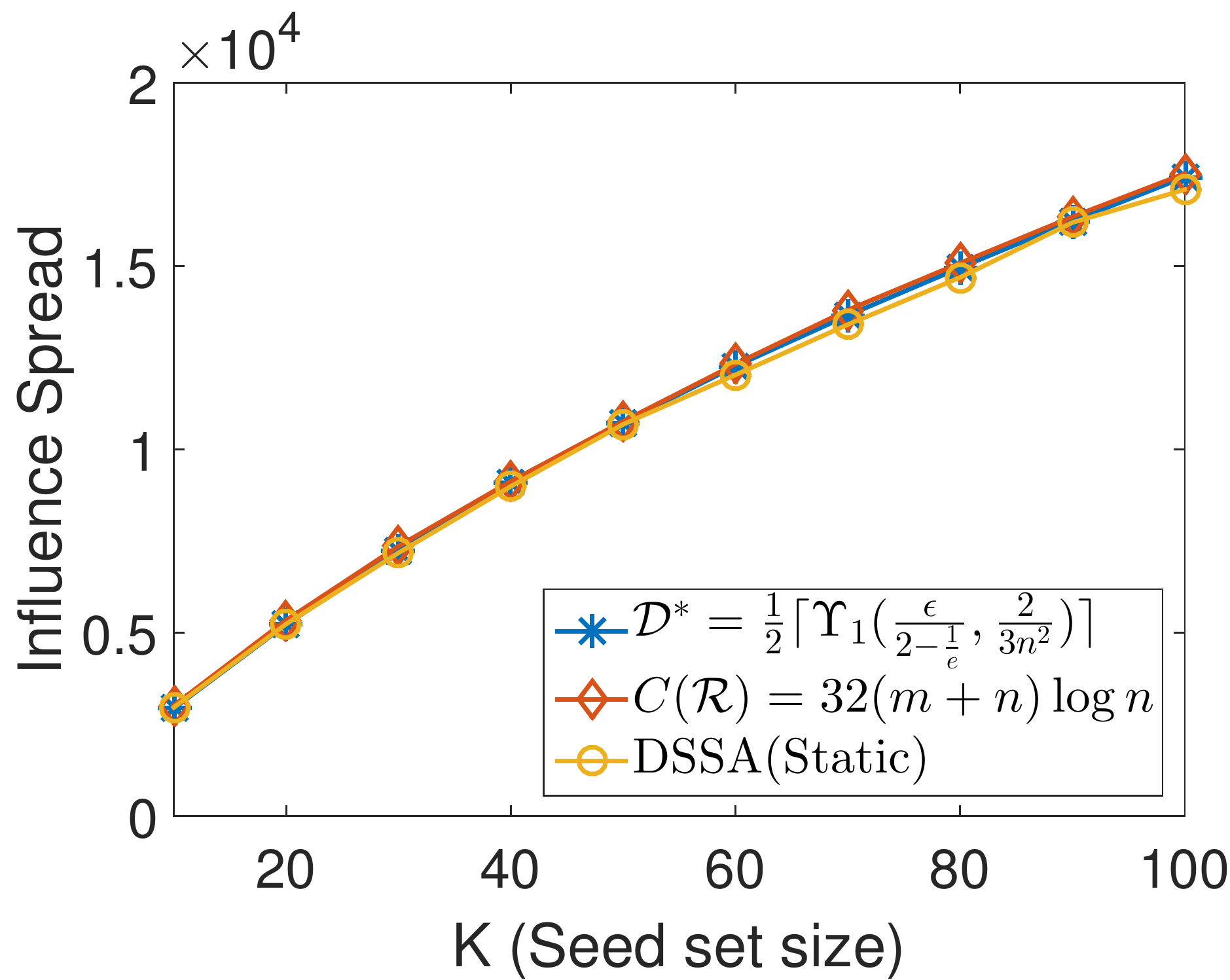}
    }
    \subfigure[soc-Pokec (LT)]{
      \includegraphics[width=.17\textwidth]{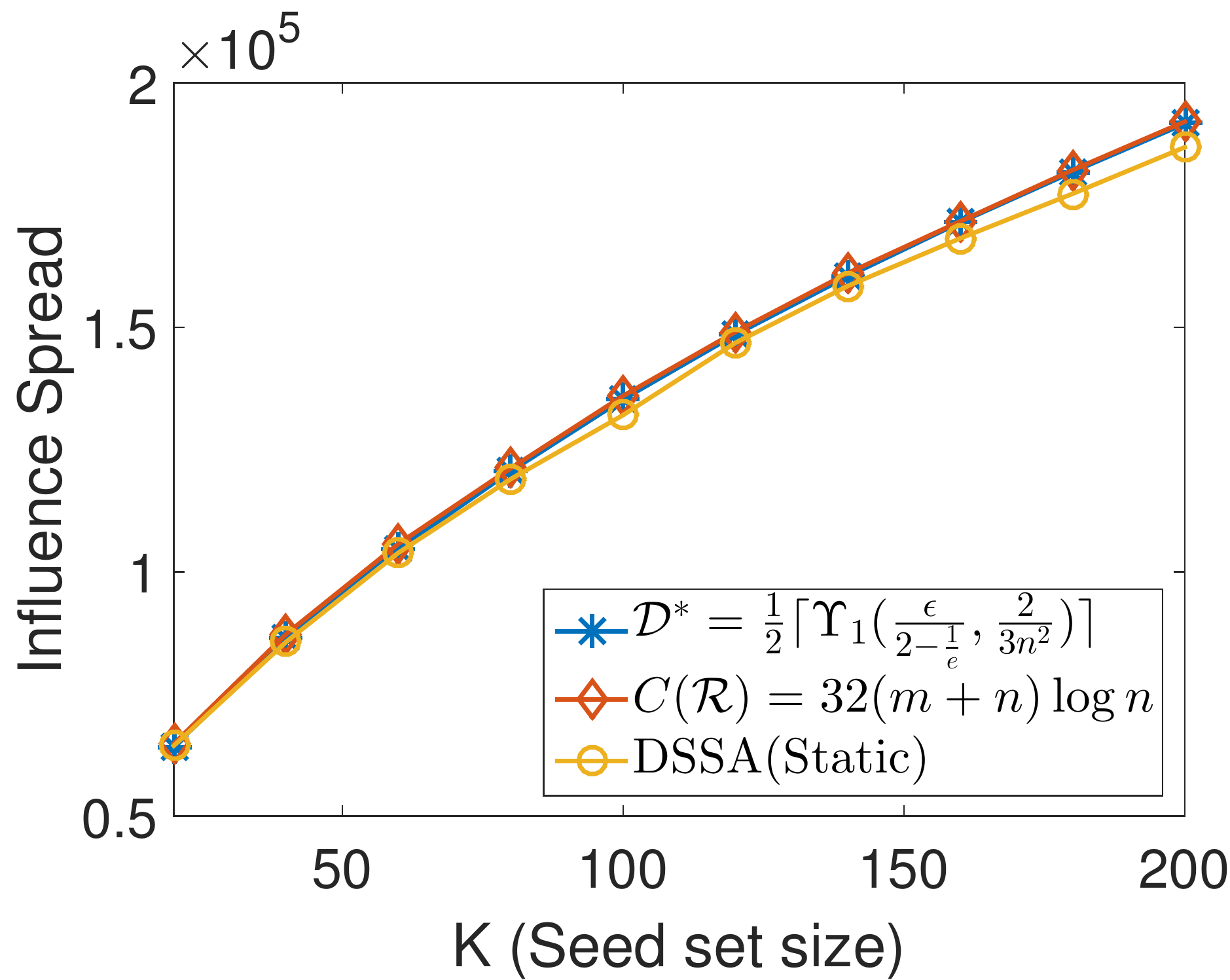}
    }
    \subfigure[flickr-growth (LT)]{
      \includegraphics[width=.17\textwidth]{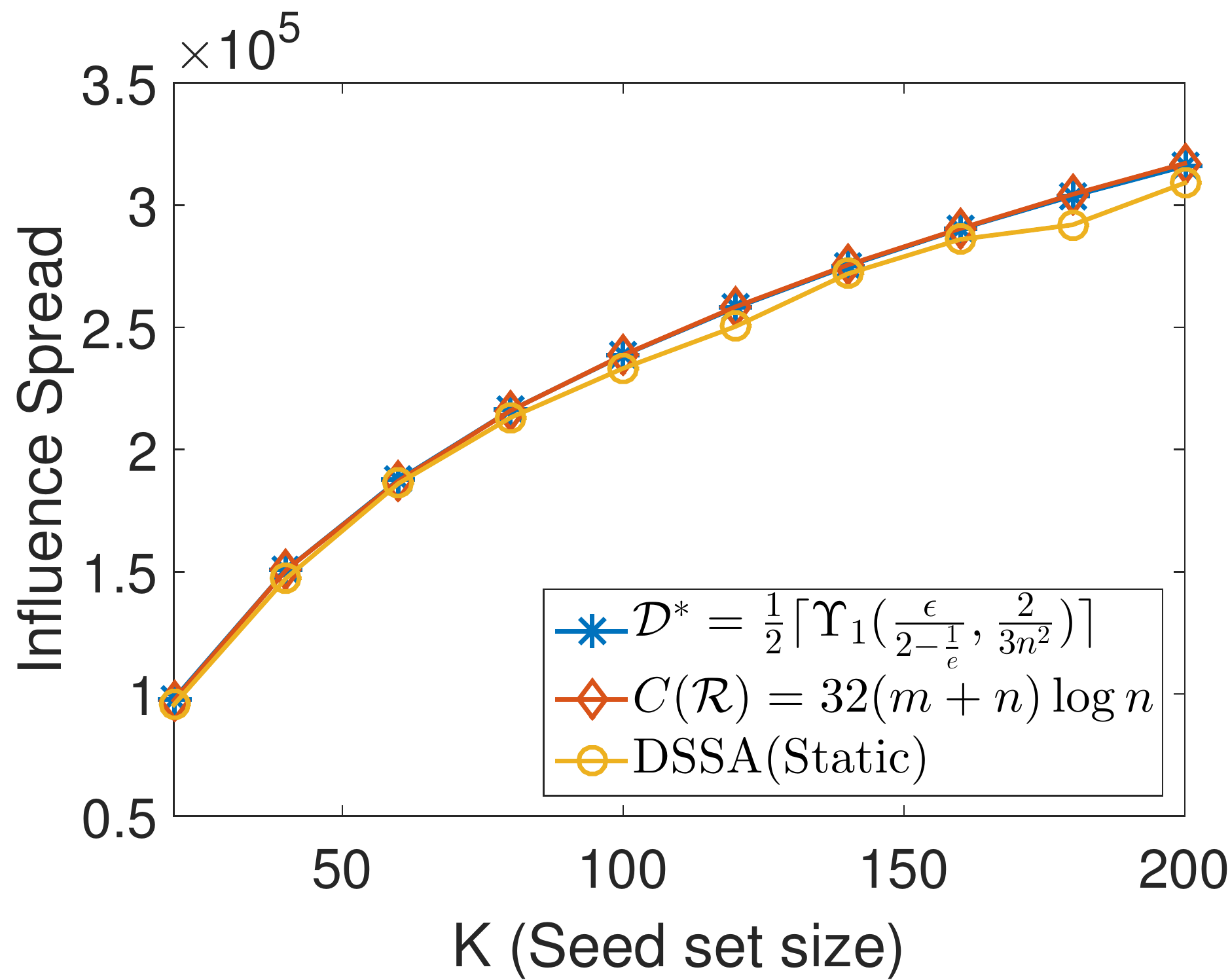}
    }
    \subfigure[Twitter (LT)]{
      \includegraphics[width=.17\textwidth]{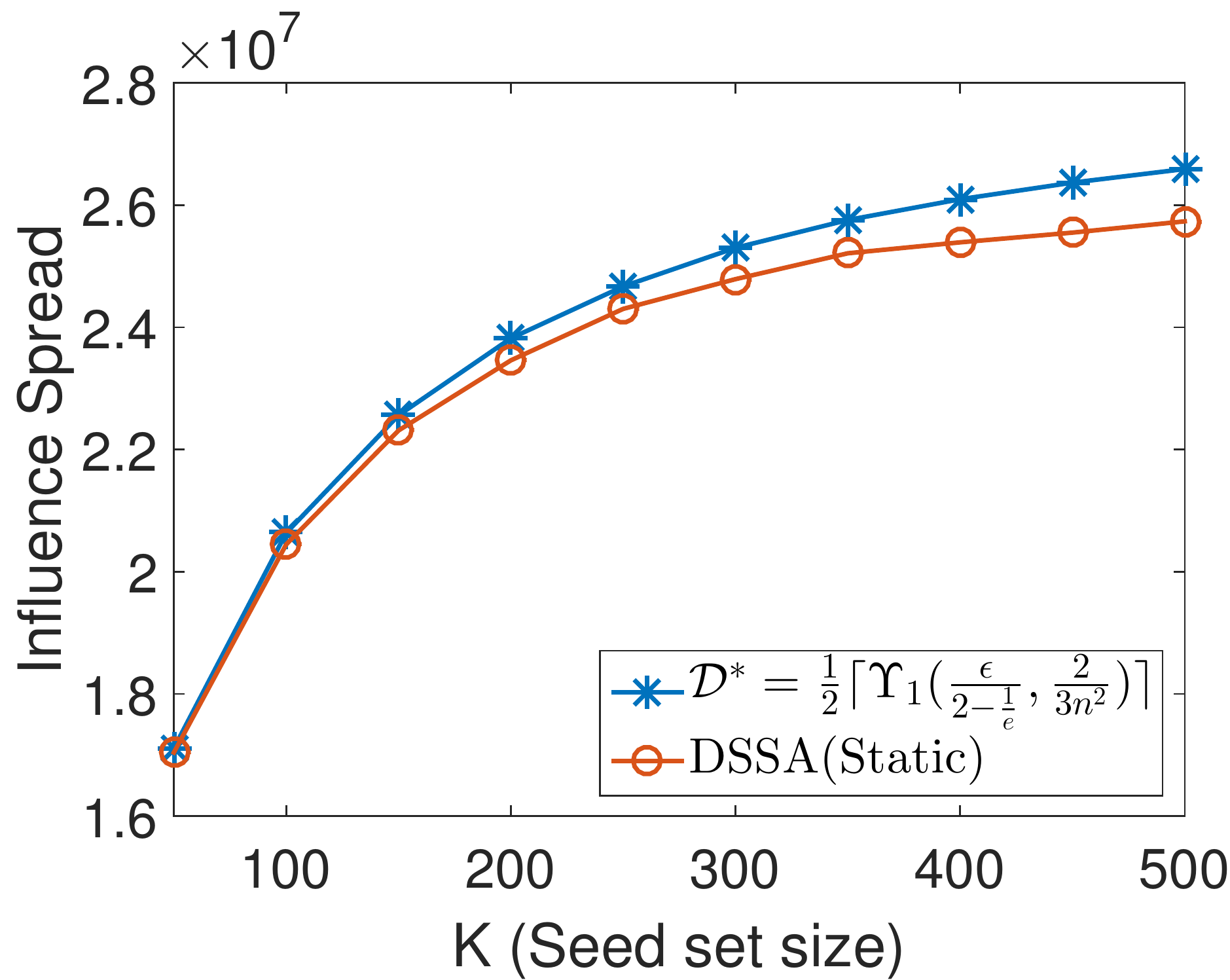}
    }
    \subfigure[wiki-Vote (IC)]{
      \includegraphics[width=.17\textwidth]{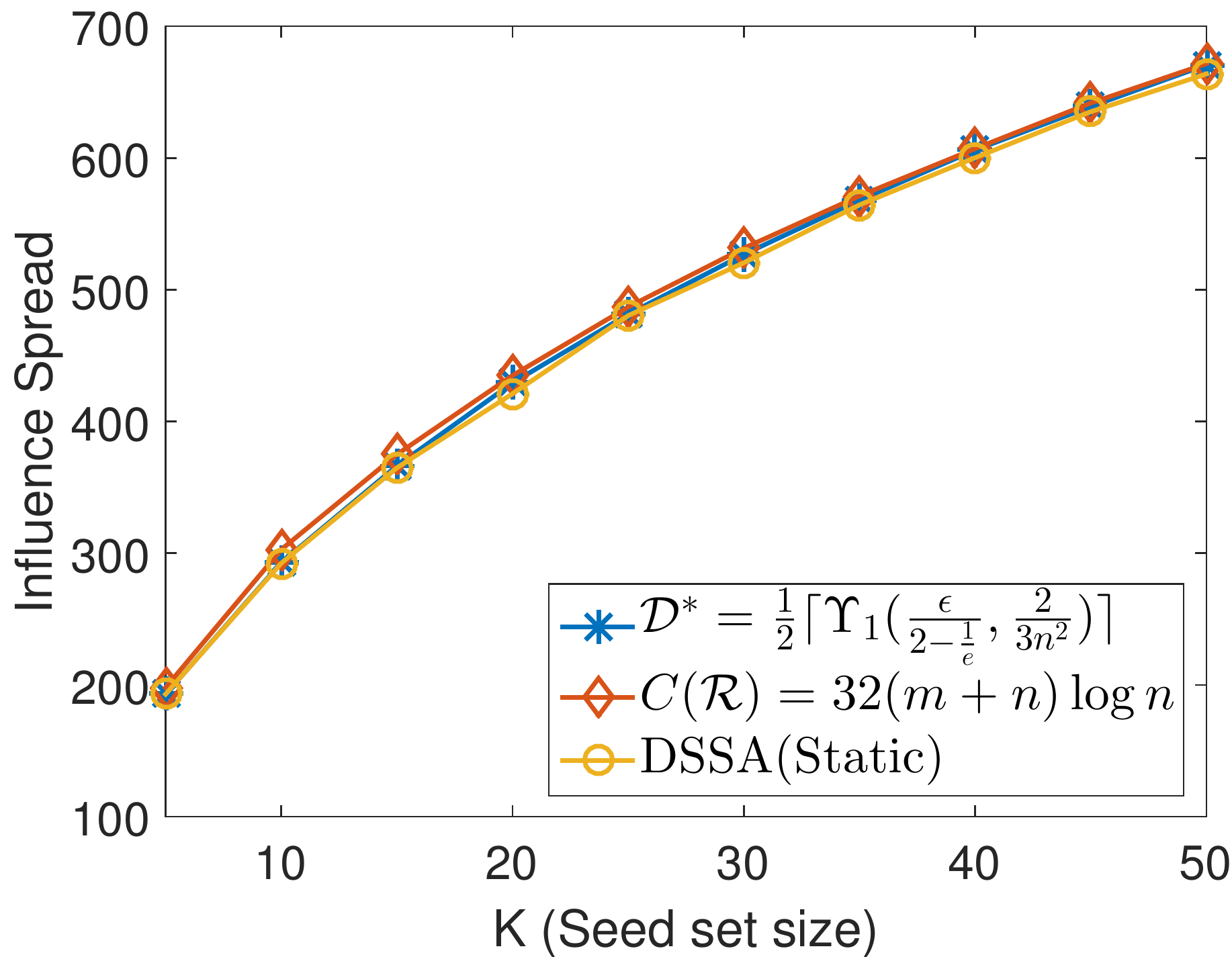}
    }
    \subfigure[Flixster (IC)]{
      \includegraphics[width=.17\textwidth]{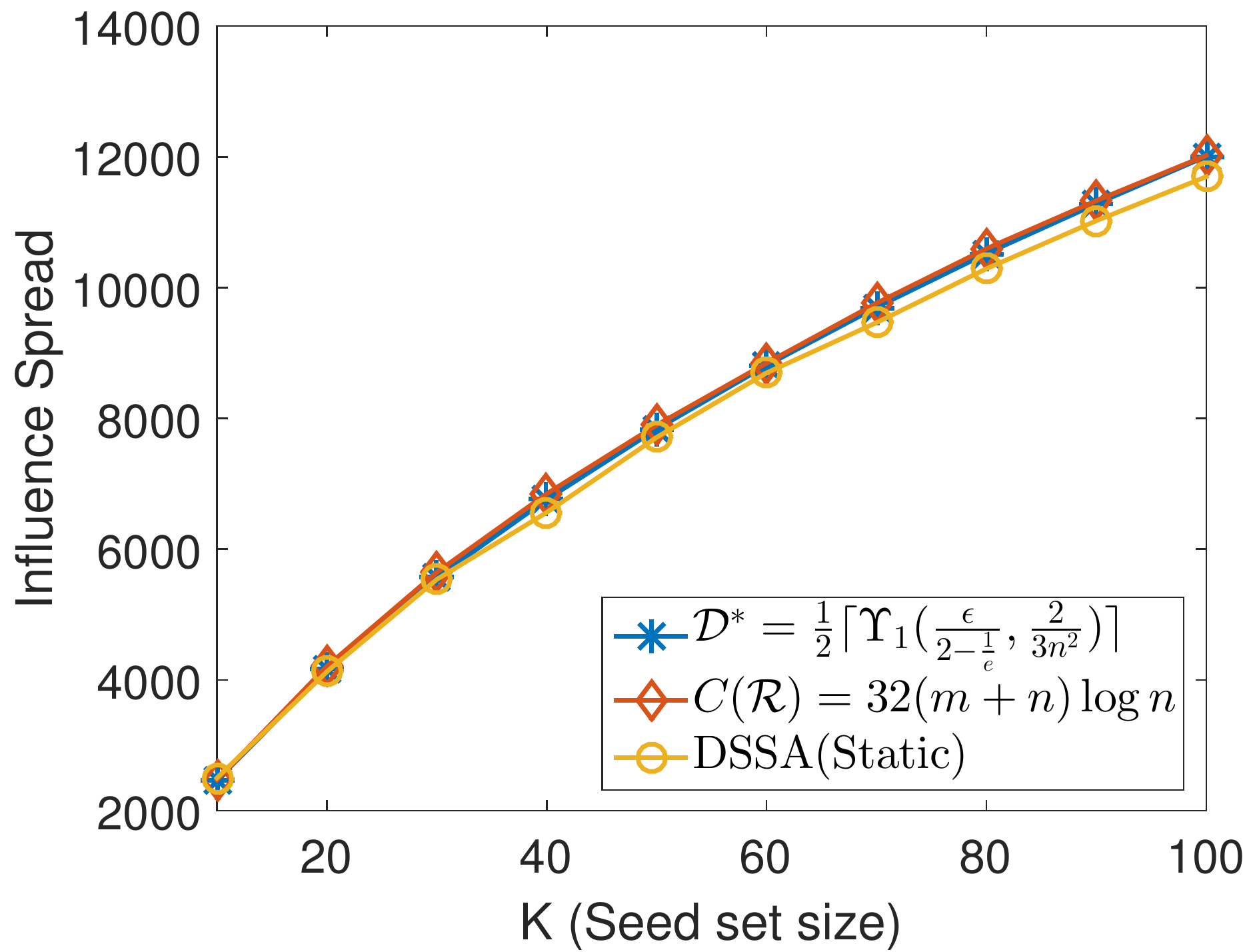}
    }
    \subfigure[soc-Pokec (IC)]{
      \includegraphics[width=.17\textwidth]{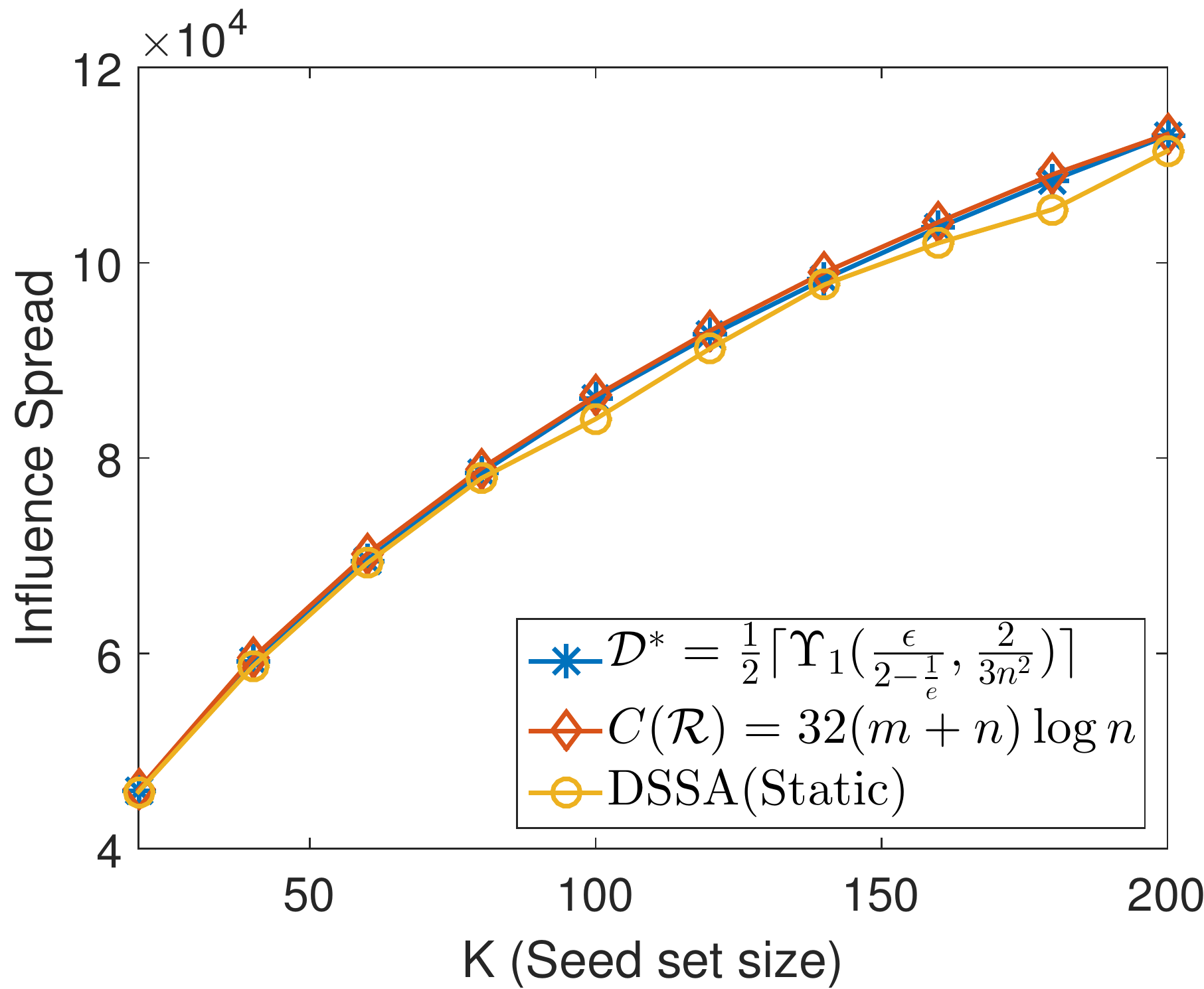}
    }
    \subfigure[flickr-growth (IC)]{
      \includegraphics[width=.17\textwidth]{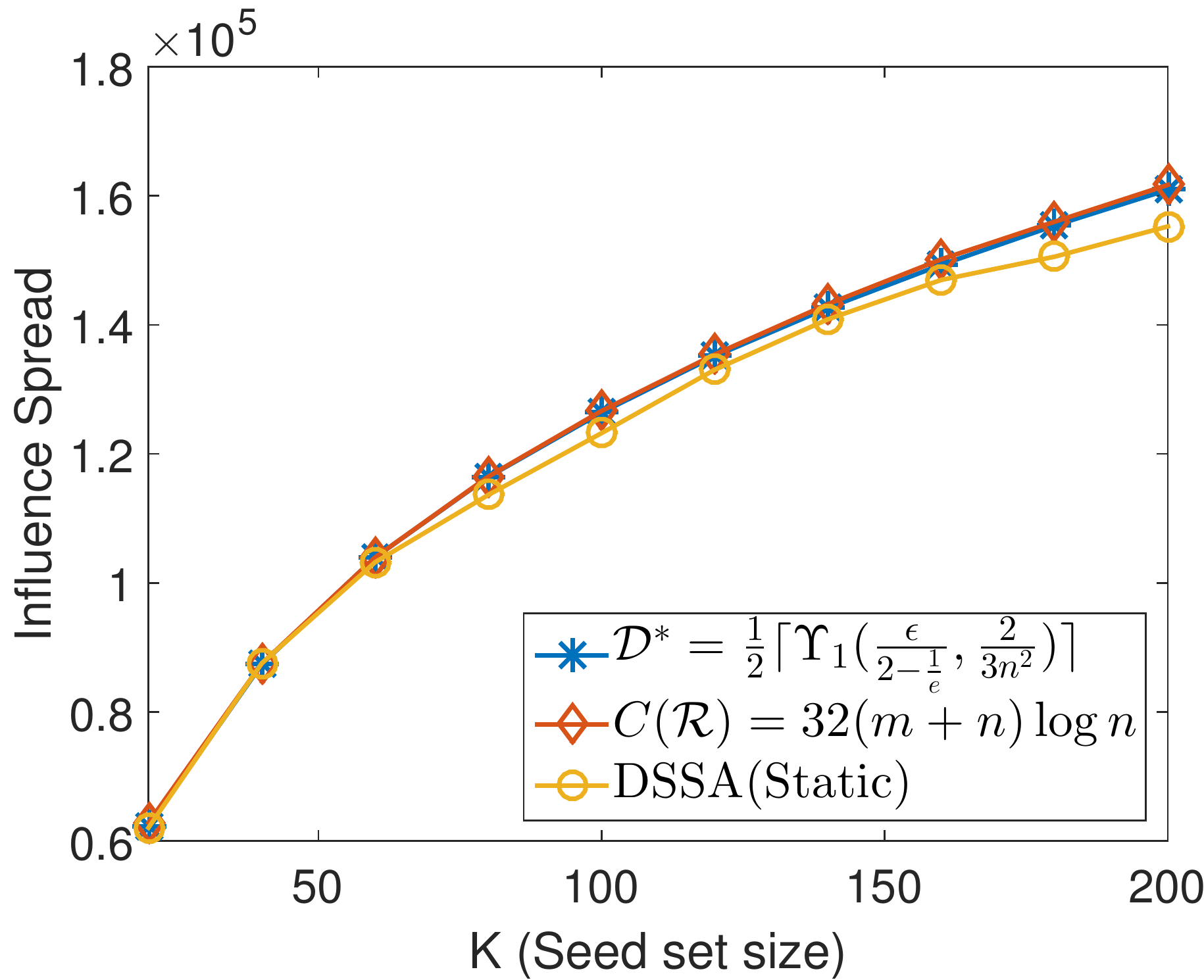}
    }
    \subfigure[Twitter (IC)]{
      \includegraphics[width=.17\textwidth]{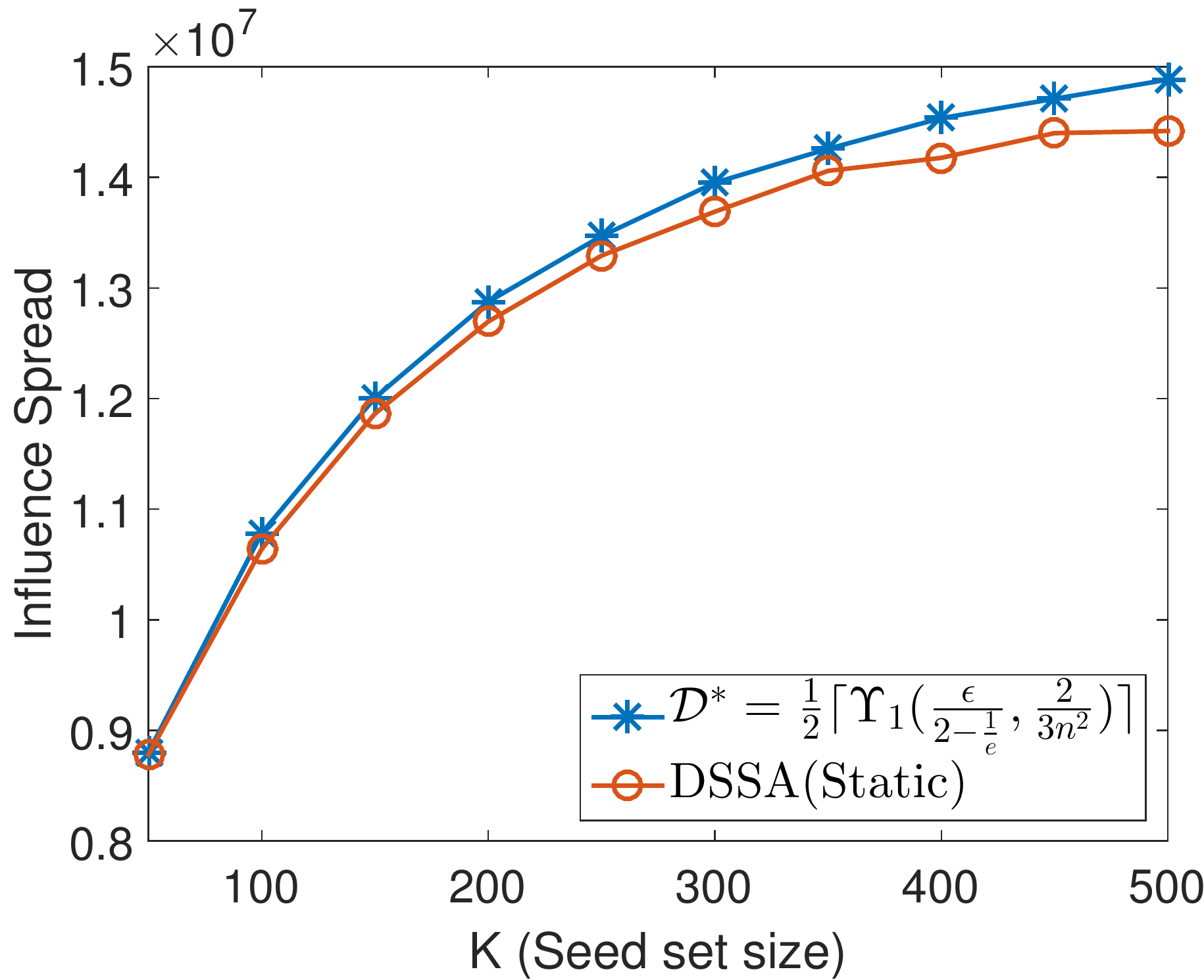}
    }
    \caption{Influence Spread v.s.\ Seed Set Size $K$.}
    \label{fig:im_inf}
\end{figure*}

\begin{figure*}[h]
  \centering
    \subfigure[wiki-Vote]{
      \includegraphics[width=.17\textwidth]{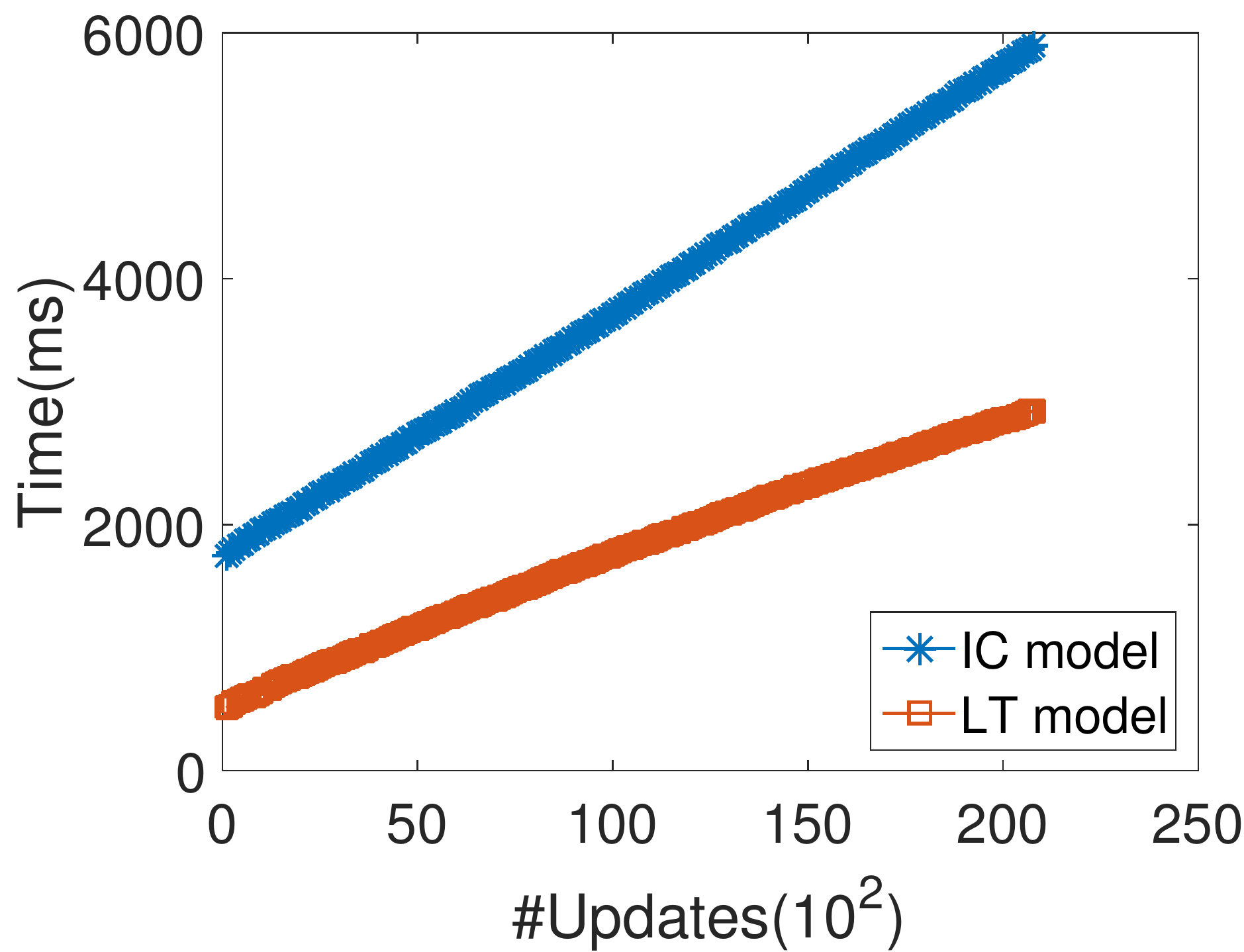}
    }
    \subfigure[Flixster]{
      \includegraphics[width=.17\textwidth]{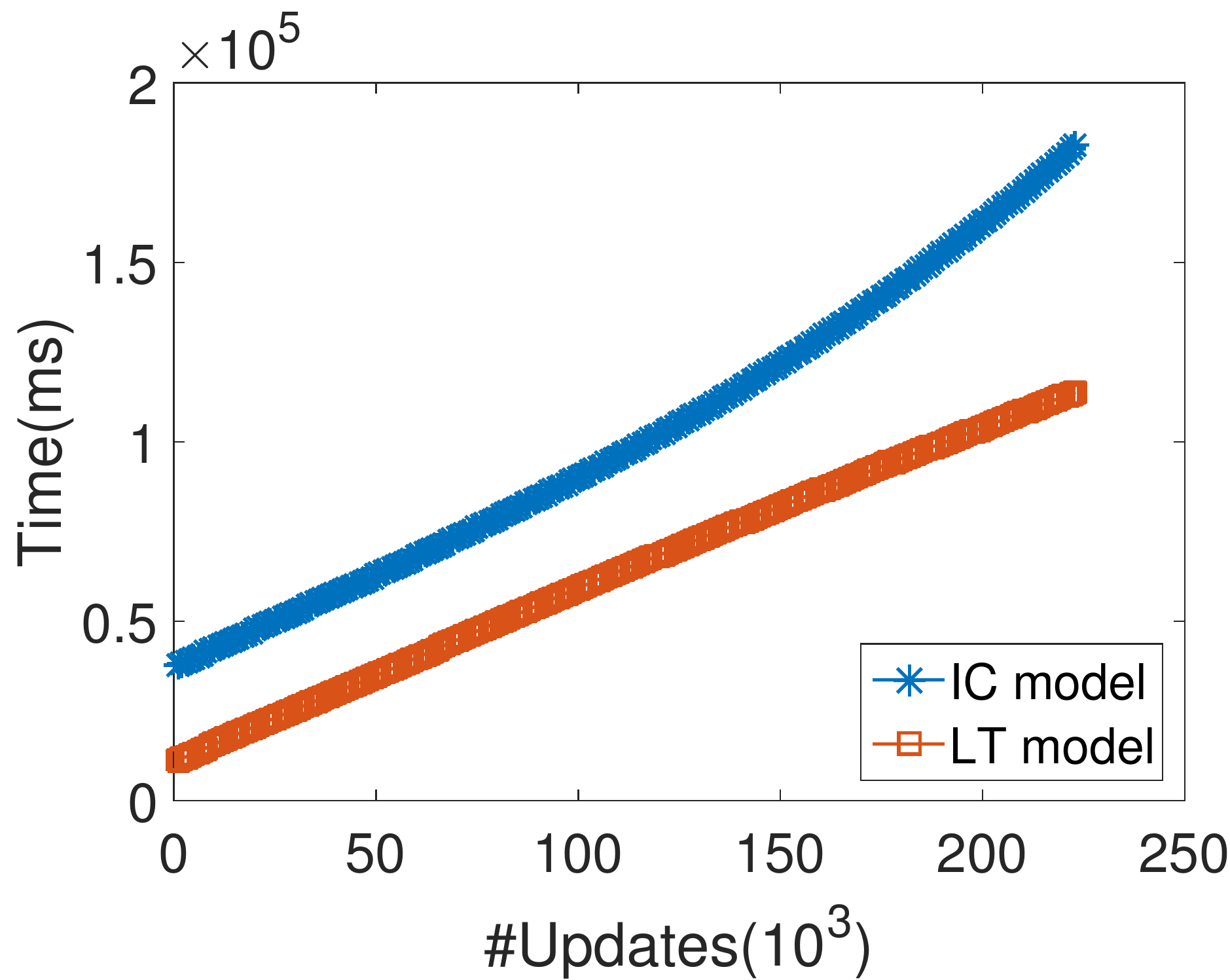}
    }
    \subfigure[soc-Pokec]{
      \includegraphics[width=.17\textwidth]{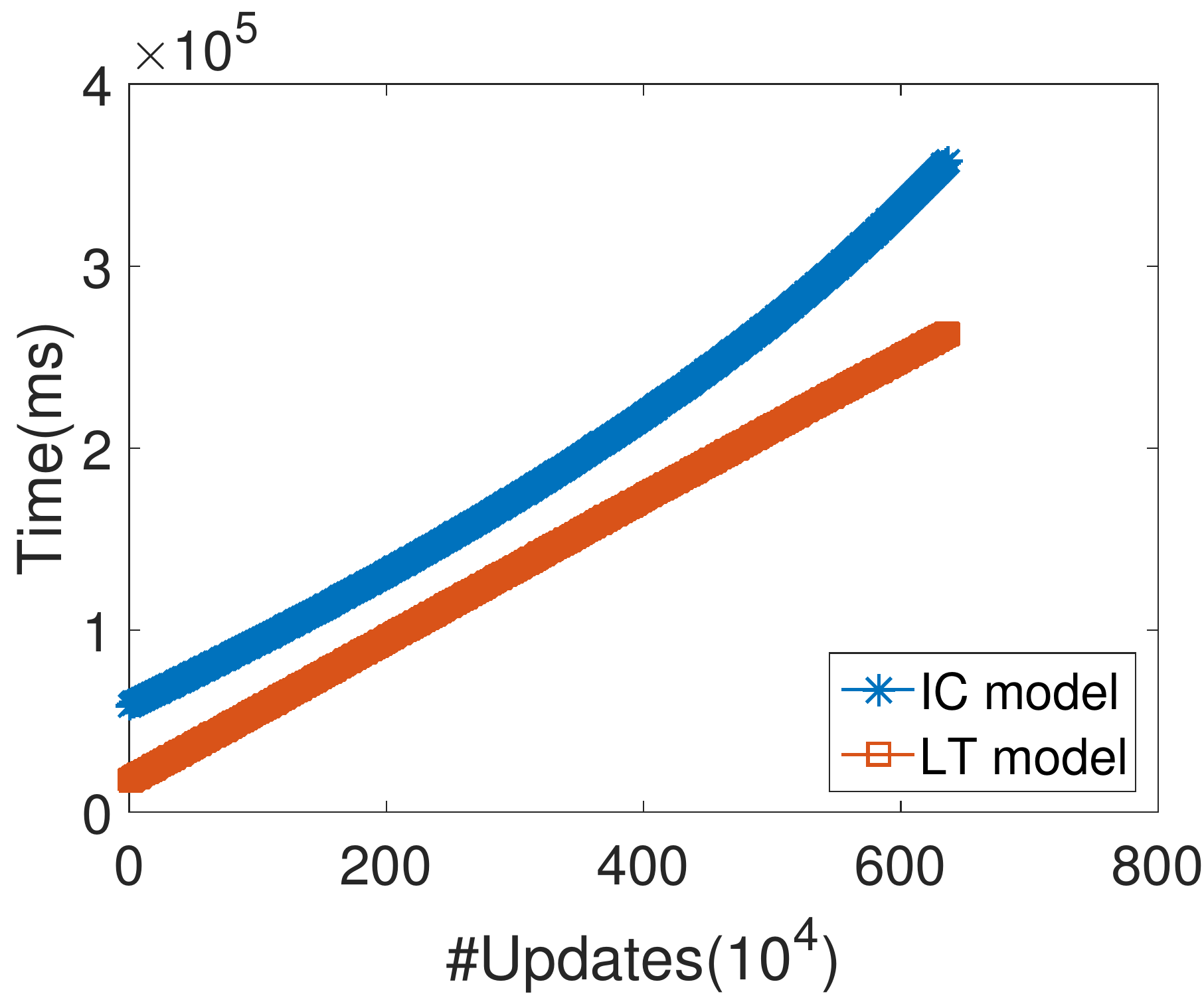}
    }
    \subfigure[flickr-growth]{
      \includegraphics[width=.17\textwidth]{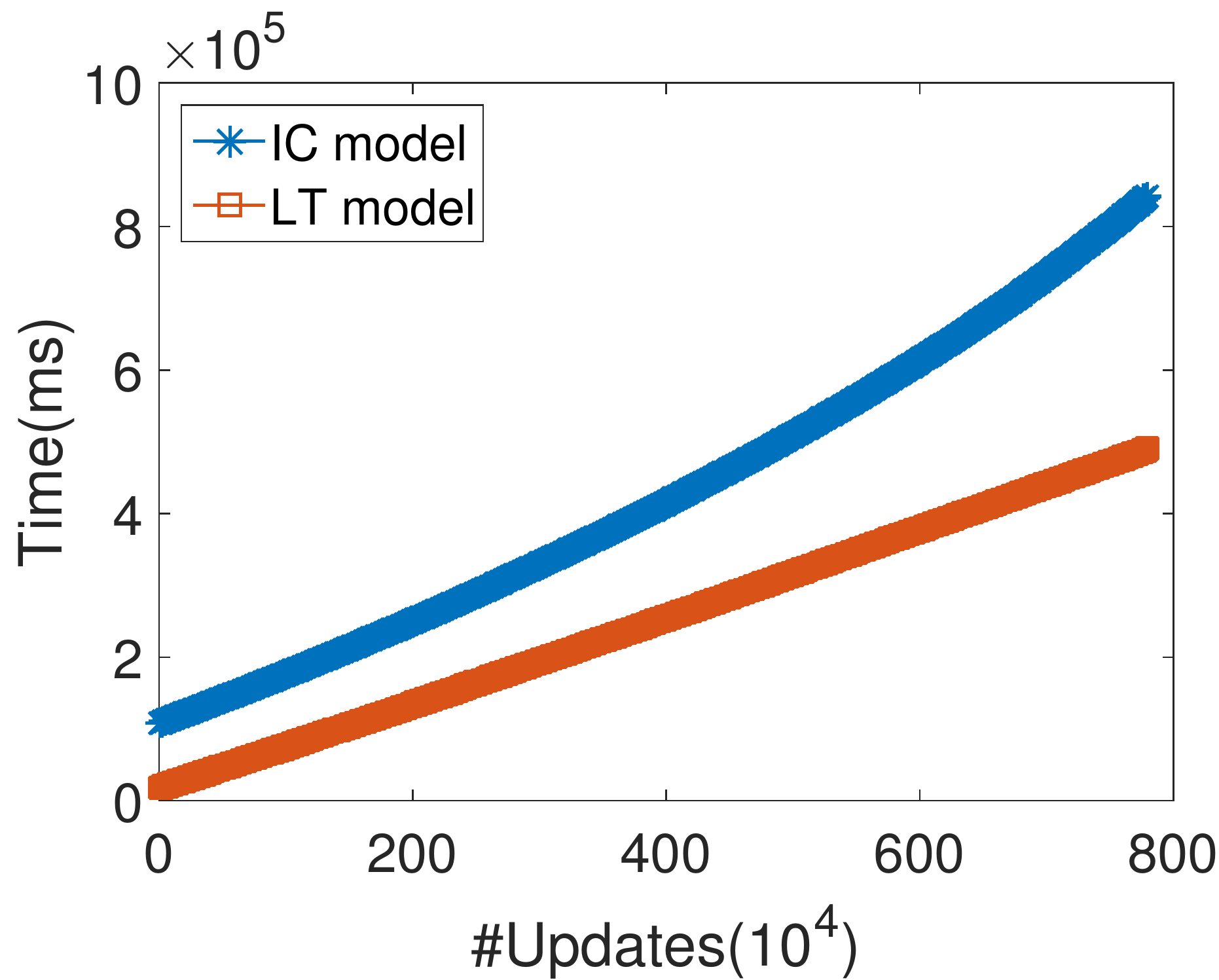}
    }
    \subfigure[Twitter]{
      \includegraphics[width=.17\textwidth]{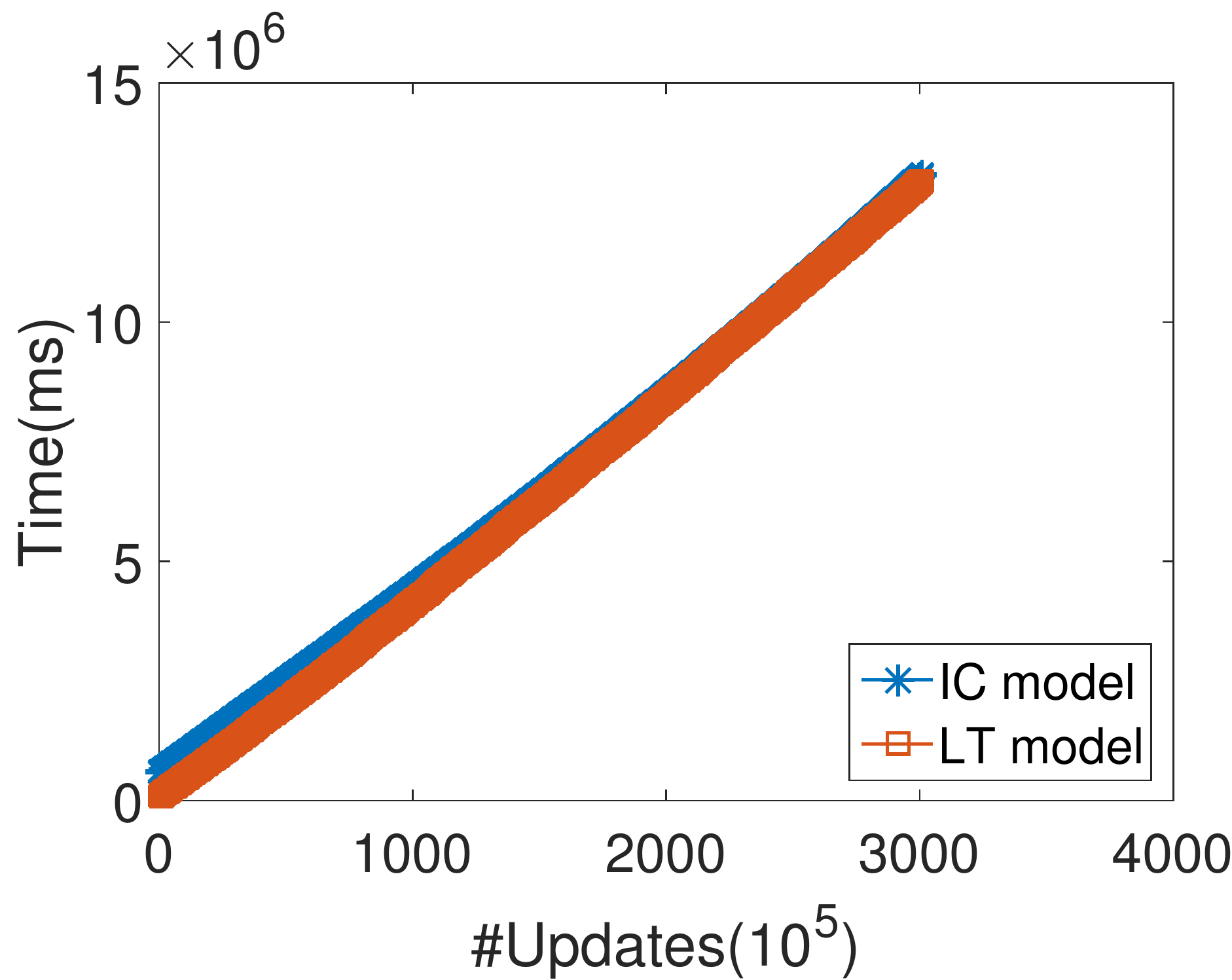}
    }
    \caption{Scalability (Influence Maximization).}
    \label{fig:im_efficiency}
\end{figure*}

\begin{figure}[t]
  \centering
   \subfigure[flickr-growth]{
      \includegraphics[width=.14\textwidth]{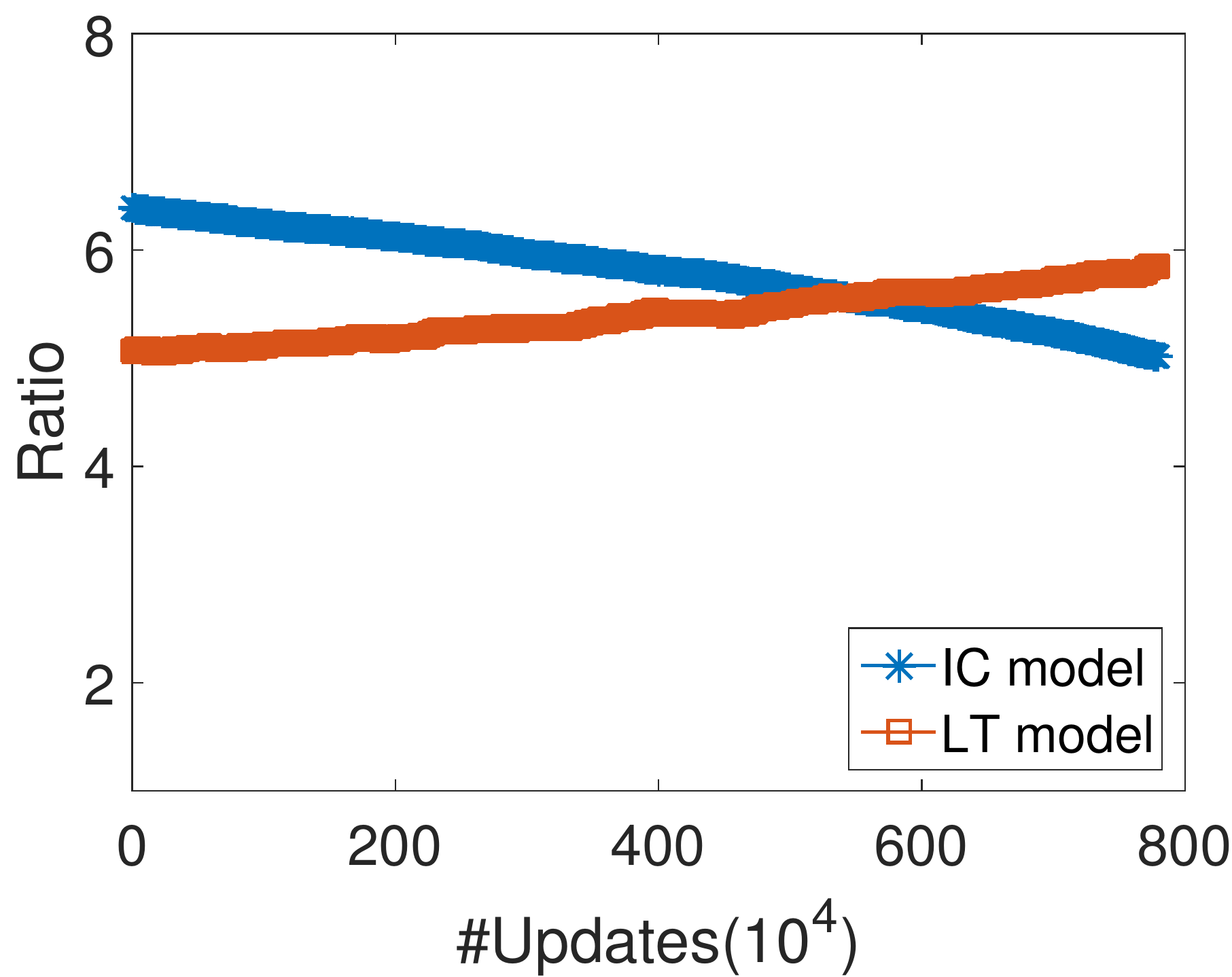}
    }
    \subfigure[soc-Pokec]{
      \includegraphics[width=.14\textwidth]{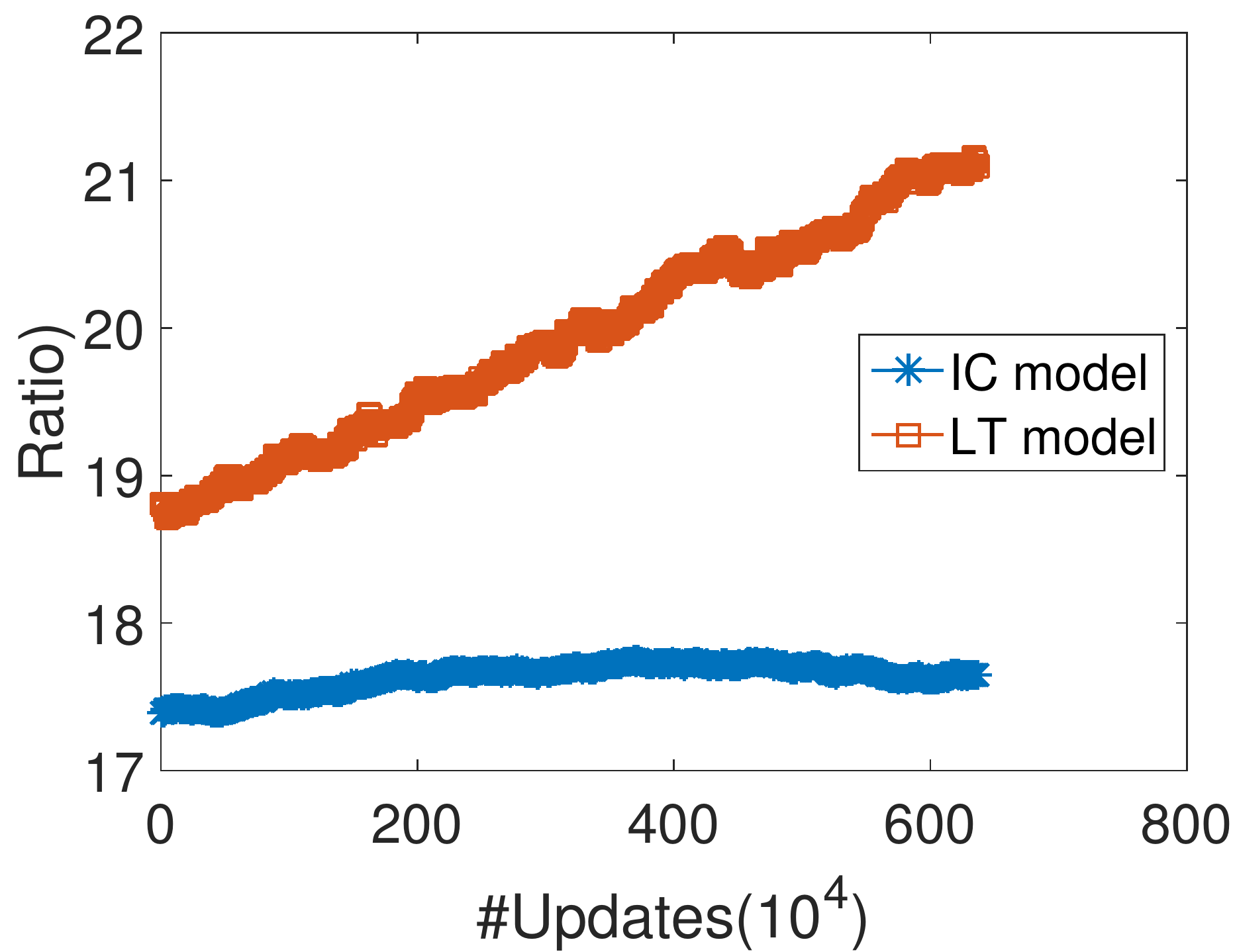}
    }
    \subfigure[Twitter]{
      \includegraphics[width=.14\textwidth]{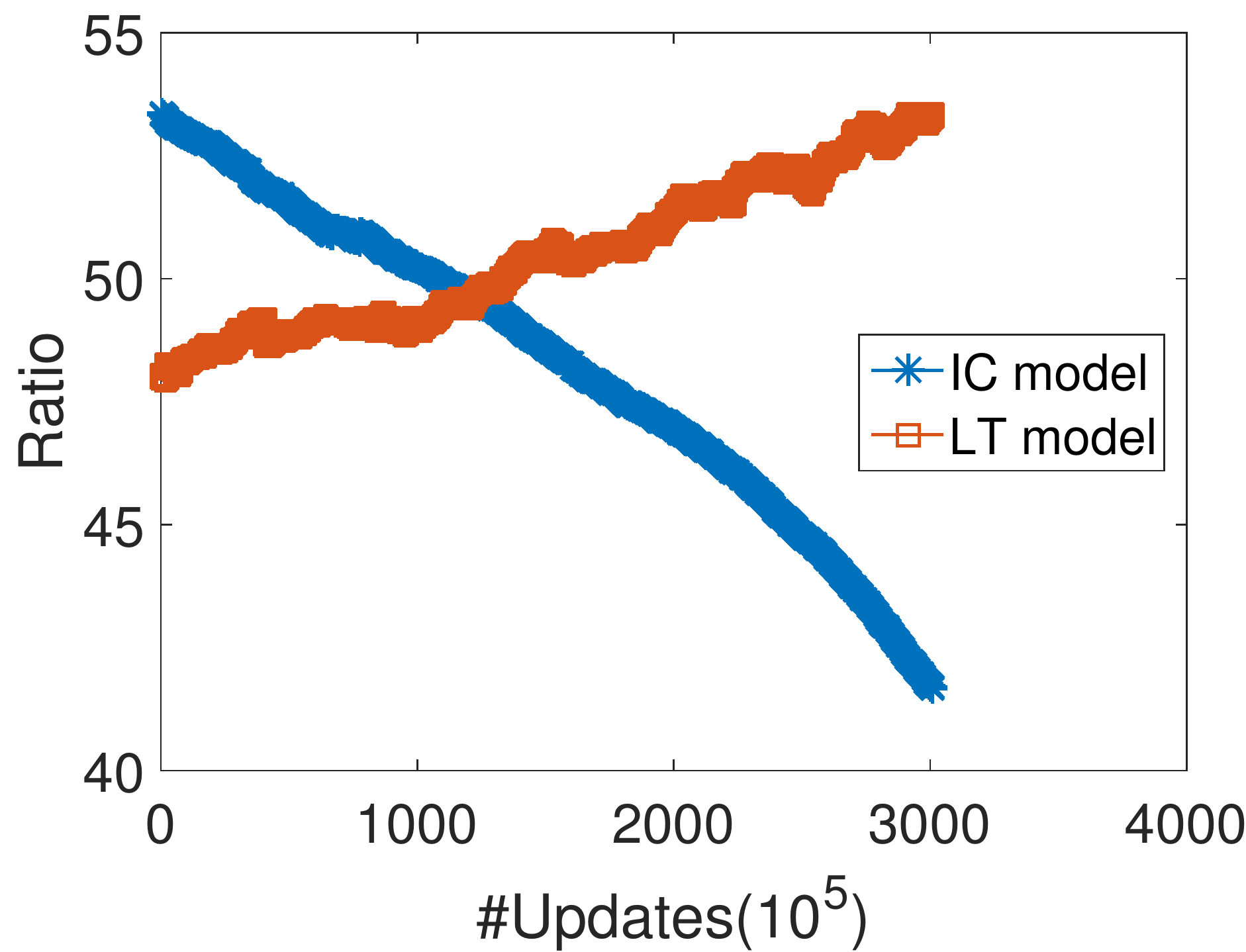}
    }
    \caption{Comparison to $32(m+n)\log{n}$.}
    \label{fig:im_ratio}
\end{figure}

\begin{figure}[t]
  \centering
    \subfigure[soc-Pokec (LT)]{
      \includegraphics[width=.14\textwidth]{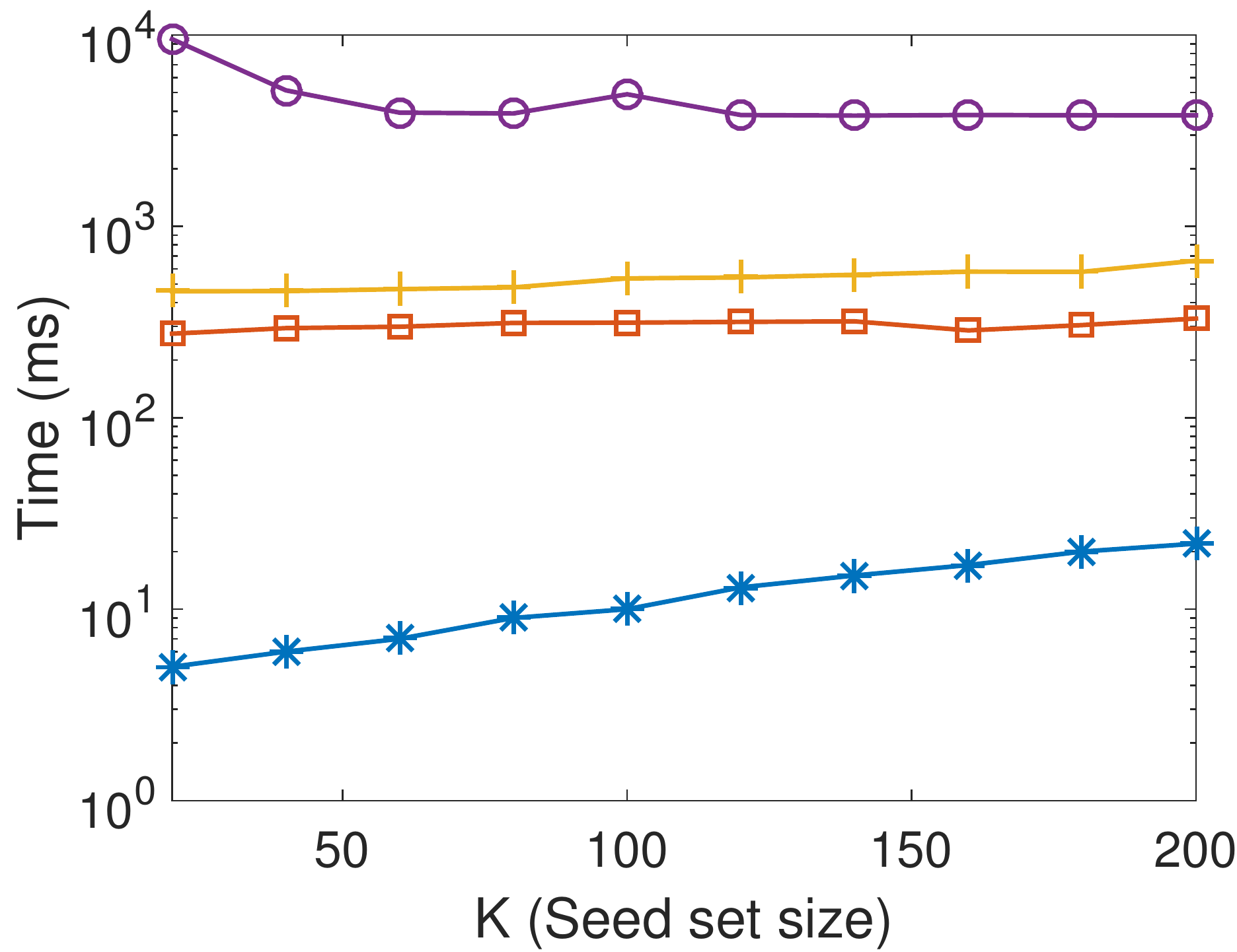}
    }
    \subfigure[flickr-growth (LT)]{
      \includegraphics[width=.14\textwidth]{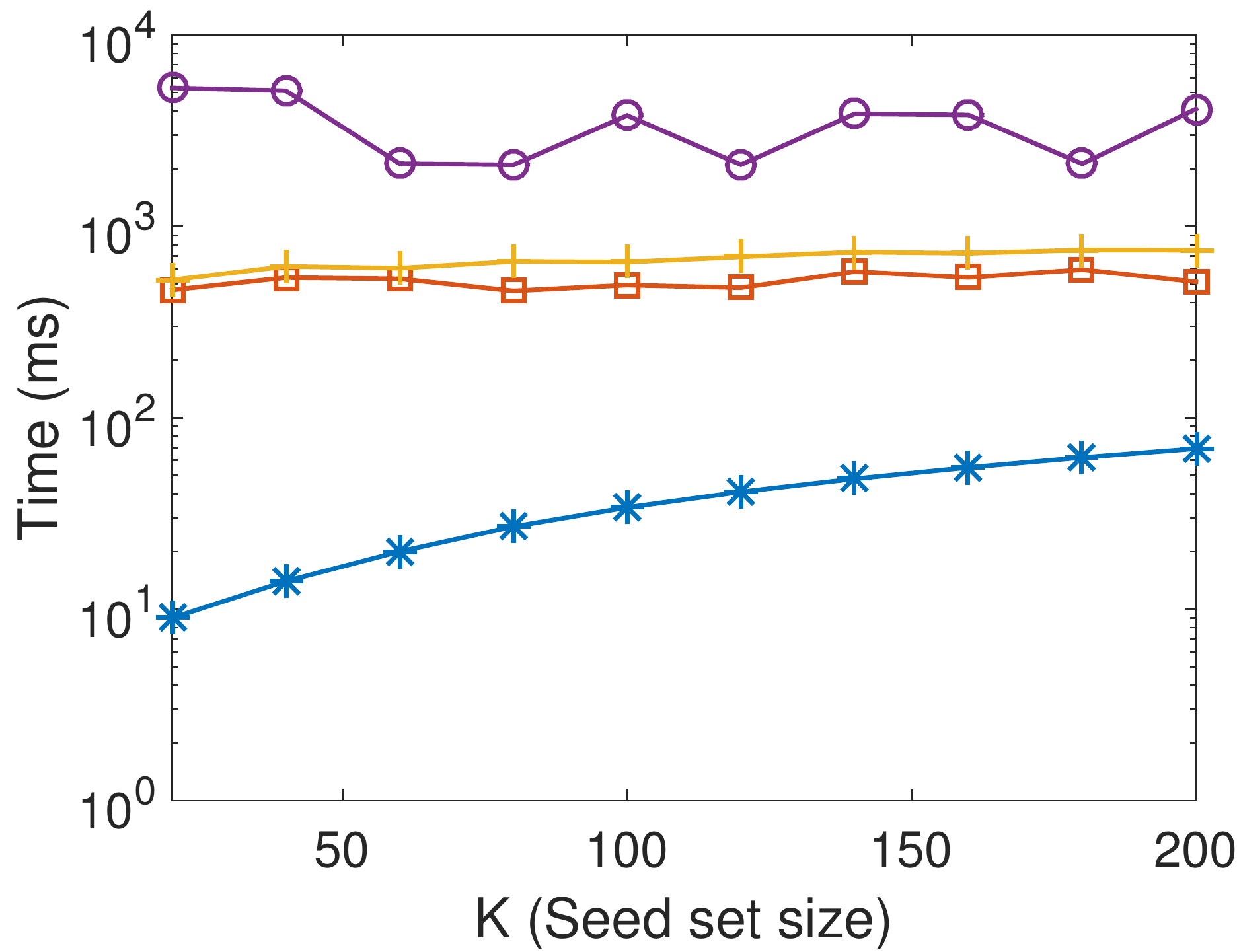}
    }
    \subfigure[Twitter (LT)]{
      \includegraphics[width=.14\textwidth]{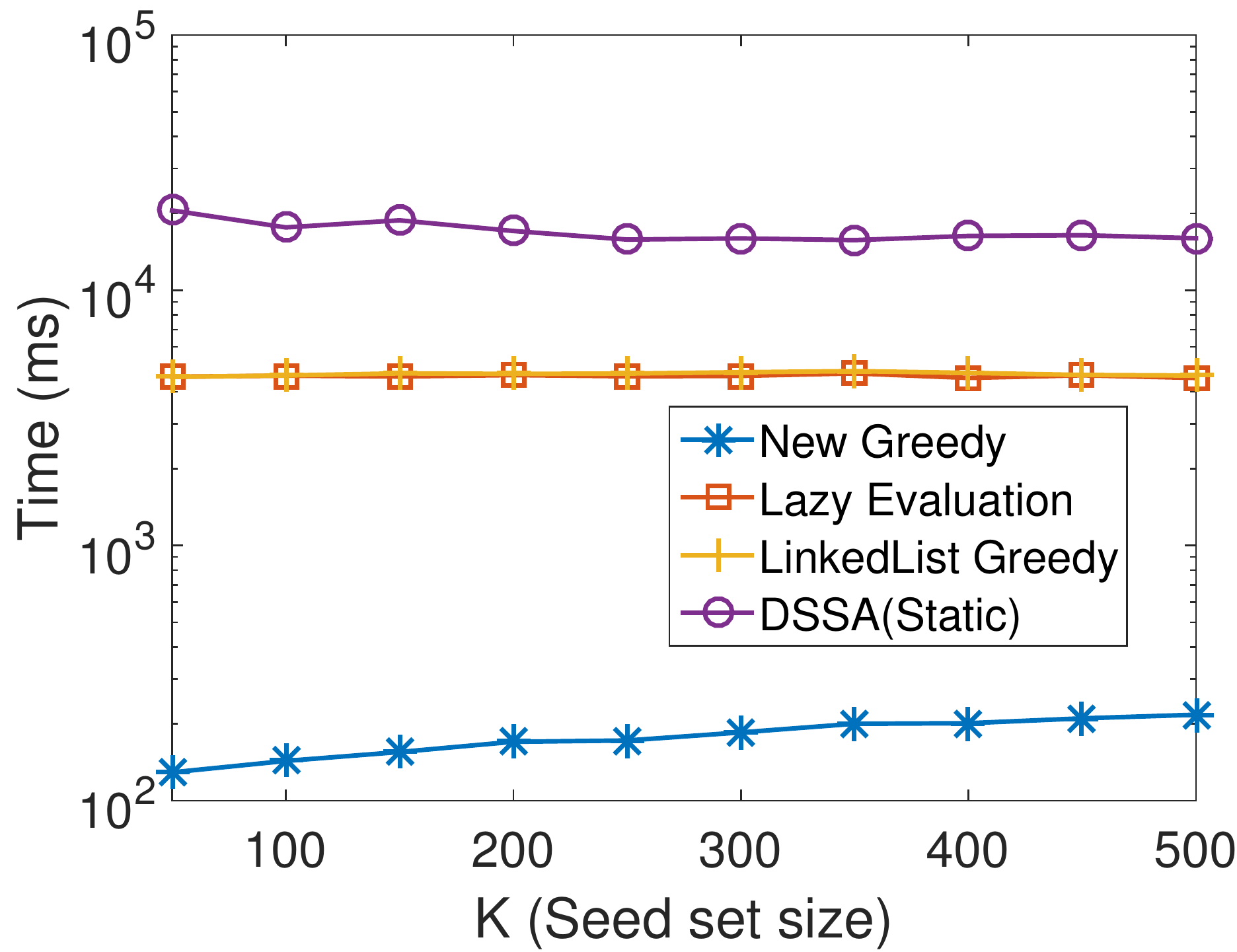}
    }
    \subfigure[soc-Pokec (IC)]{
      \includegraphics[width=.14\textwidth]{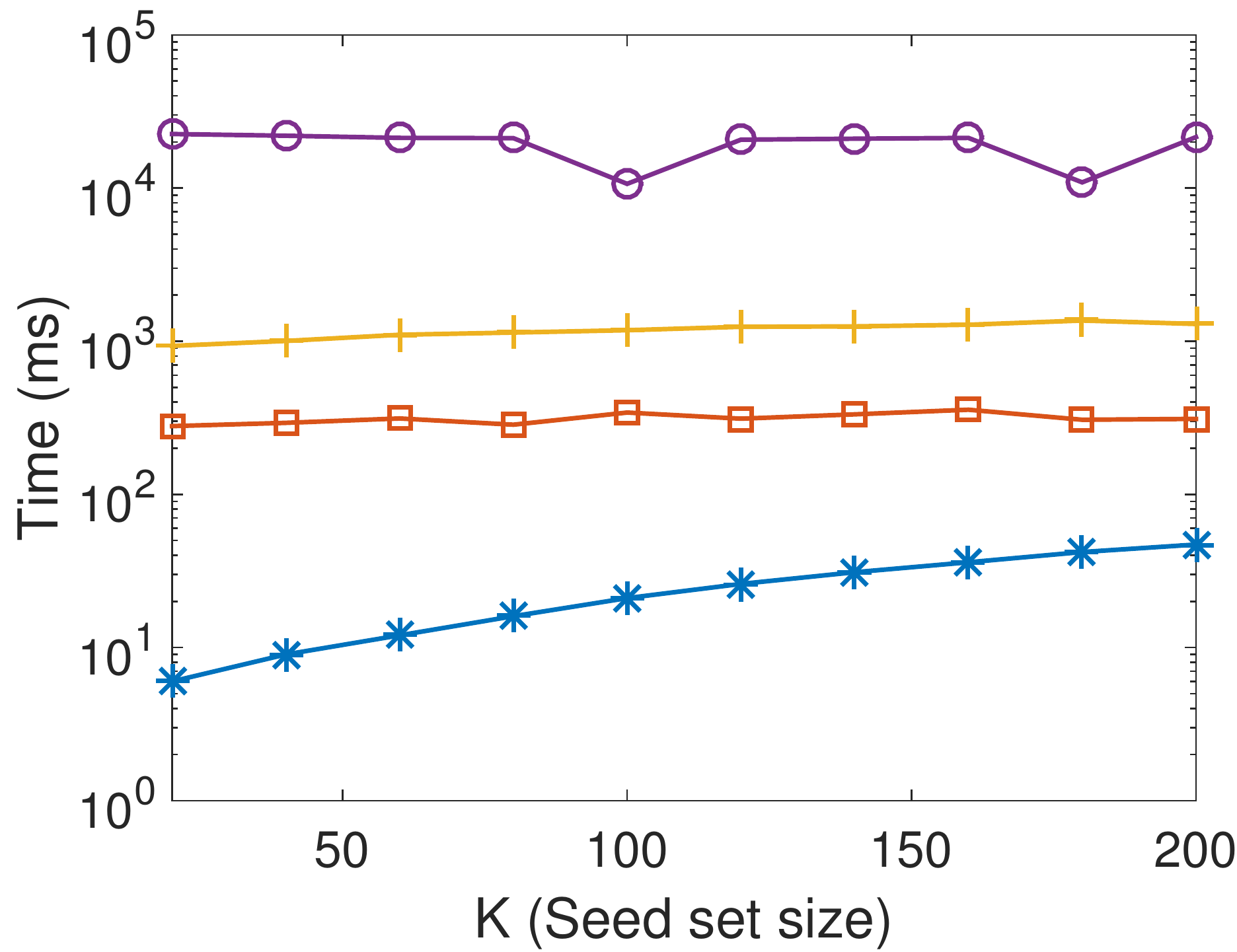}
    }
    \subfigure[flickr-growth (IC)]{
      \includegraphics[width=.14\textwidth]{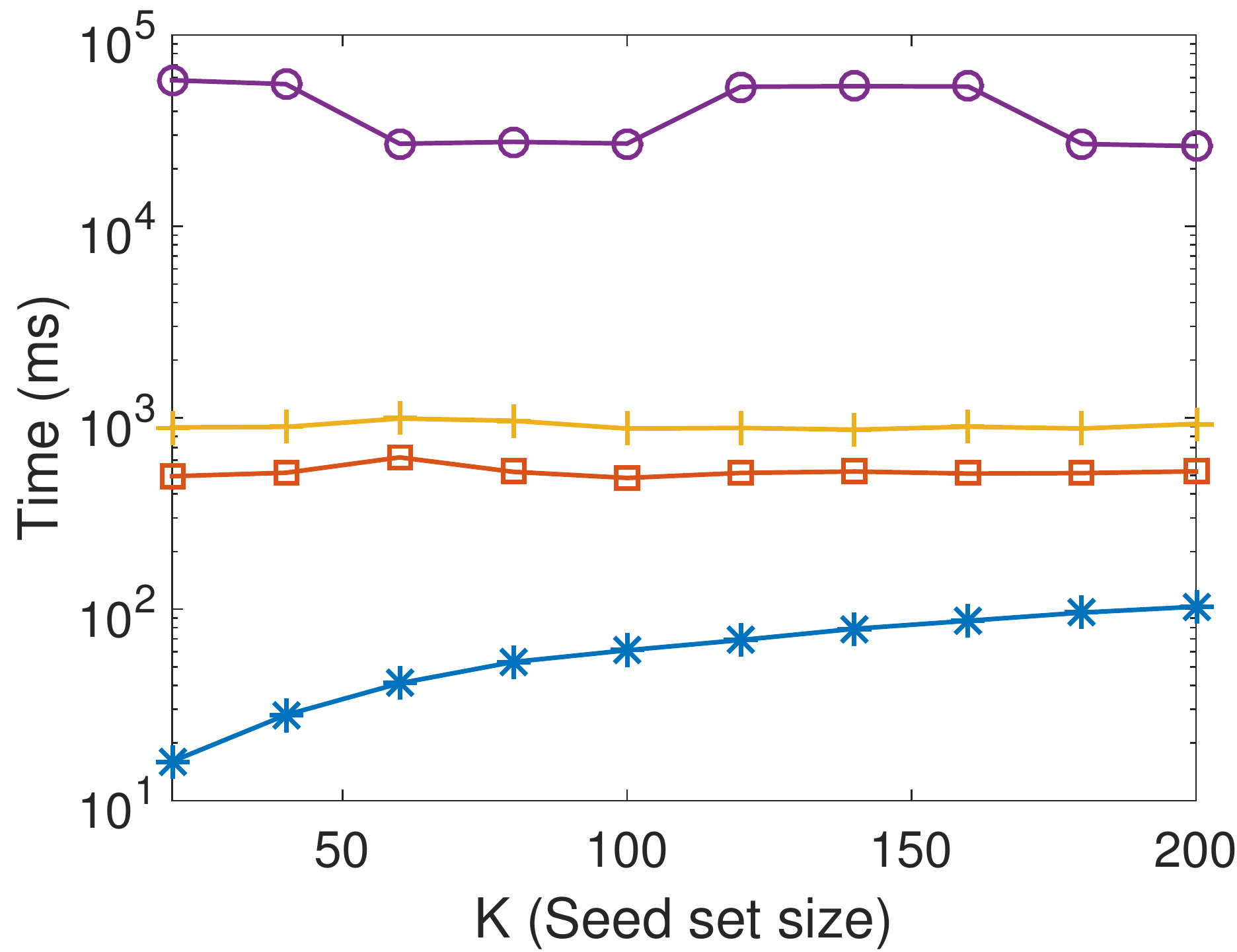}
    }
    \subfigure[Twitter (IC)]{
      \includegraphics[width=.14\textwidth]{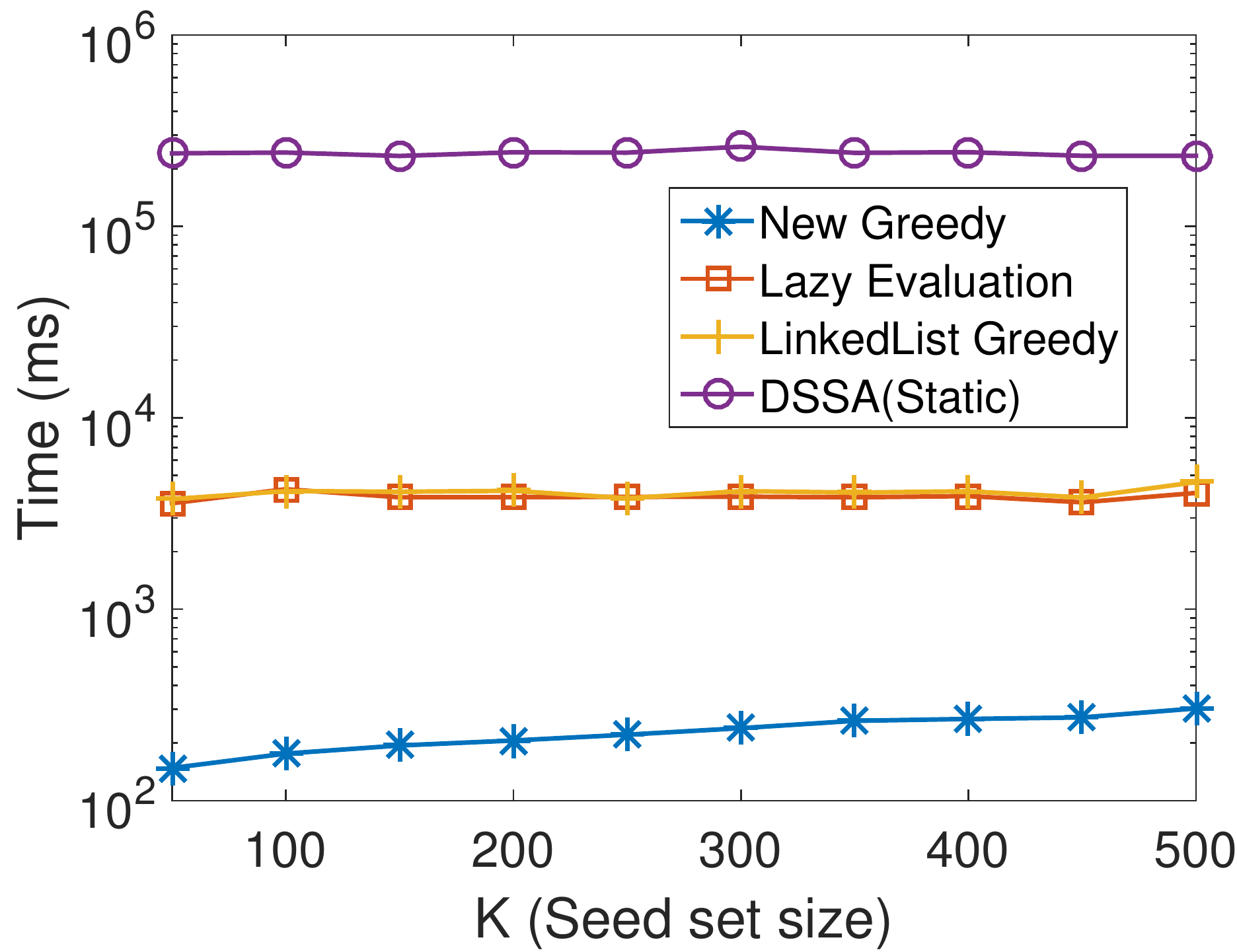}
    }
    \caption{Influence Maximization Query Time.}
    \label{fig:im_query}
\end{figure}

\subsection{Influence Maximization}
\subsubsection{Effectiveness \updates{of Sample Size}}
\updates{
We demonstrate that our practical solution that maintains $\mathcal{D}^*=\frac{1}{2} \lceil \Upsilon_1(\frac{\epsilon}{2-1/e}, \frac{2}{3n^2}) \rceil$ is effective by testing results of IM queries on the final snapshot of each data set. The algorithm that collects a large number of RR sets until $\mathcal{C(R)}=32(m+n)\log{n}$ in~\cite{ohsaka2016dynamic} and DSSA~\cite{nguyen2016stop}, the state-of-the-art IM algorithm on static networks, are compared. The parameters of DSSA are set such that DSSA is 0.5-optimal with probability $1-\frac{1}{n}$. The three algorithms in comparison decide the sample size in different ways and generate different number of RR sets. Comparing effectiveness of the three algorithms is actually comparing effectiveness of the decisions of sample size in the three algorithms. Thus, we use ``$\mathcal{D}^*=\frac{1}{2} \lceil \Upsilon_1(\frac{\epsilon}{2-1/e}, \frac{2}{3n^2}) \rceil$'' and ``$C(R)=32(m+n)\log{n}$'' to denote our practical solution and the method in~\cite{ohsaka2016dynamic}, respectively. We did not run the method ``$C(R)=32(m+n)\log{n}$'' on the twitter data set because it is too costly.} 

\updates{
To extract the seed set $S_k$ of the final snapshot, we set $T_{\mathcal{D}}=\min\{\frac{\mathcal{D}^*}{k-1}, \mathcal{D}^*-1\}$ and ran the New Greedy algorithm (Algorithm~\ref{alg:newgreedy}) on the RR sets generated by each algorithm. It turned out that by setting $T_{\mathcal{D}}=\min\{\frac{\mathcal{D}^*}{k-1}, \mathcal{D}^*-1\}$ the New Greedy Algorithm always returned $S_k$ such that $\mathcal{D}(S_k) \geq (1-\frac{1}{e})\max_{|S|=k}{\mathcal{D}(S)}$. Algorithm~\ref{alg:llgreedy} and Lazy Evaluation (the query algorithm in~\cite{ohsaka2016dynamic}) were also tested as the IM query algorithm, but the results were all very similar. Thus, we only report the results by using New Greedy as the IM query algorithm.}  

\nop{
We tested the effectiveness of our influence maximization algorithms on dynamic networks. We compared five algorithms by running influence maximization queries on the final snapshot of each dataset. The five algorithms are LinkedList Greedy (Algorithm~\ref{alg:llgreedy}), Lazy Evaluation (the query algorithm in~\cite{ohsaka2016dynamic}, that is, a special case of Algorithm~\ref{alg:newgreedy} by setting $T_{\mathcal{D}}=-1$), DSSA~\cite{nguyen2016stop} (the state-of-the-art influence maximization algorithm on static networks), \todo{Give this algorithm a better name.}  which is to collect a large number of RR sets until $\mathcal{C(R)}=32(m+n)\log{n}$ for extracting the $k$-seed set, and our New Greedy algorithm, which is to run Algorithm~\ref{alg:newgreedy} by setting $T_\mathcal{D}=\min\{\frac{\mathcal{D}^*}{k-1}, \mathcal{D}^*-1\}$ first, and if the returned $t$ is not $k+1$, we rerun Algorithm~\ref{alg:newgreedy} by setting $T_\mathcal{D}=-1$. The DSSA algorithm is implemented to fulfill $0.5$-optimal influence maximization queries with probability at least $1-\frac{1}{n}$. We did not run the algorithm $32(m+n)\log{n}$ on Twitter, the largest data set, because it is too costly.
}

To measure the effectiveness of a seed set $S_k$, we generated another collection of RR sets $\mathcal{R'}$ such that the cost $C(\mathcal{R'})$ is $32(m+n)\log{n}$ (except on Twitter dataset $C(\mathcal{R'})$ is set to $(m+n)\log{n}$ because $32(m+n)\log{n}$ is to costly). The influence of $S_k$ is estimated using $\mathcal{R'}$. By Theorem~\ref{th:sample}, we checked $\mathcal{D}'(S_k)$, the degree of $S_k$ in $\mathcal{R'}$, and found that $\mathcal{R'}$ estimated $I(S_k)$ accurately. We calculate that, with high probability, the relative error rate is at most 2\%. Fig.~\ref{fig:im_inf} shows the results where all values are averages taken on the results from 10 instances. All algorithms have pretty close performance except that DSSA is slightly worse than the others. This is because DSSA is dedicated for a specific seed set size $k$ and it always generates a smaller number of RR sets than other methods. This demonstrates that \updates{our practical solution that maintains} $\mathcal{D}^*=\frac{1}{2} \lceil \Upsilon_1(\frac{\epsilon}{2-1/e}, \frac{2}{3n^2}) \rceil$ is effective in practice, although it does not have theoretical guarantees.
 
\subsubsection{Scalability}
Fig.~\ref{fig:im_efficiency} reports the running time of maintaining RR sets for influence maximization queries over update streams. Similar to tracking influential individuals, our algorithm for maintaining RR sets scales roughly linearly under the LT model, while under the IC model in some cases it does not. Still the probable reason is our experiment setting. The maximum and the average influence do not vary much against updates under the LT model, while some updates under the IC model lead to big changes of the maximum influence or the average influence. For the largest dataset Twitter, our algorithms processed 0.3 billion updates in less than 4 hours.

To compare with the method in~\cite{ohsaka2016dynamic}, we report the ratio of the number of RR sets needed by the method in~\cite{ohsaka2016dynamic} to the number of RR sets maintained by our algorithm. This ratio reflects the improvement in efficiency of our algorithm over~\cite{ohsaka2016dynamic} because the cost of maintaining RR sets against an update and the cost of an influence maximization query are both roughly linear to the number of RR sets. The results are shown in Fig.~\ref{fig:im_ratio}. Our algorithm consistently maintains fewer RR sets than~\cite{ohsaka2016dynamic} as the ratio is consistently greater than $1$. On soc-Pokec and the largest data set Twitter, our improvement is more than an order of magnitude. The results on the two small data sets are similar and are omitted limited by space. One may find that for the LT model, the ratio in Fig.~\ref{fig:im_ratio} increases over the update stream in general, while for the IC model, in two data sets the ratio keeps decreasing. This is due to the experimental setup of updates under the LT and IC model. The effects of update streams on influence spreads of vertices are different under the LT and IC models.

We also report the efficiency of \updates{IM query algorithms (implementations of the greedy algorithm)} in Fig.~\ref{fig:im_query}. \updates{New Greedy (Algorithm~\ref{alg:newgreedy}), LinkedList Greedy (Algorithm~\ref{alg:llgreedy}), Lazy Evaluation (the IM query algorithm in~\cite{ohsaka2016dynamic}) and DSSA~\cite{nguyen2016stop} are compared. New Greedy, LinkedList Greedy and Lazy Evaluation were all ran directly on the maintained (by our practical solution that keeps $\mathcal{D}^*=\frac{1}{2} \lceil \Upsilon_1(\frac{\epsilon}{2-1/e}, \frac{2}{3n^2}) \rceil$) RR sets of the final snapshot, while DSSA first sampled a number of RR sets based on the final snapshot, and then extracted a seed set by running the greedy algorithm on the sampled RR sets.} Limited by space, we omit the results on the two small data sets, \updates{which are similar to Fig.~\ref{fig:im_query}}. The results show that our New Greedy algorithm is always the fastest query algorithm, and the larger a network, the bigger the improvement of the New Greedy algorithm over the baselines. For the largest data set Twitter, our New Greedy algorithm returns a seed set with good quality within 300ms, and it is an order of magnitude faster than the Lazy Evaluation algorithm in~\cite{ohsaka2016dynamic}, and two to three orders of magnitude faster than the DSSA algorithm~\cite{nguyen2016stop} running on the static network.

\section{Conclusion}\label{sec:con}
In this paper, we tackled two versions of tracking top-$k$ influential vertices in dynamic networks. We adopted two simple signals to decide a proper number of RR sets for the two tasks, and showed that with high probability our sample size ensures that the result has good quality guarantees. We reported a series of experiments on five real networks and demonstrated the effectiveness and efficiency of our algorithms.

Parallelizing our methods in large distributed systems is an interesting future direction. Since our solutions are based on independent sampling, they have a great potential to be parallelized for further accelerations. Also, for influence maximization task, devising methods that decide the sample size efficiently and only keep a small number of RR sets to handle big seed set size $k$ with quality guarantees still remains an open problem.

\bibliographystyle{abbrv}
\bibliography{Tracking_Top_K}

\section*{Appendix}
\textbf{Proof of Theorem~\ref{th:sample}} Suppose $M_1$ is the stopping time when the first time $\sum_{i=1}^{M_1}{Z_i}=\lceil \Upsilon_1(\epsilon,\delta) \rceil$ and $M_2$ is the stopping time when the first time $\sum_{i=1}^{M_2}{Z_i}=\lceil \Upsilon_1(\epsilon,\delta) \rceil +1$. Clearly, $M_1 \leq M < M_2$. According to the Stopping Rule theorem, we have $\textup{Pr}\{M \geq \frac{\lceil \Upsilon_1(\epsilon,\delta) \rceil}{(1+\epsilon)\mu_Z}\} \geq 1-\delta/2$ and $\textup{Pr}\{M \leq \frac{\lceil \Upsilon_1(\epsilon,\delta) \rceil+1}{(1-\epsilon)\mu_Z}\} \geq 1-\delta/2$. Since $\sum_{i=1}^M{Z_i}=\lceil \Upsilon_1(\epsilon,\delta) \rceil$, we have $\textup{Pr}\{\frac{\sum_{i=1}^M{Z_i}}{M} \leq (1+\epsilon)\mu_Z\} \geq 1-\frac{\delta}{2}$ and $\textup{Pr}\{\frac{\sum_{i=1}^M{Z_i}}{M} \geq \frac{\lceil \Upsilon_1(\epsilon,\delta) \rceil}{\lceil \Upsilon_1(\epsilon,\delta) \rceil+1}(1-\epsilon)\mu_Z\} \geq 1-\frac{\delta}{2}$.
\\

\noindent \textbf{Proof of Lemma~\ref{lemma:rnd}} The first part can be directly obtained by applying Theorem~\ref{th:sample}. We prove the second part. Note that $\Upsilon_1(\epsilon,\delta)$ is decreasing with respect to $\epsilon$. When $\epsilon \leq \frac{1}{3}$, for every $v$ such that $\epsilon_v > 0.6$, we have $\mathcal{D}_1(v) < \Upsilon_1(0.6,\frac{\delta}{n}) < \frac{1}{2}\Upsilon_1(\epsilon,\frac{\delta}{n})=\frac{1}{2}\mathcal{D}_1^k$. We prove that, with high probability, for every $v$, $\mathcal{D}_1(v) < \frac{1}{2}\Upsilon_1(\epsilon,\frac{\delta}{n})$ implies $I_v < \frac{n\Upsilon_1(\epsilon,\frac{\delta}{n})}{M_1}$.
	
If $I_v \geq \frac{n\Upsilon_1(\epsilon,\frac{\delta}{n})}{M_1}$, according Corollary~\ref{cor:c1}, we have
    \begin{equation*}
    \begin{split}
        \textup{Pr}\{\mathcal{D}_1(v) < \frac{1}{2}\Upsilon_1(\epsilon,\frac{\delta}{n}) \} &\leq \textup{Pr}\{\frac{\mathcal{D}_1(v)}{M_1} < \frac{1}{2}\frac{I_v}{n} \} \\
        &\leq \textup{exp} \Big\{ -\frac{1}{8}\frac{M_1\Upsilon_1(\epsilon,\frac{\delta}{n})}{M_1} \Big\} \\
        &\leq \textup{exp} \Big\{ -\frac{(e-2)\ln{\frac{2n}{\delta}}}{2\epsilon^2} \Big\}
    \end{split}
    \end{equation*}

    Since $\epsilon \leq \frac{1}{3}$, we have that if $I_v \geq \frac{n\Upsilon_1(\epsilon,\frac{\delta}{n})}{M_1}$, $\textup{Pr}\{\mathcal{D}_1(v) < \frac{1}{2}\Upsilon_1(\epsilon,\frac{\delta}{n}) \} \leq (\frac{\delta}{2n})^{\frac{9}{2}(e-2)} \leq \frac{\delta}{2n}$. This means if $\mathcal{D}_1(v) < \frac{1}{2}\Upsilon_1(\epsilon,\frac{\delta}{n})$, with probability at least $1-\frac{\delta}{2n}$, we have $I_v < \frac{n\Upsilon_1(\epsilon,\frac{\delta}{n})}{M_1}$.
\\

\noindent \textbf{Proof of Lemma~\ref{lemma:lower}} According to Lemma~\ref{lemma:rnd}, and applying the union bound, we have that with probability at least $1-\frac{\delta}{2}$, for all $v$ such that $\mathcal{D}_1(v)= \Upsilon_1(\epsilon_v,\frac{\delta}{n})$ and $\epsilon_v \leq 0.6$, $\frac{n\mathcal{D}_1(v)}{M_1} \leq (1+\epsilon_v)I_v$ so $I_v \geq \frac{n\mathcal{D}_1(v)}{(1+\epsilon_v)M_1}$. Clearly there are at least $k$ vertices such that $\frac{n\mathcal{D}_1(v)}{M_1} \geq \frac{n\mathcal{D}_1^k}{M_1}$. Also, $\epsilon_v$ is decreasing with respect to $\mathcal{D}_1(v)$. Thus, if $\mathcal{D}_1(v) \geq \mathcal{D}_1^k$, then $\frac{n\mathcal{D}_1(v)}{(1+\epsilon_v)M_1} \geq \frac{n\mathcal{D}_1^k}{(1+\epsilon)M_1}$. Therefore, with probability at least  $1-\frac{\delta}{2}$, there are at least $k$ vertices whose influence spreads are no smaller than $\frac{n\mathcal{D}_1^k}{(1+\epsilon)M_1}$. So we get that $\textup{Pr}\{\frac{I^k}{n} \geq \frac{\mathcal{D}_1^k}{(1+\epsilon)M_1}\} \geq 1-\frac{\delta}{2}$.
\\

\noindent \textbf{Proof of Lemma~\ref{lemma:upper}}    We still apply Lemma~\ref{lemma:rnd} and the union bound. With probability at least $1-\frac{\delta}{2}$, for all $v \in V$, $\mathcal{D}_1(v)=\Upsilon_1(\epsilon_v,\frac{\delta}{n})$,
    \begin{enumerate}
    \item If $\epsilon_v \leq 0.6$, $\frac{n\mathcal{D}_1(v)}{M_1} \geq \frac{\mathcal{D}_1(v)}{\mathcal{D}_1(v)+1}(1-\epsilon_v)I_v$ and $I_v \leq \frac{n(\mathcal{D}_1(v)+1)}{(1-\epsilon_v)M_1}$; and
    \item If $\epsilon_v > 0.6$, then $I_v < \frac{n\mathcal{D}_1^k}{M_1}$.
    \end{enumerate}
    Consider the function $f(\epsilon,\delta,n)=\frac{\Upsilon_1(\epsilon,\frac{\delta}{n})+1}{1-\epsilon}=\frac{(1+\epsilon)\frac{4(e-2)\ln{\frac{2n}{\delta}}}{\epsilon^2}+2}{1-\epsilon}$ which is decreasing with respect to $\epsilon$ in the interval $\epsilon \in (0,0.6]$, when $\delta \leq \frac{1}{4}$ and $n \geq 1$. Thus, when (1) and (2) hold for every $v \in V$, if $\Upsilon_1(0.6,\frac{\delta}{n}) \leq \mathcal{D}_1(v) \leq \mathcal{D}_1^k$, we have $I_v \leq \frac{n(\mathcal{D}_1(v)+1)}{(1-\epsilon_v)M_1} \leq \frac{n(\mathcal{D}_1^k+1)}{(1-\epsilon)M_1}$. Moreover, if $\mathcal{D}_1(v) < \Upsilon_1(0.6,\frac{\delta}{n})$, we have $I_v < \frac{n\mathcal{D}_1^k}{M} \leq \frac{n(\mathcal{D}_1^k+1)}{(1-\epsilon)M_1}$. Therefore, with probability at least $1-\frac{\delta}{2}$, there are at least $n-k+1$ vertices whose influence spreads are no greater than $\frac{n(\mathcal{D}_1^k+1)}{(1-\epsilon)M_1}$, which means $I^k \leq \frac{n(\mathcal{D}_1^k+1)}{(1-\epsilon)M_1}$.
\\

\noindent \textbf{Proof of Theorem~\ref{th:individual}} Based on Corollary~\ref{cor:bound_Ik} and Lemma~\ref{lemma:deviate}, when $\epsilon \leq \frac{1}{3}$, $\delta \leq \frac{1}{4}$, $\mathcal{D}_1^k=\lceil \Upsilon_1(\epsilon,\frac{\delta}{n}) \rceil$ and $M_2=M_1$, with probability at least $1-2\delta$, the following conditions hold.
    \begin{enumerate}
        \item $\frac{\mathcal{D}_1^k}{(1+\epsilon)M_1} \leq \frac{I^k}{n} \leq \frac{\mathcal{D}_1^k+1}{(1-\epsilon)M_1}$; \label{con:bound}
        \item For every $v$ such that $I_v \geq I^k$, $\frac{\mathcal{D}_2(v)}{M_2} \geq (1-\epsilon)\frac{I^k}{n}$; and \label{con:recall}
        \item For every $v$ such that $I_v \leq (1-2\epsilon) I^k$, $\frac{\mathcal{D}_2(v)}{M_2} \leq \frac{I_v}{n}+\epsilon\frac{I^k}{n}$. \label{con:error}
    \end{enumerate}
    Under these 3 conditions, obviously, if $I_v \geq I^k$ then $D_2(v) \geq \frac{(1-\epsilon)M_2I^k}{n} \geq \frac{1-\epsilon}{1+\epsilon}\mathcal{D}_1^k=T$. If $I_v \leq \frac{nT}{M_2}-\epsilon I^k \leq (1-2\epsilon)I^k$, then $\frac{\mathcal{D}_2(v)}{M_2} \leq \frac{I_v}{n}+\epsilon \frac{I^k}{n} \leq \frac{T}{M_2}$. Thus, if $I_v \leq \frac{nT}{M_2}-\epsilon I^k$, then $\mathcal{D}_2(v) \leq T$. Now we prove $\frac{nT}{M_2}-\epsilon I^k$ is not much smaller than $I^k$, which means when we use $T$ as a filtering threshold, the influence spread of any false positive vertex is close to the true threshold $I^k$. Let $B=4(e-2)\ln{\frac{2n}{\delta}}$, so $\mathcal{D}_1^k=\lceil \Upsilon_1(\epsilon,\frac{\delta}{n}) \rceil \geq \frac{B(1+\epsilon)}{\epsilon^2}$. When $n \geq 1$ and $\delta \leq \frac{1}{4}$, $B > 5$. Thus,
    \begin{align*}
        \frac{nT}{M_2}-\epsilon I^k &= \frac{n(1-\epsilon)\mathcal{D}_1^k}{(1+\epsilon)M_1} - \epsilon I^k \\
        &\geq [\frac{\mathcal{D}_1^k}{\mathcal{D}_1^k+1}\frac{(1-\epsilon)^2}{1+\epsilon}-\epsilon]I^k \\
        &\geq [\frac{B(1+\epsilon)}{B(1+\epsilon)+\epsilon^2}\frac{(1-\epsilon)^2}{1+\epsilon}-\epsilon]I^k
    \end{align*}
    Since $\frac{B(1+\epsilon)}{B(1+\epsilon)+\epsilon^2}=\frac{1}{1+\frac{\epsilon^2}{(1+\epsilon)B}} \geq \frac{1}{1+\frac{\epsilon^2}{B}} \geq 1-\frac{\epsilon^2}{B}$, we have
    \begin{align*}
        \frac{nT}{M_2}-\epsilon I^k
        &\geq [(1-\frac{\epsilon^2}{B})\frac{(1-\epsilon)^2}{1+\epsilon}-\epsilon]I^k \\
        &\geq (\frac{1-3\epsilon}{1+\epsilon}-\frac{\epsilon^2}{B})I^k \geq (1-\frac{4B+\epsilon}{B}\epsilon)I^k \geq (1-\frac{61}{15}\epsilon)I^k
    \end{align*}
    Our algorithm returns the set of vertices $S=\{u \mid \mathcal{D}_2(u) \geq T\}$. Summarize the above analysis, we have that with probability at least $1-2\delta$, (1) if $I_u \geq I^k$, $u \in S$, and (2) $\min_{u \in S}{I_u} \geq [(1-\frac{\epsilon^2}{B})\frac{(1-\epsilon)^2}{1+\epsilon}-\epsilon]I^k \geq (1-\frac{61}{15}\epsilon)I^k$.
\\

\noindent \textbf{Proof of Theorem~\ref{th:im}} Suppose $\mathcal{D}_1(u)=\mathcal{D}_1^*$, which means when we pick the vertex with maximum degree in $\mathcal{R}_1$, we get $u$. According to Lemma~\ref{lemma:rnd}, by applying the union bound we have $\textup{Pr}\{\frac{I_u}{n} \geq \frac{n\mathcal{D}_1(u)}{(1+\epsilon)M_1}\} \geq 1-\frac{\delta}{2}$. Thus, we have lower bound of $I^*=\max_{u \in V}{I_u}$ with high probability. Specifically, we have $\textup{Pr}\{I^* \geq \frac{n\mathcal{D}_1(u)}{(1+\epsilon)M_1}=\frac{n\mathcal{D}_1^*}{M_1}\} \geq 1-\frac{\delta}{3}$. Since $\mathcal{D}_1^*=\lceil \Upsilon_1(\epsilon,\frac{2\delta}{3n}) \rceil$, with probability at least $1-\frac{\delta}{3}$, $M_1 \geq \frac{4n(e-2)\ln{\frac{3n}{2\delta}}}{I^*\epsilon^2}$ and $M_2 \geq \frac{4n(e-2)\ln{\frac{3N}{2\delta}}}{I^*\epsilon^2}$. When $M_2 \geq \frac{4n(e-2)\ln{\frac{3N}{2\delta}}}{I^*\epsilon^2}$, for any $k$-seed set $S$, applying Corollary~\ref{cor:c1} and utilizing the fact that $I(S_k^*) \geq I^*$ and $I(S_k^*) \geq I(S)$, we have
$$
	\textup{Pr}\{\forall S \subseteq V, |S|=k, |\frac{\mathcal{D}_2(S)}{M_2}-\frac{I(S)}{n}| \leq \frac{\epsilon I(S_k^*)}{n}\} \geq 1-\frac{2\delta}{3}
$$
When $|\frac{\mathcal{D}_2(S)}{M_2}-\frac{I(S)}{n}| \leq \frac{\epsilon I(S_k^*)}{n}$ for all $k$-seed set $S$, we have
\begin{align*}
	\frac{I(S_k)}{n} &\geq \frac{\mathcal{D}_2(S_k)}{M_2}-\frac{\epsilon I(S_k^*)}{n} \\
			        &\geq (1-\frac{1}{e})\frac{\mathcal{D}_2(S_k^*)}{M_2}-\frac{\epsilon I(S_k^*)}{n} \\
			        &\geq (1-\frac{1}{e})(1-\epsilon)\frac{I(S_k^*)}{n}-\frac{\epsilon I(S_k^*)}{n} \\
			        &=[1-\frac{1}{e}-(2-\frac{1}{e}) \epsilon ]\frac{I(S_k^*)}{n}
\end{align*}
Therefore, applying the union bound, we have $\textup{Pr}\{I(S_k) \geq [1-\frac{1}{e}-(2-\frac{1}{e}) \epsilon ]I(S_k^*)\} \geq 1-\delta$.
\\

\noindent \textbf{Proof of Theorem~\ref{th:filter_greedy}} According to the submodularity of $\mathcal{D}(S)$ with respect to $S$, it is easy to find that $\mathcal{M}(u_i;S'_k) \leq \mathcal{D}(u_i)$ for every $i$. For $i \leq q-1$, $\mathcal{D}(u_i) \geq \mathcal{M}(u_i) > T_{\mathcal{D}}$. Thus, when running the greedy algorithm by copying the whole $\mathcal{D}$, in the first $q-1$ iteration of choosing seeds, $u_i$ is always the vertex with the maximum marginal gain in the $i$-th iteration, no matter only vertices in $U$ or all vertices in $V$ are considered. So we have that the first $q-1$ seeds of $S_k$ and $S'_k$ are the same. Thus, $S_k=\{u_1,u_2,...,u_{q-1},v_q,...,v_k\}$, where $v_i$ could be different from $u_i$.
	
	We prove that $\mathcal{M}(v_q;S_k) \leq T_{\mathcal{D}}$ by contradiction. If $\mathcal{M}(v_q;S_k) > T_{\mathcal{D}}$, then $\mathcal{D}_{v_q} > T_{\mathcal{D}}$ and $v_q$ should be considered in building $S'_k$. But for the $q$-th seed $u_q$, $\mathcal{M}(u_q;S'_k) \leq \mathcal{D}(\{u_1,u_2,..u_{q-1}\} \cup \{v_q\})-\mathcal{D}(\{u_1,u_2,..u_{q-1}\})$, which contradicts the fact that $u_q$ has the largest marginal gain in the $q$-th iteration of building $S'_k$.
	
	Since $\mathcal{M}(v_q;S_k) \leq T_{\mathcal{D}}$, we have $\mathcal{D}(S_k)=\mathcal{D}(\{u_1,...,u_{q-1}\})+\sum_{i=t}^k{\mathcal{M}(v_i;S_k)}$. According to the submodularity of $\mathcal{D}(S)$ and the greedy algorithm, it is easy to verify that $\mathcal{M}(v_q;S_k) \geq \mathcal{M}(v_j;S_k)$ for $j \geq q$. Thus, we have $\mathcal{D}(S_k) \leq \mathcal{D}(\{u_1,...,u_{q-1}\}) + (k-q+1)\mathcal{M}(v_q;S_k) \leq \mathcal{D}(\{u_1,...,u_{q-1}\})  + (k-q+1)T_{\mathcal{D}}$. Because $\{u_1,...,u_{q-1}\} \subseteq S'_k$, we have $\mathcal{D}(S'_k) \geq \mathcal{D}(\{u_1,...,u_{q-1}\}) \geq \mathcal{D}(S_k)-(k-q+1)T_{\mathcal{D}}$.
\\

\noindent \textbf{Proof of Corollary~\ref{cor:filter}} First, $T_\mathcal{D}=\min\{\frac{\mathcal{D}^*}{k-1},\mathcal{D}^*-1\}$ ensures that $u$ is always considered in building $S'_k$ if $\mathcal{D}(u)=\mathcal{D}^*$. Also, $T_\mathcal{D}=\min\{\frac{\mathcal{D}^*}{k-1},\mathcal{D}^*-1\} < \mathcal{D}^*$. So we have $q \geq 2$, where $t$ is the first time that a seed added to $S'_k$ has a marginal gain no greater than $T_\mathcal{D}$. Thus, if $k=1$, $\mathcal{D}(S'_k)=\mathcal{D}(S_k)$. When $k=2$, $\mathcal{D}(S'_k) \geq \mathcal{D}^*$ and $\mathcal{D}(S_k) \leq 2\mathcal{D}^*$ so $\mathcal{D}(S'_k) \geq \frac{1}{2}\mathcal{D}(S_k)$. When $k \geq 3$, $T_\mathcal{D}=\min\{\frac{\mathcal{D}^*}{k-1},\mathcal{D}^*-1\}=\frac{\mathcal{D}^*}{k-1}$ because $\mathcal{D}^* \geq 2$. Therefore, $\frac{\mathcal{D}(S'_k)}{\mathcal{D}(S_k)} \geq \frac{\mathcal{D}(S'_k)}{\mathcal{D}(S'_k)+(k-q+1)\frac{\mathcal{D}^*}{k-1}} \geq \frac{\mathcal{D}^*}{\mathcal{D}^*+(k-1)\frac{\mathcal{D}^*}{k-1}}=\frac{1}{2}$.
\\

\noindent \textbf{Remark of mistakes in~\cite{dinh2015social}} The major error made by~\cite{dinh2015social} is that the union bound is missed when bounding $\textup{Pr}\{\frac{n\mathcal{D}(S_k)}{M}  \leq (1+\epsilon_1)I(S_k)\}$. In~\cite{dinh2015social}, when the sampling phase ends, it is guaranteed that $\mathcal{D}(S_k) \geq \lceil \Upsilon_1(\epsilon_1,\delta) \rceil$, where $\epsilon_1=\frac{\epsilon}{2(1-1/e)-\epsilon}$. Dinh~\textit{et~al.}~\cite{dinh2015social} made a claim that $\mathcal{D}(S_k) \geq \lceil \Upsilon_1(\epsilon_1, \delta) \rceil$ implies that $\textup{Pr}\{\frac{n\mathcal{D}(S_k)}{M}  \leq (1+\epsilon_1)I(S_k)\} \geq 1-\frac{\delta}{2}$. However, we cannot directly apply Theorem~\ref{th:sample} on $S_k$ to get this conclusion, because $S_k$ is deliberately picked by the greedy algorithm, where multiple candidate sets are involved. To better understand this issue, let us recall our proof of Lemma~\ref{lemma:lower} and Lemma~\ref{lemma:upper}. In our proof, no matter the sampled RR sets are, we always look at every vertex one by one and then apply the union bound. But $S_k$ is deliberately picked by the greedy algorithm on $\mathcal{R}$, it is easy to find that $S_k$ depends on the sampled RR sets. This introduces extra uncertainty. Also, this issue is similar to the overfitting issue in machine learning~\cite{mohri2012foundations}. We relate the sampled RR sets $\mathcal{R}$ to the training data, $S$ to a classifier and $\frac{\mathcal{D}(S)}{M}$ to $S$'s accuracy on $\mathcal{R}$. Then the parameter of a classifier $S$ is the vertices in $S$. The parameter learning process (the greedy algorithm) returns only one $S$ but it involves multiple other candidate sets. If the number of candidate sets (corresponds to the size of parameter space in machine learning) is huge, picking the parameters (a seed set $S_k$) that has a very high accuracy on the training data may lead to overfitting.
\\

\end{document}